\documentclass[10pt,journal,compsoc]{IEEEtran}

\ifCLASSOPTIONcompsoc
  \usepackage[nocompress]{cite}
\else
  \usepackage{cite}
\fi

\usepackage{graphicx}
\usepackage{amsmath}
\usepackage{amssymb}
\usepackage{url}
\usepackage{hyperref}
\usepackage{booktabs}
\usepackage{multirow}
\usepackage{array}
\usepackage{adjustbox}
\usepackage{float}
\usepackage{xcolor}
\usepackage{colortbl}
\usepackage{tikz}
\usetikzlibrary{positioning}
\usetikzlibrary{shapes.geometric}
\usepackage{comment}
\usepackage{stfloats}

\renewcommand{\arraystretch}{1.3}
\DeclareMathSizes{5.7}{5.7}{5}{5}

\definecolor{colR}{RGB}{245,158,11}
\definecolor{colI}{RGB}{239,68,68}
\definecolor{colS}{RGB}{59,130,246}

\graphicspath{{./}}
\begin{document}
\setcounter{secnumdepth}{3}

\title{Finding Patient Zero in Temporal Contact Networks: A Comparative Study}

\author{Nathan~Cincera~Brack, Martin~Sterchi, Lorenz~Hilfiker, and~Abraham~Bernstein%
\IEEEcompsocitemizethanks{%
\IEEEcompsocthanksitem N.~Cincera~Brack is with the Department of Informatics, University of Zurich, Zurich, Switzerland, and also with the University of Applied Sciences and Arts Northwestern Switzerland, Olten, Switzerland (e-mail: \mbox{nathan.brack@uzh.ch}).
\IEEEcompsocthanksitem M.~Sterchi and L.~Hilfiker are with the University of Applied Sciences and Arts Northwestern Switzerland, Olten, Switzerland (e-mail: \mbox{martin.sterchi@fhnw.ch}).
\IEEEcompsocthanksitem A.~Bernstein is with the Department of Informatics, University of Zurich, Zurich, Switzerland.}%
}

\IEEEtitleabstractindextext{%
\begin{abstract}
Identifying the source of an epidemic, i.e., the patient zero, in a temporal contact network is important for effective outbreak containment and resource allocation. While several methods have been proposed for this task, they have rarely been evaluated under comparable conditions, and no systematic comparison exists for temporal contact networks. We address this gap with a comparative study of seven major methods, namely Steiner Tree~(ST), Individual-Based Approximation~(IBA), Dynamic Message Passing~(DMP), Monte Carlo Mean-Field~(MCMF), Soft Margin Estimator~(SME), Backtracking Network~(BN, a graph neural network), and Belief Propagation~(BP). We unify all methods under a single problem formulation and evaluate them on eight diverse temporal networks. In the main experiment, we compare source detection performance for varying basic reproduction numbers. Three complementary experiments on a subset of networks study (i) learning curves, (ii) partial observability, and (iii) parameter misspecification. We find that BP and BN are the strongest source detectors overall. SME is strong under partial observation but mediocre otherwise. IBA, DMP, and MCMF share a common weakness rooted in their factorized-likelihood assumption, leading to overconfident predictions. We identify scaling these methods to larger networks as the most pressing direction for future work.
\end{abstract}

}

\IEEEpubid{\parbox{\textwidth}{\centering\footnotesize This work has been submitted to the IEEE for possible publication. Copyright may be transferred without notice, after which this version may no longer be accessible.}}
\maketitle

\IEEEdisplaynontitleabstractindextext
\IEEEpeerreviewmaketitle

\section{Introduction}

Identifying the source of an epidemic in a contact network is a methodological challenge of great practical importance. Locating the patient zero enables health authorities to reconstruct transmission chains, allocate containment resources, and prevent future outbreaks~\cite{bradshawBidirectionalContactTracing2021, endoImplicationBackwardContact2020}. The challenge can be formulated analogously in other domains, such as computer virus forensics~\cite{shahDetectingSourcesComputer2010} or misinformation tracing~\cite{sattarovDetectionRumorsTheir2025}. In all these domains, a single undetected source can cause damage across large parts of the network.

Early studies focused on static networks where no temporal information about contacts is available and proposed centrality-based measures as a first solution, such as rumor centrality~\cite{shahDetectingSourcesComputer2010} (derived via maximum likelihood under a tree assumption) or Jordan centrality~\cite{luoHowIdentifyInfection2014}. Since then, source detection has received considerable attention and undergone development along two parallel lines of research.

The first line has moved the state of the art well beyond centrality measures toward more advanced approaches, several of which also apply to temporal networks. These include the Steiner Tree~(ST) method~\cite{rozenshteinReconstructingEpidemicTime2016}, Individual-Based Approximation~(IBA)~\cite{rochaIndividualbasedApproachEpidemic2016}, Dynamic Message Passing~(DMP)~\cite{lokhovInferringOriginEpidemic2014, koherContactBasedModelEpidemic2019}, Belief Propagation~(BP)~\cite{braunsteinInferenceCausalityEpidemics2016}, and Monte Carlo simulation-based approaches, namely Monte Carlo Mean-Field~(MCMF)~\cite{sterchiActiveQueryingApproach2023} and the Soft Margin Estimator~(SME)~\cite{antulov-fantulinIdentificationPatientZero2015a}. More recently, graph neural networks have been proposed as a deep learning alternative, using Monte Carlo simulations as training data~\cite{sterchiGraphNeuralNetworks2025, ruInferringPatientZero2023, dongMultipleRumorSource2019, shaSourceDetectionNetworks2021}. Among them, the Backtracking Network~(BN)~\cite{ruInferringPatientZero2023} was specifically developed for temporal networks.

In the second line of research, studies have increasingly used temporal networks, where timestamps of contacts are explicitly recorded~\cite{holmeTemporalNetworks2012, holmeModernTemporalNetwork2015, holmeTemporalNetworkTheory2023}. Timestamps carry important information about epidemic spreading and constrain which transmission paths are possible: for example, if a contact between A and B precedes a contact between B and C, a pathogen can propagate from A to C via B, but not in the reverse direction. Taking into account the temporal order of contacts often captures disease propagation more realistically, and several source detection methods have been developed specifically for temporal networks.

However, a systematic comparison of the proposed methods on temporal networks is still lacking. Such a comparative evaluation was conducted for source detection methods on \textit{static} networks nearly a decade ago~\cite{jiangIdentifyingPropagationSources2017}. More recently, Sterchi et al.~\cite{sterchiGraphNeuralNetworks2025} reviewed graph neural network approaches for source detection on static networks. However, no comparable assessment exists for \textit{temporal} networks. 

The lack of such a study has two major consequences for the field. First, methods are rarely evaluated on the same temporal networks, as Table~\ref{tab:used_networks} illustrates, let alone on a sufficiently large or diverse set of networks. As a consequence, results may reflect idiosyncratic network characteristics rather than general performance. The contact network of sex workers and their customers in Brazil is the most frequently used benchmark, with four methods evaluated on it; all other networks appear in at most one or two studies. We refer to this issue as \textit{heterogeneous network choices}.

\begin{table*}[t]
\centering
\caption{Comparative overview of source detection methods and the empirical datasets used for their evaluation.}
\label{tab:used_networks}
\small
\begin{tabular}{lcccccccc}
\hline
 & \begin{tabular}[b]{@{}c@{}}Brazil\\Escort\end{tabular} & \begin{tabular}[b]{@{}c@{}}Hospital\\Ward\end{tabular} & \begin{tabular}[b]{@{}c@{}}Malawi\\Village\end{tabular} & \begin{tabular}[b]{@{}c@{}}Scientific\\Conference\end{tabular} & \begin{tabular}[b]{@{}c@{}}High-\\school\end{tabular} & \begin{tabular}[b]{@{}c@{}}German\\Livestock\end{tabular} & \begin{tabular}[b]{@{}c@{}}Swiss\\Pig\end{tabular} & \begin{tabular}[b]{@{}c@{}}Others$^*$\end{tabular} \\ \hline
Rozenshtein et al. (2016)~\cite{rozenshteinReconstructingEpidemicTime2016} &  &  &  &  &  &  &  & $\times$ \\
Rocha \& Masuda (2016)~\cite{rochaIndividualbasedApproachEpidemic2016} &  & $\times$ &  & $\times$ &  &  &  &  \\
Koher et al. (2019)~\cite{koherContactBasedModelEpidemic2019} &  &  &  & $\times$ &  & $\times$ &  &  \\
Sterchi et al. (2023)~\cite{sterchiActiveQueryingApproach2023} & $\times$ &  & $\times$ &  &  &  & $\times$ &  \\
Antulov-Fantulin et al. (2015)~\cite{antulov-fantulinIdentificationPatientZero2015a} & $\times$ &  &  &  &  &  &  &  \\
Braunstein \& Ingrosso (2016)~\cite{braunsteinInferenceCausalityEpidemics2016} & $\times$ &  &  &  & $\times$ &  &  &  \\
Ru et al. (2023)~\cite{ruInferringPatientZero2023} & $\times$ & $\times$ &  &  &  &  &  & $\times$ \\
\hline
\multicolumn{9}{p{0.92\textwidth}}{\footnotesize$^*$Multiple networks without physical contact: both Ru et al.~\cite{ruInferringPatientZero2023} and Rozenshtein et al.~\cite{rozenshteinReconstructingEpidemicTime2016} use an online-messaging network; Rozenshtein et al.\ further use Facebook posts and Tumblr quotes; Ru et al.\ further use European email and a Bitcoin trust network.} \\ \hline
\end{tabular}
\end{table*}

Second, baseline selection is inconsistent across studies: some studies compare the proposed methods only against simple heuristics, or against a weak random-node-selection baseline. Others benchmark against more competitive methods such as MCMF, IBA, or DMP, but do so selectively. Specifically, Rozenshtein et al.~\cite{rozenshteinReconstructingEpidemicTime2016} compare against a one-hop cascade baseline; Rocha and Masuda~\cite{rochaIndividualbasedApproachEpidemic2016}, Sterchi et al.~\cite{sterchiActiveQueryingApproach2023}, and Antulov-Fantulin et al.~\cite{antulov-fantulinIdentificationPatientZero2015a} each compare only against a random baseline; Koher et al.~\cite{koherContactBasedModelEpidemic2019} benchmark against IBA; Ru et al.~\cite{ruInferringPatientZero2023} compare against centrality measures and SME; and Braunstein and Ingrosso~\cite{braunsteinInferenceCausalityEpidemics2016} include no benchmark at all. We refer to this second issue as \textit{inconsistent baseline selection}.

Beyond these two core issues, studies also differ substantially in problem formulations, underlying assumptions, and choices of epidemic parameters, complicating direct comparisons even when results on the same networks are available. Computational efficiency and algorithmic complexity, central concerns for practical deployment, also receive insufficient attention in the literature. Together, these issues make it hard to draw reliable conclusions about the relative merits of the proposed approaches.

In this paper, we address this gap by conducting a systematic comparison of major source detection methods on \textit{temporal} networks. We evaluate ST, IBA, DMP, MCMF, SME, BN, and BP under a unified problem formulation and a common experimental framework. Specifically, our contributions are: (i) a unified problem formulation that makes all methods directly comparable; (ii) an analysis of methodological differences, underlying assumptions, and algorithmic complexity; (iii) experiments across networks of varying size and density, assessing performance as a function of the basic reproduction number $R_0$ and outbreak size; and (iv) three complementary experiments studying (a) learning curves, i.e., performance as a function of the amount of training data, (b) performance when only part of the snapshot is observed, and (c) robustness to misspecified epidemic parameters.

The remainder of this paper is structured as follows. Section~\ref{sec:background} introduces the mathematical background. Section~\ref{sec:problem} formally states the source detection problem. Section~\ref{sec:methods} presents the baselines, followed by each of the methods, including an analysis of their assumptions and algorithmic complexity. Section~\ref{sec:comparative} describes the comparative study, including the networks used, the experimental setup for the main experiment as well as the three complementary experiments, and the evaluation metrics. Section~\ref{sec:results} presents the results of all four experiments. Section~\ref{sec:discussion} discusses the findings, their implications, and the limitations of this study. Finally, Section~\ref{sec:conclusion} concludes the paper.

\section{Background}
\label{sec:background}

\paragraph{\textit{Temporal networks}} We are given a temporal contact network, that is, a graph $G_T = (V, E_T)$ with timestamped edges $(u, v, t) \in E_T$, where $u, v \in V$ and $t \in T$ is a timestamp, with $T \subseteq \mathbb{N}_0$ a finite, ordered set of non-negative integers. We assume \textit{undirected} timestamped edges; however, the Monte Carlo simulations of epidemic outbreaks as well as the source detection methods support directed networks as well. By convention, the nodes (individuals) are enumerated as integers, i.e.,\ $V = \{0, 1, 2, \cdots, |V|-1\} \subset \mathbb{N}_0$, where $|V|$ is the number of nodes, and the starting time of the contact network is $\min (T) = 0$.

\paragraph{\textit{Static projections}} We focus primarily on temporal networks, but for some of the presented methods, static projections of the temporal networks are needed. We define the static projection $G = (V, E)$ of a natively temporal network as a standard static graph with edge set

\[ E = \bigcup_{t \in T} \{ (u,v) \mid (u,v,t) \in E_T \}, \]
i.e., taking the unique set of edges that result from ignoring the temporal information.

\paragraph{\textit{Spreading models}} The epidemic dynamics on a temporal network are typically modeled using compartmental spreading models, which specify how nodes transition between discrete states. We consider the SIR model, a direct extension of the SI model, both of which are commonly adopted choices for source detection. In the SI model, each node is either susceptible ($S$) or infected ($I$). At each discrete time step, an infected node can transmit the disease to a susceptible neighbor if there exists a contact. In the SI model, an infected node remains in this state indefinitely. The SIR model extends this by introducing a recovered state ($R$): an infected node has an independent chance to recover at each time step, after which it is immune to reinfection.

\paragraph{\textit{Observed snapshot}} In addition to the temporal contact network, we observe a snapshot $\mathbf{x} = \bigl(x_v\bigr)_{v\in V}$ of epidemic states $x_v \in \{S, I, R\}$ at time $T_1 > T_0$, where $T_0 \geq 0$ denotes the start of the epidemic.\footnote{Note that $T_0, T_1 \in T$ are scalars. We use capitalized letters rather than $t_0$ and $t_1$, since the subscripted notation $t_v$ is reserved for the infection time of node $v$ in BP (Section~\ref{sec:bp}).} The snapshot may be complete, meaning the state of every node $v \in V$ is known, or partial, meaning only the states of a subset of nodes are observed. In the latter case, $x_v \in \{S, I, R, U\}$, where $U$ denotes an unobserved state.
Such a snapshot is obtained from a simulation that proceeds as follows: at $T_0$, all nodes except the source are susceptible. At each subsequent time step $t \leq T_1$, an infected node $v$ transmits the infection to a susceptible neighbor $u$ with probability $\beta$ if a contact $(u,v,t) \in E_T$ occurs. In the SIR model, each infected node additionally recovers at each time step with probability $\mu$.

\section{Problem Statement}
\label{sec:problem}

The \emph{source detection problem} can be stated as follows: given a temporal contact network $G_T = (V, E_T)$ and an epidemic snapshot $\mathbf{x}$, identify the true source $s \in V$ of the epidemic. More generally, this is an instance of a node-selection task: a multi-class classification task, where each class corresponds to one node. 

Since a given snapshot is typically compatible with many candidate source nodes, the true source is generally not uniquely identifiable. Consider, for instance, a network of just two people who were in contact and both end up infected: given only the snapshot, there is no way of knowing who was infected first. Thus, rather than best-guessing a \textit{single} node $v \in V$, many methods take a probabilistic approach and quantify how likely each node is to be the source. The goal is therefore to approximate the true underlying posterior distribution $p_{\mathbf{x}}(v) = P(s = v \mid \mathbf{x})$ as closely as possible.

Thus, given a snapshot $\mathbf{x}$, a method typically returns an estimated \emph{probability distribution} $\hat{p}_{\mathbf{x}}$ over candidate source nodes $v \in V$,
\[
    \hat{p}_{\mathbf{x}}(v) = \hat{P}(s = v \mid \mathbf{x}), \qquad \sum_{v \in V} \hat{P}(s = v \mid \mathbf{x}) = 1.
\]
Alternatively, a method may return an estimated \emph{likelihood} $\mathcal{L}$ of observing the snapshot $\mathbf{x}$, given that candidate $v \in V$ is the source, i.e.,
\[
    \mathcal{L}(\mathbf{x}, v) = \hat{P}(\mathbf{x} \mid s = v),
\]
or, similarly, a \emph{log-likelihood}~$\ell$, i.e.,
\[
    \ell(\mathbf{x}, v) = \log \hat{P}(\mathbf{x} \mid s = v).
\]
To keep the notation uncluttered, we do not explicitly condition on the temporal network and the spreading model in all mathematical expressions. Using Bayes' theorem and assuming a uniform prior, the likelihood can be transformed into a probability distribution by normalization.

Methods lacking a probabilistic interpretation altogether (only the centrality measures and ST, cf.\ Sections~\ref{sec:baselines} and~\ref{sec:steiner-tree}) may output any quantity that induces a ranking over $V$, such as a real-valued cost $\hat{f}(v) \in \mathbb{R}$. We use the convention that a lower cost indicates a more likely source.

\section{Methods}
\label{sec:methods}

This section presents the key methods that have been proposed in the literature to solve the source detection problem on temporal networks, beginning with the baselines. Table~\ref{tab:methods} gives an overview of core assumptions, key features, and algorithmic complexity of each method.

\begin{table*}[t]
\centering
\caption{Overview of source detection methods: core assumptions, key properties, and algorithmic complexity. For MCMF and SME, the cost of simulating the training data dominates the cost of inference, so the reported complexity is that of the simulations. For BN, training the network computationally dominates both data simulation and inference, so only the training cost is reported. Here, $K$ = number of training snapshots per node ($K \cdot |V|$ total), $L$ = number of GNN layers, $d_h$ = hidden dimension, $M$ = number of training epochs. For BP, $R$ = number of iterations until convergence.}
\label{tab:methods}
\small
\setlength{\tabcolsep}{3pt}
\begin{tabular}{p{1.5cm}p{7.7cm}ccccp{4.6cm}}
\hline
\textbf{Method}
  & \textbf{Core Idea and Assumptions}
  & \textbf{SIR$^{\mathrm{a}}$}
  & \textbf{FL$^{\mathrm{b}}$}
  & \textbf{MC$^{\mathrm{c}}$}
  & \textbf{TE$^{\mathrm{d}}$}
  & \textbf{Algorithm Complexity} \\ \hline
ST~\cite{rozenshteinReconstructingEpidemicTime2016}
  & Non-probabilistic, purely graph-based: the source is the root of a minimum temporal Steiner tree.
  & & & & $\times$ & $\mathcal{O}(|V| \cdot |E_T|)$ \\[4pt]
IBA~\cite{rochaIndividualbasedApproachEpidemic2016}
  & Update equations for node state probabilities, assuming that the infections of a node by different neighbors occur independently.
  & $\times$ & $\times$ & & & $\mathcal{O}(|V| \cdot |E_T|) \subseteq \mathcal{O}(|V| \cdot |T| \cdot |E|)$ \\[4pt]
DMP~\cite{koherContactBasedModelEpidemic2019}
  & Temporal adaptation of Lokhov et al.~\cite{lokhovInferringOriginEpidemic2014}: node state probabilities via message passing, assuming the branches around a cavity node are independent.
  & $\times$ & $\times$ & & $\times$ & $\mathcal{O}(|T| \cdot |E|^2) \subseteq \mathcal{O}(|V|^2 \cdot |T| \cdot |E|)$ \\[4pt]
MCMF~\cite{sterchiActiveQueryingApproach2023}
  & Estimates node state probabilities from simulated outbreaks; no assumptions beyond SIR and FL.
  & $\times$ & $\times$ & $\times$ & & \parbox[t]{4.5cm}{$\mathcal{O}\!\left(K \cdot |V| \cdot |E| \cdot \log \frac{|E_T||V|}{|E|}\right)$\\$\subseteq \mathcal{O}\!\left(K \cdot |V| \cdot |E| \cdot \log(|T||V|)\right)$} \\[4pt] \noalign{\vskip 6pt}
SME~\cite{antulov-fantulinIdentificationPatientZero2015a}
  & Scores candidates by snapshot similarity, assuming similar outbreaks stem from the same source. Designed for SI; discards recovery information.
  & $(\times)$ & & $\times$ & & \parbox[t]{4.5cm}{$\mathcal{O}\!\left(K \cdot |V| \cdot |E| \cdot \log \frac{|E_T||V|}{|E|}\right)$\\$\subseteq \mathcal{O}\!\left(K \cdot |V| \cdot |E| \cdot \log(|T||V|)\right)$} \\[4pt]
BN~\cite{ruInferringPatientZero2023}
  & GNN trained on the static projection $G$; temporal contacts encoded as binary per-timestep edge features.
  & & & $\times$ & & \parbox[t]{4.5cm}{$\mathcal{O}\!\left(M \cdot K \cdot |V| \cdot |E| \cdot \right.$\\$\hphantom{\mathcal{O}\!(}\,(|T| \cdot d_h + L \cdot d_h^2))$} \\[4pt]
BP~\cite{braunsteinInferenceCausalityEpidemics2016}
  & Belief propagation over individual infection times; no assumptions beyond SIR. Explicit source prior.
  & $\times$ & & & $\times$ & $\mathcal{O}(R \cdot |T| \cdot |E|)$ \\[4pt]
  \hline
\multicolumn{7}{p{\dimexpr\textwidth-2\tabcolsep\relax}}{
  $^{\mathrm{a}}$ \textit{SIR} indicates whether the method assumes an SIR spreading model; parentheses mark partial applicability (SME assumes SI and discards recovery information when applied to SIR data).\newline
  $^{\mathrm{b}}$ \textit{Factorized likelihood (FL)} assumes node states to be conditionally independent given the source, so that the likelihood factorizes over nodes. In practice, this is usually violated: frequently interacting nodes tend to be co-infected.\newline
  $^{\mathrm{c}}$ \textit{Monte Carlo (MC)} indicates that the method requires Monte Carlo simulated outbreaks as data, either as training data (BN and other deep learning approaches) or to empirically estimate node state probabilities (MCMF) or outbreak similarity (SME).\newline
  $^{\mathrm{d}}$ \textit{Tree-exact (TE)} indicates that the method's approximation becomes exact on tree-structured networks, i.e., that no approximation error remains relative to the quantity the method is designed to compute.} \\ \hline
\end{tabular}
\end{table*}

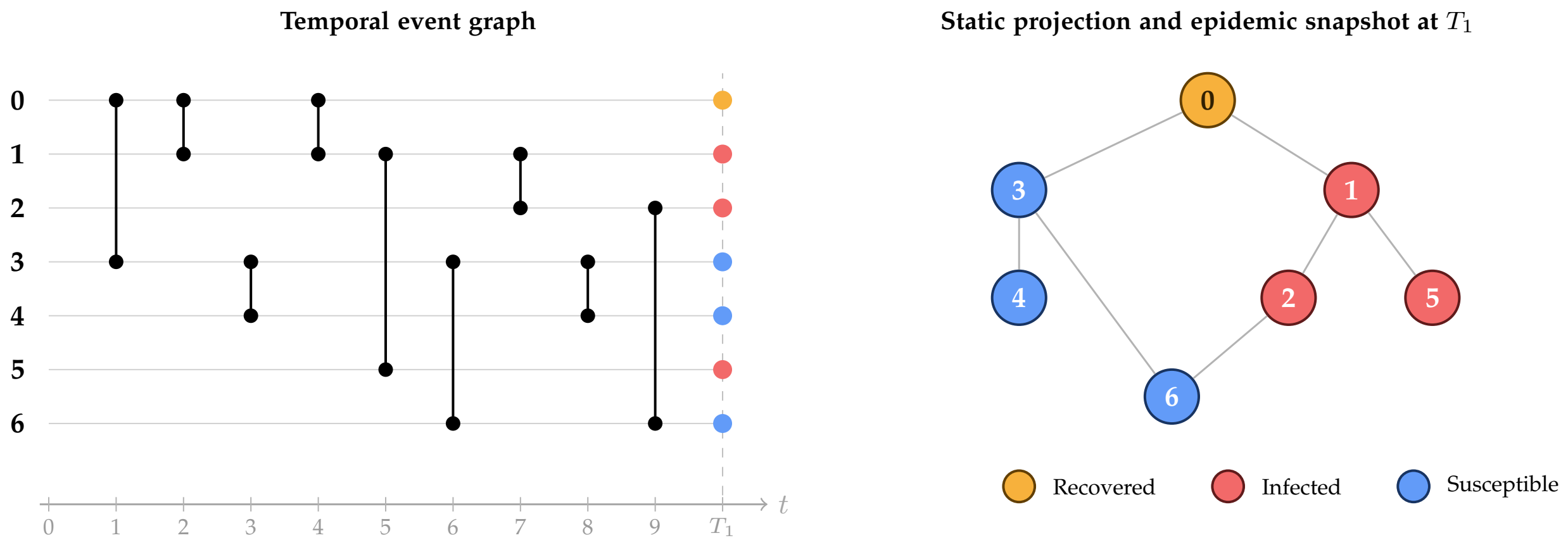
\begin{figure*}[t]
\centering
\resizebox{0.9\textwidth}{!}{%
\begin{tikzpicture}[
  snode/.style={circle, draw, thick, minimum size=0.6cm,
                inner sep=1pt, font=\small\bfseries},
  recovered/.style={snode, fill=colR!80, draw=colR!40!black, text=colR!20!black},
  infected/.style={snode,  fill=colI!80, draw=colI!40!black, text=white},
  susceptible/.style={snode, fill=colS!80, draw=colS!40!black, text=white},
]

\begin{scope}[xshift=10cm]
  \node[above] at (2.9,4.6) {\footnotesize\textbf{Static projection and epidemic snapshot at $T_1$}};

  \node[recovered]   (n0) at (2.9, 4.0) {0};
  \node[infected]    (n1) at (4.5, 3.0) {1};
  \node[infected]    (n2) at (3.8, 1.8) {2};
  \node[susceptible] (n3) at (0.8, 3.0) {3};
  \node[susceptible] (n4) at (0.8, 1.8) {4};
  \node[infected]    (n5) at (5.4, 1.8) {5};
  \node[susceptible] (n6) at (2.5, 0.7) {6};

  \draw[semithick, gray!60] (n0) -- (n1);
  \draw[semithick, gray!60] (n1) -- (n2);
  \draw[semithick, gray!60] (n0) -- (n3);
  \draw[semithick, gray!60] (n3) -- (n4);
  \draw[semithick, gray!60] (n1) -- (n5);
  \draw[semithick, gray!60] (n2) -- (n6);
  \draw[semithick, gray!60] (n3) -- (n6);

  \node[recovered, minimum size=0.35cm, inner sep=0pt]
    (lR) at (0.8, -0.3) {};
  \node[right=2pt of lR, font=\scriptsize] (lRtext) {Recovered};
  \node[infected, minimum size=0.35cm, inner sep=0pt, right=14pt of lRtext]
    (lI) {};
  \node[right=2pt of lI, font=\scriptsize] (lItext) {Infected};
  \node[susceptible, minimum size=0.35cm, inner sep=0pt, right=14pt of lItext]
    (lS) {};
  \node[right=2pt of lS, font=\scriptsize] {Susceptible};
\end{scope}

\begin{scope}[xshift=0cm]
  \node[above] at (4.0,4.6) {\footnotesize\textbf{Temporal event graph}};

  \pgfmathsetmacro{\tscale}{0.75}

  \foreach \nd/\y in {0/4.0, 1/3.4, 2/2.8, 3/2.2, 4/1.6, 5/1.0, 6/0.4}{
    \node[font=\small\bfseries, anchor=east] at (-0.15, \y) {\nd};
    \draw[gray!35, thin] (0, \y) -- (10*\tscale+0.1, \y);
  }

  \draw[->, semithick, gray!70] (-0.1, -0.5)
    -- (10*\tscale + 0.5, -0.5)
    node[right, font=\small] {$t$};
  \foreach \t in {1,...,9}{
    \draw[gray!60] (\t*\tscale, -0.42) -- (\t*\tscale, -0.58);
    \node[font=\scriptsize, gray!80] at (\t*\tscale, -0.75) {\t};
  }
  \draw[gray!60] (0, -0.42) -- (0, -0.58);
  \node[font=\scriptsize, gray!80] at (0, -0.75) {0};
  \draw[gray!60] (10*\tscale, -0.42) -- (10*\tscale, -0.58);
  \node[font=\scriptsize, gray!80] at (10*\tscale, -0.75) {$T_1$};
  \draw[dashed, gray!50, thin] (10*\tscale, -0.4) -- (10*\tscale, 4.3);

  \foreach \nd/\y/\col in {0/4.0/colR, 1/3.4/colI, 2/2.8/colI, 3/2.2/colS, 4/1.6/colS, 5/1.0/colI, 6/0.4/colS}{
    \fill[\col!80] (10*\tscale, \y) circle (3pt);
  }

  \draw[black, thick] (1*\tscale, 2.2) -- (1*\tscale, 4.0);
  \fill[black] (1*\tscale, 4.0) circle (2.3pt);
  \fill[black] (1*\tscale, 2.2) circle (2.3pt);

  \draw[black, thick] (2*\tscale, 3.4) -- (2*\tscale, 4.0);
  \fill[black] (2*\tscale, 4.0) circle (2.3pt);
  \fill[black] (2*\tscale, 3.4) circle (2.3pt);

  \draw[black, thick] (3*\tscale, 1.6) -- (3*\tscale, 2.2);
  \fill[black] (3*\tscale, 2.2) circle (2.3pt);
  \fill[black] (3*\tscale, 1.6) circle (2.3pt);

  \draw[black, thick] (4*\tscale, 3.4) -- (4*\tscale, 4.0);
  \fill[black] (4*\tscale, 4.0) circle (2.3pt);
  \fill[black] (4*\tscale, 3.4) circle (2.3pt);

  \draw[black, thick] (5*\tscale, 1.0) -- (5*\tscale, 3.4);
  \fill[black] (5*\tscale, 3.4) circle (2.3pt);
  \fill[black] (5*\tscale, 1.0) circle (2.3pt);

  \draw[black, thick] (6*\tscale, 0.4) -- (6*\tscale, 2.2);
  \fill[black] (6*\tscale, 2.2) circle (2.3pt);
  \fill[black] (6*\tscale, 0.4) circle (2.3pt);

  \draw[black, thick] (7*\tscale, 2.8) -- (7*\tscale, 3.4);
  \fill[black] (7*\tscale, 3.4) circle (2.3pt);
  \fill[black] (7*\tscale, 2.8) circle (2.3pt);

  \draw[black, thick] (8*\tscale, 1.6) -- (8*\tscale, 2.2);
  \fill[black] (8*\tscale, 2.2) circle (2.3pt);
  \fill[black] (8*\tscale, 1.6) circle (2.3pt);

  \draw[black, thick] (9*\tscale, 0.4) -- (9*\tscale, 2.8);
  \fill[black] (9*\tscale, 2.8) circle (2.3pt);
  \fill[black] (9*\tscale, 0.4) circle (2.3pt);

\end{scope}
\end{tikzpicture}}
\caption{Left: temporal event graph~\cite{saramakiWeightedTemporalEvent2019}; each vertical segment connects two nodes at the moment of a contact, with colored markers at $T_1$ showing the epidemic snapshot (recovered, infected, susceptible). Right: static projection of a temporal contact network with epidemic snapshot at $T_1$, with colors indicating node states (recovered, infected, susceptible). Temporal topology constrains valid source candidates: node~2 cannot be the true source, since the only path from node~2 to the recovered node~0 runs via node~1, but all contacts between nodes~0 and~1 occur at $t \in \{2, 4\}$, before node~1 is reachable from node~2 at $t=7$. By the same reasoning, node~5 cannot be the source. The only viable candidates are therefore node~0 and node~1, with node~0 being more likely given that it has already recovered.}
\label{fig:temporal_example}
\end{figure*}

Before we begin, we define the \emph{candidate set} $\mathcal{C} \subseteq V$ as all nodes that cannot be ruled out as the source. In a full snapshot, candidates are nodes observed to be infected or recovered, while in a partial snapshot, their states may also be unobserved. Crucially, in temporal networks (but not in a static, connected network), the candidate set can sometimes be further narrowed by temporal topology (cf.\ Figure~\ref{fig:temporal_example}). Nodes ruled out in this way are assigned zero probability or likelihood. For log-likelihood scores, this is represented by $-\infty$ by convention and for a real-valued cost $\hat{f}$ by $+\infty$. All methods introduced below operate on this candidate set and with the mentioned conventions.

\subsection{Baselines}
\label{sec:baselines}

Before introducing the source detection methods, we establish two baselines as reference points.

The most trivial baseline is the \emph{uniform baseline} (UB), which assigns equal probability to each candidate $c$,
\begin{equation}
\hat{p}_{\mathbf{x}}(c) = \frac{1}{|\mathcal{C}|} \mathbf{1}_{\{c \in \mathcal{C}\}}.
\label{eq:uniform_baseline}
\end{equation}
To be able to apply rank-based evaluation metrics such as top-$k$, we induce a ranking over $\mathcal{C}$ for the uniform baseline as well. To that end, ties are broken at random, resulting in a random permutation of $\mathcal{C}$.

More advanced baselines are given by common \emph{centrality measures}. We choose to employ Jordan Centrality (JC), which has previously been applied to the source detection problem, though only in the static case~\cite{zhuCatchEmAllLocating2017, luoHowIdentifyInfection2014, luoUniversalityJordanCenters2017}. Adapting JC to temporal networks is straightforward, since it relies only on shortest paths, which are equally well-defined on temporal networks, although with varying terminology~\cite{holmeTemporalNetworks2012}. We use \emph{shortest} time-respecting paths, i.e., those with the shortest distance in terms of the number of contacts, as opposed to \emph{fastest} ones, i.e., those with the shortest duration~\cite{scholtesHigherorderAggregateNetworks2016}. Moreover, two conventions exist in the literature: shortest paths are either restricted to the subgraph induced by infected, recovered, and unobserved nodes~\cite{shahDetectingSourcesComputer2010}, or allowed to pass through any node, including susceptible ones~\cite{luoHowIdentifyInfection2014}. We adopt the latter, since the ST method (Section~\ref{sec:steiner-tree}), which also relies on shortest paths, follows this convention~\cite{rozenshteinReconstructingEpidemicTime2016}.

Let $d(c,v)$ denote the length of the shortest temporal path from candidate $c$ to an infected or recovered node $v$, starting at $T_0$ (with $d(c,v) = \infty$ if $v$ is unreachable from $c$), and define the \emph{eccentricity} of $c$ as its largest such distance,
\begin{equation}
e(c) = \max_{v :\, x_v \in \{I, R\},\, v \neq c} d(c,v).
\label{eq:eccentricity}
\end{equation}
Temporal Jordan Centrality (TJC) then uses this eccentricity as a candidate's cost, $\hat{f}(c) = e(c)$. The so-called \textit{Jordan Center}, the node of smallest eccentricity, thus has the lowest cost.

\subsection{Steiner Tree (ST)}
\label{sec:steiner-tree}

Rozenshtein et al.~\cite{rozenshteinReconstructingEpidemicTime2016} reconstruct the epidemic as a tree connecting a candidate source to all infected or recovered nodes, and rank candidates by the total weight of this tree. More formally, let a temporal Steiner tree $\mathcal{T} \subset G_T$ rooted at $c$ be a set of contacts containing a unique time-respecting path from $c$ to every node in a given set of terminals, with no contact entering $c$~\cite{rozenshteinReconstructingEpidemicTime2016}. For candidate $c$, we then seek the temporal Steiner tree with terminals $\{v : x_v \in \{I,R\}\}$ that minimizes
\begin{equation}
\mathcal{T}_c = \operatorname*{arg\,min}_{\mathcal{T} \subset G_T} w(\mathcal{T}),
\label{eq:steiner_tree}
\end{equation}
where $w(\mathcal{T})$ denotes the weight of tree $\mathcal{T}$. We use a uniform weighting of each contact in the tree, so that $w(\mathcal{T})$ is simply the number of contacts in $\mathcal{T}$. The minimum Steiner tree $\mathcal{T}_c$ can be understood as an approximation of the transmission tree that would explain the observed infections and recoveries if $c$ were the true source.

Computing an exact minimum Steiner tree is NP-hard~\cite{rozenshteinReconstructingEpidemicTime2016}, and existing approximation algorithms remain expensive. Rozenshtein et al.\ therefore propose a computationally cheap approximation, $\hat{\mathcal{T}}_c$, simply given by the union of the shortest paths from $c$ to every node in $\{v : x_v \in \{I,R\}\}$, yielding $\hat{f}(c) = w(\hat{\mathcal{T}}_c)$.

A key advantage of the ST method is that it is spreading-model agnostic: it needs no SI or SIR assumption, unlike most other methods in this study. In fact, the ST method arguably makes the fewest assumptions of all methods presented here: given our uniform weighting scheme, its only assumption is that shorter infection paths---i.e., with fewer hops---are more likely than longer ones.

However, our formulation departs from Rozenshtein et al.'s original in four significant respects, which we make explicit here. First, they weight each contact by how far the report times of the involved nodes lie from the contact time, giving more confidence to reports close to the contact. In our setting, where all nodes are observed at once in a single snapshot at $T_1$, this weight degenerates to $T_1 - t$ and would merely favor late contacts. The authors note, however, that their formulation remains meaningful for uniform weights~\cite{rozenshteinReconstructingEpidemicTime2016}, which is the variant we adopt. Second, their formulation only distinguishes active from inactive nodes, so we treat both infected and recovered nodes as active. Third, their method is primarily used to reconstruct the epidemic's transmission tree, but also identifies the source, as their algorithm returns the one candidate source (and its associated tree) that best explains the observed infections. Instead of returning only this single best candidate, however, we retain the tree cost of every candidate $c \in \mathcal{C}$ and rank candidates by it. Fourth, their method extends to the multi-source setting via Steiner forests, which we do not use since we consider a single source only.

Furthermore, in our single-source setting, the ST method is closely related to temporal closeness centrality. Closeness centrality is defined as the inverse of \textit{farness}, where farness is defined as the sum of distances to infected or recovered nodes, i.e., $\sum_{v \,:\, x_v \in \{I,R\}} d(c,v)$. The key difference to the ST method is that, since these shortest paths can overlap, this sum counts shared contacts multiple times, once per path, whereas the Steiner tree counts them once. In other words, the Steiner tree cost is the farness minus a discount for path overlap. To be precise, this relation holds for the classical, inverse-of-sum formulation of closeness centrality, not for the harmonic-style, sum-of-inverses variant sometimes used for temporal networks, e.g., by Scholtes et al.~\cite{scholtesHigherorderAggregateNetworks2016}.

Lastly, since the computation of shortest paths dominates the rest, the complexity of our variant of ST is the same as for Temporal Jordan Centrality or temporal closeness centrality, which are likewise shortest-path based: for every candidate $c \in \mathcal{C}$, we need the shortest temporal paths from $c$ to all other nodes, which are required for the approximation $\hat{\mathcal{T}}_c$ above. With $|\mathcal{C}| = \mathcal{O}(|V|)$, this incurs a one-time cost of $\mathcal{O}(|V| \cdot |E_T|)$ time~\cite{rozenshteinReconstructingEpidemicTime2016}. Assembling each candidate's tree from these precomputed paths is cheaper, thus $\mathcal{O}(|V| \cdot |E_T|)$ is already the total complexity.

\subsection{Individual-Based Approximation (IBA)}
IBA tracks node-level probabilities for susceptibility, infection, and recovery over time. Rocha and Masuda introduced IBA for temporal networks and discrete-time epidemics and applied it to source detection~\cite{rochaIndividualbasedApproachEpidemic2016}. Crucially, for the derivation of its update equations, IBA relies on an \emph{independence assumption} of node states, as shown below.

More formally, let $S_v(t), I_v(t)$ and $ R_v(t)$ denote the probability of node $v$ to be in the respective SIR state at time $t$. As a helper function, let us introduce the probability $p_{uv}(t)$ that $v$ is not infected by $u$ at time $t$. It is given by $p_{uv}(t) = 1-\beta(1-\mu)\, I_u(t-1)$ if there is a contact between $u$ and $v$ at time $t$, i.e.,\ if $(u,v,t) \in E_T$, and $p_{uv}(t) = 1$ otherwise. 

The node state probability update equation for susceptibility is
\begin{equation}
    S_v(t)=S_v(t-1)\prod_{u\in V} p_{uv}(t).
\end{equation}
Note that the product of the $p_{uv}(t)$ is essentially taken over all neighbors $u$ of $v$ at time $t$, that is, over $\{u \mid (u,v,t)\in E_T\}$, since $p_{uv}(t)$ is set to $1$ otherwise. Here the \emph{independence assumption} is crucial: the probability factorizes into a product only if (i)~the events that each neighbor $u$ does not infect $v$ are independent of each other, and (ii)~the state of $v$ is independent of the states of its neighbors, so that $p_{uv}(t)$ can be computed from the marginal infection probability of $u$. The other node state probability update equations follow:
\begin{equation}
    I_v(t)= (1-\mu)I_v(t-1) + S_v(t-1)\left(1-\prod_{u\in V} p_{uv}(t)\right),
\end{equation}
\begin{equation}
    R_v(t)=1-S_v(t)-I_v(t).
\end{equation}
Given the start and end time of the epidemic, $T_0$ and $T_1$, these update equations are computed iteratively for $t=T_0+1, T_0+2, \dots, T_1$, for each candidate source node $c\in \mathcal{C}$. The additional superscript $c$ marks $S_v^c(t), I_v^c(t), R_v^c(t)$ as computed under the assumption that $c$ is the source, with initialization $I_c^c(T_0) = 1$ and $S_v^c(T_0) = 1$ for $v \neq c$.

Finally, for each candidate source $c$ and its estimated node state probabilities $S_v^c(T_1), I_v^c(T_1), R_v^c(T_1)$ for all $v \in V$, the likelihood $\mathcal{L}(\mathbf{x}, c)$ or log-likelihood $\ell(\mathbf{x}, c)$ of observing a given snapshot $\mathbf{x} = \bigl(x_v\bigr)_{v\in V}$ at time $T_1$ is computed. Here, a \emph{second} independence assumption is invoked so that the likelihood factorizes (just like DMP and MCMF, cf.\ Table~\ref{tab:methods}): treating the observed node states as conditionally independent given the source, it holds that $\hat{P}(\mathbf{x}\mid s = c) = \prod_{v} \hat{P}(x_v \mid s = c)$. Then, the (log-)likelihood can be computed as follows:
\begin{align*}
    \ell(\mathbf{x}, c) &= \log \hat{P}(\mathbf{x}\mid s = c) = \sum_{v \in V} \log \hat{P}(x_v \mid s = c) \\
    &= \sum_{v \in V} \Bigl(\mathbf{1}_{\{x_v=S\}} \log S_v^c(T_1) + \mathbf{1}_{\{x_v=I\}} \log I_v^c(T_1) \\
    &\qquad + \mathbf{1}_{\{x_v=R\}} \log R_v^c(T_1)\Bigr).
\end{align*}
Lastly, to establish the algorithmic complexity of IBA, let $\bar{d} = |E_T| / (|V| \cdot |T|)$ denote the average number of contacts per node and per time step. Rocha and Masuda~\cite{rochaIndividualbasedApproachEpidemic2016} report a complexity of $\mathcal{O}(|T| \cdot |V| \cdot \bar{d})$ for a single run of the update equations. For source detection, we run it once per candidate source, which adds a factor of $|V|$. This gives a total complexity of $\mathcal{O}(|V| \cdot |T| \cdot |V| \cdot \bar{d})$, which simplifies to $\mathcal{O}(|V| \cdot |E_T|)$ using $\bar{d} = |E_T| / (|V| \cdot |T|)$. Computing the likelihood adds $\mathcal{O}(|V|^2)$ for summing over all nodes per candidate source, but this is dominated by the update equations and does not change the total complexity.

\subsection{Dynamic Message Passing (DMP)}
\label{sec:dmp}

DMP takes an \emph{edge-centric} perspective, in contrast to the \emph{node-centric} view of IBA. Like IBA, it tracks node state probabilities $S_v(t), I_v(t), R_v(t)$ for each node, but additionally maintains auxiliary quantities for both directions of each edge of the static projection $G = (V, E)$. It is the explicit temporal evolution of these edge-level quantities that the word \emph{dynamic} refers to. These quantities allow correlations between neighboring node states to be captured without invoking the independence assumption of IBA. The message passing approach was originally proposed for static networks by Lokhov et al.~\cite{lokhovInferringOriginEpidemic2014} and extended to temporal networks by Koher et al.~\cite{koherContactBasedModelEpidemic2019}, which is the setting we consider here.

The central idea is the \emph{cavity construction}. When computing the probability that node $v$ avoids infection, $v$ is treated as if it cannot retransmit the disease to its neighbors (it is placed in the \emph{cavity state}). This prevents the \emph{echo chamber effect} inherent in IBA: probability can flow from an infected node $u$ to a neighbor $v$ and back to $u$, causing mutual reinfection even in a simple two-node graph and inflating infection probabilities. By suppressing the outward probability flow from $v$, different branches of $G$ remain independent as long as $v$ is susceptible. The approximation is exact on temporal networks whose static projection $G$ is a tree.

The key auxiliary quantity in DMP is $\theta_{u \to v}(t)$, defined for each directed edge $(u, v) \in E$ as the probability that $u$ has not transmitted the disease to $v$ up to time $t$, with initial condition $\theta_{u \to v}(T_0) = 1$. Node-level susceptibility is then obtained by factorizing over incoming edges:
\begin{equation}
    S_v(t) = S_v(T_0) \prod_{u \in V} \theta_{u \to v}(t),
    \label{eq:S_node}
\end{equation}
where the product is effectively over incoming neighbors of $v$ in $G$, since $\theta_{u \to v}(t) = 1$ for $(u,v) \notin E$. Under the tree assumption on $G$ this factorization is exact; on graphs with loops it serves as an approximation.

To describe how $\theta_{u \to v}(t)$ evolves, we note that it decreases only upon a contact $(u,v,t) \in E_T$ if $u$ is infected but has not yet transmitted to $v$. We therefore introduce a second auxiliary quantity $I_{u \to v}(t)$, defined as the probability that $u$ is infected at time $t$ and has not transmitted the disease to $v$ at any prior time step, which gives:
\begin{equation}
    \theta_{u \to v}(t+1) = \theta_{u \to v}(t)
    - \beta \cdot \mathbf{1}_{\{(u,v,t) \in E_T\}} \cdot I_{u \to v}(t).
    \label{eq:theta}
\end{equation}
To close the system, we introduce a third auxiliary quantity $S_{u \to v}(t)$, which is the probability that $u$ is susceptible under the assumption that $v$ is in the cavity state (but \emph{not} the marginal probability $S_u$ that $u$ is susceptible). With initial condition $I_{u \to v}(T_0) = 1 - S_u(T_0)$, the full set of update equations is (see~\cite{koherContactBasedModelEpidemic2019} for a detailed derivation):
\begin{equation}
    S_{u \to v}(t+1) = S_u(T_0)
    \prod_{w \in V \setminus \{v\}} \theta_{w \to u}(t+1),
    \label{eq:S_cavity}
\end{equation}
\begin{align}
    I_{u \to v}(t+1) &=
    (1-\mu)\bigl[1 - \beta \cdot \mathbf{1}_{\{(u,v,t) \in E_T\}}\bigr]
    \cdot I_{u \to v}(t) \notag \\
    &\quad + S_{u \to v}(t) - S_{u \to v}(t+1).
    \label{eq:I_cavity}
\end{align}
Once again, note that the product in Eq.~\eqref{eq:S_cavity} runs essentially over all neighbors of $u$ in $G$ since $\theta_{w \to u}(t) = 1$ for $(w,u) \notin E$, but excluding the cavity node $v$.

Plugging Eq.~\eqref{eq:S_cavity} into Eq.~\eqref{eq:I_cavity}, the system is fully determined by the update equations for $\theta_{u \to v}$ and $I_{u \to v}$ alone, Eqs.~\eqref{eq:theta} and \eqref{eq:I_cavity}. This yields $2 \times 2|E|$ update equations per time step: two auxiliary quantities for each of the $2|E|$ directed edges in $G$, with each node acting as the cavity node for its incoming edges. Finally, the node-level state probabilities are obtained from the auxiliaries. The susceptibility $S_v(t)$ is already given by Eq.~\eqref{eq:S_node}, complemented by:
\begin{equation}
    R_v(t+1) = R_v(t) + \mu I_v(t),
\end{equation}
\begin{equation}
    I_v(t+1) = 1 - S_v(t+1) - R_v(t+1).
\end{equation}
As for IBA, these equations are iterated with initial conditions for each candidate source $c \in \mathcal{C}$ with $I_c^c(T_0) = 1$ and $S_v^c(T_0) = 1$ for $v \neq c$. The log-likelihood of the observed snapshot $\mathbf{x}$ at $T_1$ is again computed via the factorized product over node states.

Lastly, regarding algorithmic complexity, let $\tilde{d} = 2|E| / |V|$ be the average degree of a node in the static projection $G$. Since the product in Eq.~\eqref{eq:S_cavity}, at a cost of $\mathcal{O}(\tilde{d})$ per edge, dominates the auxiliary updates (and the auxiliary updates dominate both the computation of the node state probabilities and the likelihood computation), one full run of DMP for a candidate $c$ costs $\mathcal{O}(|T| \cdot |E| \cdot \tilde{d}) = \mathcal{O}(|T| \cdot |E|^2 / |V|)$. This yields a total complexity, over all $|\mathcal{C}| = \mathcal{O}(|V|)$ candidate sources, of $\mathcal{O}(|T| \cdot |E|^2)$.

\subsection{Monte Carlo Mean-Field (MCMF)}
\label{sec:mcmf}

MCMF, like IBA and DMP, computes individual node state probabilities $S_v^c(T_1)$, $I_v^c(T_1)$, and $R_v^c(T_1)$ at time $T_1$, but does so by averaging over $K$ simulated snapshots $\mathbf{x}^{c,k}$ initiated from candidate source $c$ at time $T_0$~\cite{sterchiActiveQueryingApproach2023}:
\begin{align*}
  S_v^c(T_1) &= \frac{1}{K} \sum_{k=1}^{K} \mathbf{1}_{\{x_v^{c,k} = S\}}, \\
  I_v^c(T_1) &= \frac{1}{K} \sum_{k=1}^{K} \mathbf{1}_{\{x_v^{c,k} = I\}}, \\
  R_v^c(T_1) &= \frac{1}{K} \sum_{k=1}^{K} \mathbf{1}_{\{x_v^{c,k} = R\}}.
\end{align*}
MCMF thereby avoids both the independence assumption of IBA and the tree assumption of DMP. Given the correctly specified spreading model with known parameters $T_0$, $\beta$, and $\mu$, the simulated probabilities are unbiased estimators of the true node state probabilities and converge to the latter as $K \to \infty$. Nevertheless, just like IBA and DMP, MCMF relies on a factorized likelihood. The underlying independence assumption is sometimes also called a \emph{mean-field}~(MF) assumption, hence the name MCMF.

With an efficient event-driven implementation~\cite{holmeFastPrincipledSimulations2021} of the Monte Carlo simulation, the algorithmic complexity can be reduced substantially. Each simulation run initializes $c$ as infected at $T_0$ and all other nodes as susceptible. Scheduled infections are maintained in a binary min-heap ordered by infection time. Since a binary heap is a binary tree structure, inserting and removing nodes only costs $\mathcal{O}(\log|V|)$, much less than a full scan over all currently infected nodes, which would cost $\mathcal{O}(|V|)$, leading to a major speed-up overall.

With that in mind, the complexity of a single simulation run can be derived as follows. An infection happens at most $|V|$ times per simulation run. The next node to be infected is always the root $r$ of the heap. Accessing, deleting, and restoring it via heap-down costs $\mathcal{O}(\log|V|)$. Upon infecting $r$, we iterate over its neighbors on the static projection in order to find future infection events. On average, $r$ has $\tilde{d} = 2|E|/|V|$ such neighbors. For each neighbor $v$ of $r$, a binary search over the temporal contacts between $r$ and $v$ costs $\mathcal{O}(\log \tilde{m})$, where $\tilde{m} = |E_T|/|E|$ is the average number of contacts between two nodes over the whole time-span. And each newly scheduled infection event is then inserted into the heap in $\mathcal{O}(\log|V|)$ time. Thus, the computational complexity of processing one infection event is
\[
  \mathcal{O}\!\left(\log|V| + \tilde{d} \cdot \bigl(\log \tilde{m} + \log|V|\bigr)\right) = \mathcal{O}\!\left(\tilde{d} \cdot \log(\tilde{m}|V|)\right),
\]
as long as $\tilde{d}\geq1$ and $\tilde{m}\geq1$. A single simulation run therefore costs
\[
  \mathcal{O}\!\left(|V| \cdot \tilde{d} \cdot \log(\tilde{m}|V|)\right)
  = \mathcal{O}\!\left(|E| \cdot \log \frac{|E_T||V|}{|E|}\right),
\]
and running $K$ simulations for each candidate source gives a total complexity of
\[
  \mathcal{O}\!\left(K \cdot |V| \cdot |E| \cdot \log
  \frac{|E_T||V|}{|E|}\right).
\]
The cost of computing the sample means and the factorized likelihood is dominated by this term and therefore does not affect the overall complexity of MCMF.

\subsection{Soft Margin Estimator (SME)}

SME estimates the source probability directly by comparing simulated outbreaks to the observed snapshot. Thus, unlike MCMF, it approximates the true posterior $p_{\mathbf{x}}(c) = P(s = c \mid \mathbf{x})$ directly, without a detour via node state probabilities and a factorized likelihood. Before introducing SME in detail, we first describe \emph{Direct Monte Carlo} (as opposed to MCMF), since SME is best understood as an approximation to it.

Given the correctly specified spreading model with known parameters $T_0$, $\beta$, and $\mu$ and $K$ simulated outbreaks starting from each candidate source $c$, the true probability distribution is given by the Direct Monte Carlo estimator
\begin{equation}
p_{\mathbf{x}}(c) = P(s = c \mid \mathbf{x}) = \lim_{K \to \infty} \frac{\sum_{k=1}^{K} \mathbf{1}_{\{\mathbf{x}^{c,k} = \mathbf{x}\}}}{\sum_{c'} \sum_{k=1}^{K} \mathbf{1}_{\{\mathbf{x}^{c',k} = \mathbf{x}\}}}.
\label{eq:direct_mc}
\end{equation}
Intuitively, Direct Monte Carlo estimates $p_{\mathbf{x}}(c)$ by counting, for each candidate source $c$, how often a simulated outbreak exactly matches the observed snapshot $\mathbf{x}$, requiring exact equality between the simulated snapshots and the observed one. However, using Direct Monte Carlo is computationally infeasible: there are $3^{|V|}$ possible snapshot configurations $\mathbf{x}$ in the SIR case (or $2^{|V|}$ in the SI case), so the number of simulations needed for a good approximation grows exponentially in $|V|$.

SME mitigates this by not requiring exact equality $\mathbf{x}^{c,k} = \mathbf{x}$, but only similarity (hence the name \emph{soft margin}). It approximates the Direct Monte Carlo estimator using the \emph{Jaccard similarity index}
\begin{equation}
\varphi(\mathbf{x}, \mathbf{x}^{c,k}) = \frac{\left| \left\{ v : x_v \neq S \right\} \cap \left\{ v : x_v^{c,k} \neq S \right\} \right|}{\left| \left\{ v : x_v \neq S \right\} \cup \left\{ v : x_v^{c,k} \neq S \right\} \right|},
\label{eq:jaccard}
\end{equation}
which equals $1$ if $\mathbf{x}^{c,k}$ and $\mathbf{x}$ agree perfectly on the set of non-susceptible nodes, and $0$ if they share none. Using Gaussian weighting $w_a(x) = \exp(-(x-1)^2/a^2)$ with parameter $a$, the SME likelihood becomes
\begin{align}
\hat{P}(\mathbf{x} \mid s = c) &= \frac{1}{K} \sum_{k=1}^{K} w_a\!\left(\varphi(\mathbf{x}, \mathbf{x}^{c,k})\right) \notag \\
&= \frac{1}{K} \sum_{k=1}^{K} \exp\!\left[-\frac{\left(\varphi(\mathbf{x}, \mathbf{x}^{c,k})-1\right)^2}{a^2}\right]. \label{eq:sme_likelihood}
\end{align}
The parameter $a$ is chosen by Antulov-Fantulin et al.\ via a convergence condition~\cite{antulov-fantulinIdentificationPatientZero2015a}. As $a \to 0$, $w_a(x)$ approaches a hard indicator $\mathbf{1}_{\{x = 1\}}$, and in the case of the SI model the SME estimator converges to the Direct Monte Carlo estimator. Importantly, SME discards the distinction between $I$ and $R$ in the Jaccard index, treating both as non-susceptible. As a consequence, the convergence to the Direct Monte Carlo estimator does not extend to the SIR model, which is a major limitation.

Regarding complexity, computing the Jaccard similarity for a single $\mathbf{x}^{c,k}$ costs $\mathcal{O}(|V|)$. Over all $K \cdot |V|$ simulations this gives $\mathcal{O}(K \cdot |V|^2)$, but this is dominated by the cost of generating the simulations. The total complexity of SME is therefore the same as for MCMF.

\subsection{Backtracking Network (BN)}

The Backtracking Network (BN)~\cite{ruInferringPatientZero2023} is a graph neural network~(GNN) on the static projection $G$ that encodes temporal contacts as binary per-timestep edge activation features.

More formally, each node $u$ is initialized with a one-hot encoding of its SIR state $\mathbf{h}_u^{(0)} \in \{0,1\}^3$; each static edge $(u,v)$ carries a binary activation vector $\mathbf{g}_{(u,v)}^{(0)} \in \{0,1\}^{|T|}$ recording at which time steps the contact was active. Across $L$ layers, two interdependent update equations propagate information:
\begin{align}
    \mathbf{g}_{(u,v)}^{(l+1)} &= f^{\mathrm{edge}}\!\left(\mathbf{g}_{(u,v)}^{(l)},\, \mathbf{h}_u^{(l)}\right), \label{eq:bn_edge}\\
    \mathbf{h}_u^{(l+1)} &= f^{\mathrm{node}}\!\left(\mathbf{h}_u^{(l)}\right) + \sum_{v \in \mathcal{N}(u)} \mathbf{g}_{(u,v)}^{(l+1)}, \label{eq:bn_node}
\end{align}
where $f^{\mathrm{edge}}$ and $f^{\mathrm{node}}$ are independent fully-connected layers with ReLU activations and hidden dimension $d_h$. After $L$ layers, a linear readout layer projects each node embedding to a scalar, and a softmax over all nodes gives $\hat{p}_{\mathbf{x}}(u)$. As for MCMF and SME, the training data consists of $K \cdot |V|$ simulated snapshots, i.e., $K$ per source node. It is trained with a cross-entropy loss and the Adam optimizer.

To benchmark BN against simpler architectures, we evaluate two ablations. The first is a \emph{Static Multi-Layer Perceptron (SMLP)} with no message passing or graph structure. However, unlike a naive per-node MLP that would process each node's state vector in isolation, we flatten the entire snapshot into one input vector, so that the SMLP can still capture correlations \textit{between} nodes. The second is a standard convolutional \emph{Static Graph Neural Network (SGNN)} as outlined in~\cite{sterchiGraphNeuralNetworks2025}, applied to the static projection $G$, using the same one-hot node features but without temporal edge information.

Training BN costs
\[
\mathcal{O}\!\left(M \cdot K \cdot |V| \cdot |E| \cdot (|T| \cdot d_h + L \cdot d_h^2)\right),
\]
where $M$ is the number of training epochs. This follows from tracing the cost of Eq.~\eqref{eq:bn_edge} per edge and per layer, disregarding Eq.~\eqref{eq:bn_node} since it is computationally dominated by the edge update. The raw edge feature $\mathbf{g}_{(u,v)}^{(0)} \in \{0,1\}^{|T|}$ is $|T|$-dimensional, but from layer $l \geq 1$ onward, $\mathbf{g}_{(u,v)}^{(l)}$ is $d_h$-dimensional. Consequently, only the first application of $f^{\mathrm{edge}}$ costs $\mathcal{O}(|T| \cdot d_h)$, while each of the remaining $L-1$ layers costs the usual $\mathcal{O}(d_h^2)$ of a fully-connected layer. Summing over all $L$ layers, all $|E|$ edges, all $K \cdot |V|$ training snapshots, and all $M$ epochs yields the stated complexity. Finally, since $|T| \cdot d_h + L \cdot d_h^2 \gg \log(|E_T||V|/|E|)$ for typical values of $|T|$, $L$, and $d_h$, training dominates data generation. Moreover, training cost further dominates inference cost, hence we report only training cost in Table~\ref{tab:methods}.

\subsection{Belief Propagation (BP)}\label{sec:bp}

BP takes a fundamentally different approach from all methods covered so far: it directly computes the posterior marginal distribution over infection times for each node~\cite{altarelliBayesianInferenceEpidemics2014, braunsteinInferenceCausalityEpidemics2016}. It relies neither on a factorized likelihood nor on simulated outbreaks as training data.

Given a small $\epsilon>0$, let $t_v \in T \cup \{T_0-\epsilon,\infty\}$ denote the infection time of node $v$, where $t_v=T_0-\epsilon$ if $v=s$ is the source, and $t_v = \infty$ if $v$ is never infected. Let $g_v$ denote the recovery delay of node $v$, i.e.\ the number of time steps $v$ remains infected before recovering, drawn from a geometric distribution $\mathcal{G}_v(g_v) = \mu(1-\mu)^{g_v}$. Let $q_{vu}$ denote the absolute time at which $v$ would transmit to $u$, i.e.\ $t_v$ plus the number of time steps after infection of $v$ until $v$ infects $u$. The corresponding transmission delay follows a geometric distribution over the contacts between $v$ and $u$, $\omega_{vu}(q_{vu}|t_v,g_v) = \beta(1-\beta)^{n_{vu}(t_v,q_{vu})} \mathbf{1}_{\{0 < q_{vu} - t_v < g_v\}} \mathbf{1}_{\{(v,u,\, q_{vu}) \in E_T\}}$, where $n_{vu}(t_v,q_{vu}) = |\{t' : t_v < t' < q_{vu},\, (v,u,t') \in E_T\}|$ counts the earlier contacts that failed to transmit, and transmission can only occur while $v$ is still infected and a contact between $v$ and $u$ exists at time $q_{vu}$, with $q_{vu} = \infty$ if no transmission occurs. The infection times, transmission delays and recovery delays of all nodes can be collected in vectors $(\mathbf{t}, \mathbf{g}, \mathbf{q})$ whose joint distribution given $\mathbf{x}$ can be expressed using Bayes' theorem:
\begin{equation}
P(\mathbf{t}, \mathbf{g}, \mathbf{q} \mid \mathbf{x}) \propto P(\mathbf{x} \mid \mathbf{t}, \mathbf{g}, \mathbf{q}) \cdot P(\mathbf{t}, \mathbf{g}, \mathbf{q}).
\end{equation}
Crucially, given all the infection times and recovery delays $(\mathbf{t}, \mathbf{g})$, the node states $\mathbf{x}$ are fully determined, i.e., $P(\mathbf{x} \mid \mathbf{t}, \mathbf{g}, \mathbf{q}) = P(\mathbf{x} \mid \mathbf{t}, \mathbf{g}) = \prod_{v \in V} \zeta_v(x_v,t_v,g_v)$, where $\zeta_v(x_v,t_v,g_v) = \mathbf{1}_{\{x_v = S,\; T_1 < t_v\}} + \mathbf{1}_{\{x_v = I,\; t_v \leq T_1 < t_v + g_v\}} + \mathbf{1}_{\{x_v = R,\; t_v + g_v \leq T_1\}}$ enforces that the state of $v$ inferred from $(t_v, g_v)$ is consistent with the observation $x_v$.

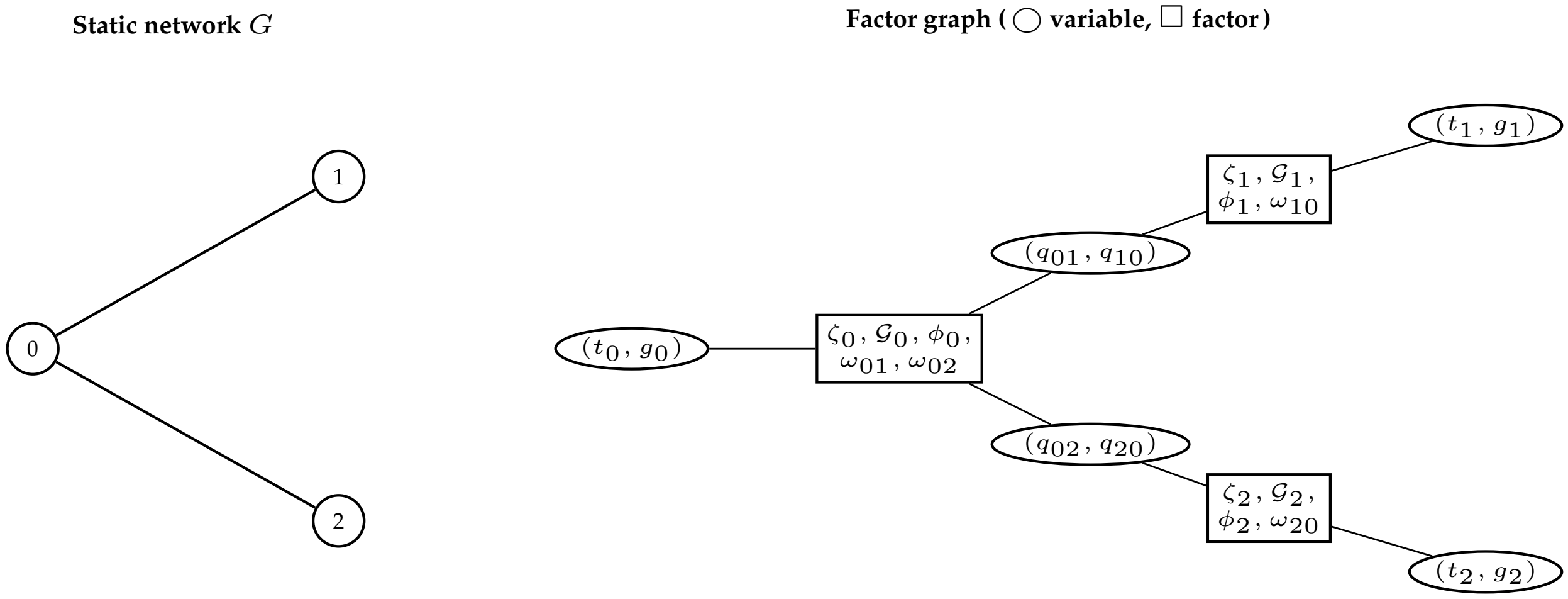
\begin{figure*}[t]
\centering
\resizebox{0.9\textwidth}{!}{%
\begin{tikzpicture}[
  font=\fontsize{5}{6}\selectfont,
  varnode/.style={circle, draw, semithick, minimum size=0.34cm, inner sep=0.5pt, font=\fontsize{5}{6}\selectfont},
  facnode/.style={rectangle, draw, semithick, minimum size=0.4cm, inner sep=2pt, font=\fontsize{5}{6}\selectfont, align=center},
]

\begin{scope}[xshift=0cm]
  \node[above] at (1.4,4.0) {\fontsize{5.7}{6.8}\selectfont\textbf{Static network $G$}};
  \node[varnode, minimum size=0.4cm] (n0) at (0.3, 1.65) {0};
  \node[varnode, minimum size=0.4cm] (n1) at (2.7, 3.0) {1};
  \node[varnode, minimum size=0.4cm] (n2) at (2.7, 0.3) {2};
  \draw[semithick] (n0) -- (n1);
  \draw[semithick] (n0) -- (n2);
\end{scope}

\begin{scope}[xshift=6.6cm]
  \node[above] at (1.75,4.0) {\fontsize{5.7}{6.8}\selectfont\textbf{Factor graph (\,$\bigcirc$ variable, $\square$ factor\,)}};

  \node[facnode] (f0) at (0.5, 1.65) {$\zeta_0,\mathcal{G}_0,\phi_0,$\\$\omega_{01},\omega_{02}$};
  \node[varnode, ellipse, minimum width=0.85cm, minimum height=0.32cm] (tg0) at (-1.6, 1.65) {$(t_0,g_0)$};
  \draw (f0) -- (tg0);

  \node[facnode] (f1) at (3.4, 2.9) {$\zeta_1,\mathcal{G}_1,$\\$\phi_1,\omega_{10}$};
  \node[varnode, ellipse, minimum width=0.85cm, minimum height=0.32cm] (tg1) at (5.1, 3.4) {$(t_1,g_1)$};
  \draw (f1) -- (tg1);

  \node[facnode] (f2) at (3.4, 0.4) {$\zeta_2,\mathcal{G}_2,$\\$\phi_2,\omega_{20}$};
  \node[varnode, ellipse, minimum width=0.85cm, minimum height=0.32cm] (tg2) at (5.1, -0.1) {$(t_2,g_2)$};
  \draw (f2) -- (tg2);

  \node[varnode, ellipse, minimum width=0.85cm, minimum height=0.32cm] (q01) at (2.0, 2.4) {$(q_{01},q_{10})$};
  \draw (f0) -- (q01);
  \draw (f1) -- (q01);

  \node[varnode, ellipse, minimum width=0.85cm, minimum height=0.32cm] (q02) at (2.0, 0.9) {$(q_{02},q_{20})$};
  \draw (f0) -- (q02);
  \draw (f2) -- (q02);

\end{scope}
\end{tikzpicture}}
\caption{Minimal example of the factor graph underlying BP, for a 3-node network with node 0 connected to nodes 1 and 2. Variables: $t_v$ = infection time of node $v$; $g_v$ = recovery delay of node $v$, bundled together as $(t_v,g_v)$; $q_{vu}$ = the time at which $v$ would transmit to $u$, if $v$ gets infected, with both directions of an edge bundled together as $(q_{uv},q_{vu})$. Factors, combined into one box per node: $\zeta_v$ checks $v$'s observed state against $(t_v,g_v)$; $\mathcal{G}_v$ is the distribution over the recovery delay $g_v$; $\phi_v$ checks that $v$'s infection time matches the earliest transmission time among its neighbors, i.e., $v$ gets infected by whoever reaches it first; $\omega_{vu}$ is the distribution over $q_{vu}$, conditioned on $g_v$ and $t_v$. Note that the factor graph's topology mirrors that of the static network; contact timestamps enter the model through the definition of $q_{vu}$ and $\omega_{vu}$. That is, transmission can only occur at times when an actual contact takes place, so $\omega_{vu}$ assigns zero probability to any $q_{vu}$ that does not coincide with a real contact in $E_T$.}
\label{fig:factor_graph_example}
\end{figure*}

For the second factor, the chain rule gives $P(\mathbf{t}, \mathbf{g}, \mathbf{q}) = P(\mathbf{t}, \mathbf{q} \mid \mathbf{g})\cdot P(\mathbf{g}) $, where $P(\mathbf{g}) = \prod_{v \in V} \mathcal{G}_v(g_v)$. Furthermore, $P(\mathbf{t}, \mathbf{q} \mid \mathbf{g}) = {\prod_{v \in V} \phi_v(t_v,\mathbf{q}_{:,v} )} {\prod_{(v,u) \in E} \omega_{vu}(q_{vu}|t_v,g_v)\,\omega_{uv}(q_{uv}|t_u,g_u)}$, where $\mathbf{q}_{:,v}$ denotes the restriction of $\mathbf{q}$ to incoming transmissions at node $v$, and $\phi_v(t_v,\mathbf{q}_{:,v} ) = \mathbf{1}_{\{t_v = \min_{u \in V\setminus \lbrace v \rbrace} q_{uv}\}}$ enforces that $t_v$ is consistent with the incoming $q_{uv}$. In addition, $\phi_v$ allows $t_v = T_0 - \epsilon$, representing the case where $v$ itself is the source and has no infector. Substituting all factors yields (with arguments omitted for brevity):
\begin{equation}
P(\mathbf{t}, \mathbf{g}, \mathbf{q} \mid \mathbf{x}) \propto \prod_{v \in V} \zeta_v\, \phi_v\, \mathcal{G}_v \prod_{(v,u) \in E} \omega_{vu}\,\omega_{uv}.
\end{equation}
This expression gives rise to a \emph{factor graph} whose topology mirrors that of $G$ (see Figure~\ref{fig:factor_graph_example} for an example). On that, we can perform the well-known (loopy) belief propagation algorithm~\cite{pearlProbabilisticReasoningIntelligent1988, murphyLoopyBeliefPropagation1999} to iteratively approximate marginal probabilities by passing messages from the variables $(\mathbf{t}, \mathbf{g}, \mathbf{q})$ to the factors $(\zeta_v, \phi_v, \mathcal{G}_v, \omega_{vu})$ and vice versa. BP is exact on trees and approximate on loopy networks.

For the purpose of source detection, we are specifically interested in the marginal probability distribution of the infection time $t_c$ of a potential source $c$:
\begin{equation}
P(t_c \mid \mathbf{x}) = \sum_{\mathbf{t} \backslash \{t_c\}, \mathbf{g}, \mathbf{q}} P(\mathbf{t}, \mathbf{g}, \mathbf{q} \mid \mathbf{x}).
\end{equation}
The approximated probability for $c$ being the source is then $\hat{p}_{\mathbf{x}}(c) = \hat{P}(c=s \mid \mathbf{x}) = \hat{P}(t_c = T_0 - \epsilon\mid \mathbf{x})$.

Regarding complexity, BP costs $\mathcal{O}(R \cdot |T| \cdot |E|)$, where $R$ is the number of iterations needed for convergence. Since $t_v$, $g_v$, and $q_{vu}$ are discrete-time-valued, each message is a length-$|T|$ distribution, costing $\mathcal{O}(|T|)$ to update per edge, and $\mathcal{O}(|T| \cdot |E|)$ per iteration over all $|E|$ edges. Note that the authors' original algorithm also supports a semi-continuous-time variant, where contact times need not be discretized onto a fixed grid $T$, but can instead take the raw, real-valued timestamps directly from the contact data~\cite{braunsteinInferenceCausalityEpidemics2016}.

\section{Comparative Study}
\label{sec:comparative}
In this section, we describe the methodology of our comparative study. As motivated in the introduction, existing methods have been evaluated on widely different networks, and even when the same network is used (e.g., the Brazilian Escort dataset), preprocessing and epidemic parameters vary considerably. Furthermore, no consensus has emerged on which evaluation metric to use. We therefore standardize these methodological choices to ensure a fair comparison. Section~\ref{sec:networks} introduces the networks, Section~\ref{sec:expsetup} the setup of the main experiment, Section~\ref{sec:expsetup_complementary} the three complementary experiments, and Section~\ref{sec:evaluation} the evaluation metrics.

\subsection{Networks}
\label{sec:networks}

Our goal is to use empirical temporal networks on which epidemic spreading can be modeled most realistically. We therefore focus on physical \emph{contact networks}, where nodes are individuals and edges are interactions between them. Due to privacy constraints, such temporal networks remain rather scarce; most of the publicly available ones have been collected by the SocioPatterns collaboration\footnote{See \url{https://www.sociopatterns.org/}}. Table~\ref{tab:networks} summarizes the eight networks we select, which we describe below.

We exclude well-known temporal networks of human interaction from SocioPatterns that span only 2--5 days (primary school~\cite{gemmettoMitigationInfectiousDisease2014}, high school~\cite{mastrandreaContactPatternsHigh2015}, hospital ward~\cite{vanhemsEstimatingPotentialInfection2013}, academic conference~\cite{isellaWhatsCrowdAnalysis2011}), since they cover too few days to meaningfully model an epidemic. The remaining SocioPatterns networks that we include are the office network in Saint-Maurice, France (near Paris)~\cite{genoisDataFacetofaceContacts2015} and the village network in Malawi~\cite{ozellaUsingWearableProximity2021}, both spanning more than 10 days.

We further include two Bluetooth-based student networks, both recording proximity between smartphones carried by the participants. The first is the Copenhagen network~\cite{stopczynskiMeasuringLargeScaleSocial2014}, covering roughly 1,000 undergraduate students at the Technical University of Denmark. The second is the Reality MIT network~\cite{eagleRealityMiningSensing2006}, covering roughly 100 students and staff in Boston over an academic year.

We also include a network of baboon interactions~\cite{gelardiMeasuringSocialNetworks2020}, the only non-human network in our study and by far the smallest (19 nodes). The baboons dataset records behavioral events between individuals; we keep only events with behavior ``Grooming'', a common and meaningful social activity.

In addition, we include the Brazilian Escort dataset~\cite{rochaSimulatedEpidemicsEmpirical2011}, a network of sexual contacts that is the most frequently used benchmark in prior work (Table~\ref{tab:used_networks}) and much larger and sparser than most temporal networks. We exclude nodes that appear in fewer than five contacts, since otherwise too many uninvolved nodes would remain in the network. The first roughly 1,000 days of the dataset are rather irregular, so we exclude this period from the epidemic simulations.

Beyond these existing datasets, we build two temporal contact networks ourselves from raw Foursquare check-in data~\cite{dingqiyangModelingUserActivity2015}, which we refer to as the New York check-ins and Tokyo check-ins networks. Since this requires several preprocessing choices, we describe the pipeline in Section~\ref{sec:foursquare_appendix} of the Supplementary Material.

\begin{table*}
\centering
\caption{Empirical temporal networks used in our experiments, in ascending order of number of nodes $|V|$. $|E|$ denotes the number of edges in the static network projection (the number of node pairs that have ever been in contact during the epidemic interval $[T_0,T_1]$), and $|E_T|$ denotes the total number of contacts in the epidemic interval. The static density is computed as $2|E|/(|V|(|V|-1))$. The static global clustering coefficient (GCC) of $G$ is the fraction of closed triplets among all connected triplets. Note that all reported statistics depend on the choice of epidemic duration and interval, as specified in Table~\ref{tab:epidemic_params}: nodes that do not appear within the relevant time span are excluded. Moreover, we use only the largest static connected component of the resulting network, so that the reported network size reflects only the nodes that could be reached by the epidemic.}
\label{tab:networks} 
\small
\begingroup
\setlength{\tabcolsep}{2pt}
\begin{tabular}{p{3.3cm}p{2.8cm}p{2.0cm}p{1.8cm}>{\centering\arraybackslash}p{0.9cm}>{\centering\arraybackslash}p{1.0cm}>{\centering\arraybackslash}p{1.2cm}>{\centering\arraybackslash}p{1.3cm}%
  >{\centering\arraybackslash}p{1.3cm}
  }
\hline
\textbf{Network}
  & \textbf{\begin{tabular}[b]{@{}c@{}}Data\\Collection\end{tabular}}
  & \textbf{\begin{tabular}[b]{@{}c@{}}Used Time\\Granularity\end{tabular}}
  & \textbf{\begin{tabular}[b]{@{}c@{}}Original\\Start\end{tabular}}
  & \textbf{$|V|$}
  & \textbf{$|E|$}
  & \textbf{$|E_T|$}
  & \textbf{\begin{tabular}[b]{@{}c@{}}Static\\Density\end{tabular}}
  & \textbf{\begin{tabular}[b]{@{}c@{}}Static\\GCC\end{tabular}}
  \\ \hline
Baboon Troop & Direct observation & hours & 2019-06-13 & 19 & 91 & 265 & 0.532 & 0.575 \\[3pt]
Village in Malawi & Proximity sensor & hours & 2019-12-22 & 82 & 308 & 6,004 & 0.093 & 0.372 \\[3pt]
Boston Students & Bluetooth & hours & 2004-09-14 & 87 & 1,055 & 8,706 & 0.282 & 0.595 \\[3pt]
Office in France & Proximity sensor & hours & 2013-06-24 & 91 & 657 & 1,771 & 0.160 & 0.363 \\[3pt]
New York Check-ins & Foursquare data & days & 2012-04-03 & 508 & 2,288 & 2,609 & 0.018 & 0.306 \\[3pt]
Copenhagen Students & Bluetooth & hours & 2012-10-01$^*$ & 676 & 52,331 & 218,250 & 0.229 & 0.406 \\[3pt]
Tokyo Check-ins & Foursquare data & days & 2012-04-03 & 886 & 3,242 & 4,081 & 0.008 & 0.202 \\[3pt]
Brazilian Escort & Web scraping & days & 2002-09-01 & 2,133 & 4,998 & 6,486 & 0.002 & 0.000$^{**}$ \\
\hline
\multicolumn{9}{p{\dimexpr 3.3cm+2.8cm+2.0cm+1.8cm+1.3cm+1.3cm+0.9cm+1.0cm+1.2cm-2\tabcolsep\relax}}{\footnotesize $^*$Original data spans almost a full year 2012-10-01--2013-09-01; we use a $\sim$28-day subset whose exact start date is unknown.\newline $^{**}$The Escort network is bipartite (sex workers and buyers), hence it contains no triangles and its clustering coefficient is 0.} \\ \hline
\end{tabular}
\endgroup
\end{table*}

\subsection{Main Experiment: Setup}
\label{sec:expsetup}

The main experiment of this comparative study focuses on the performance of methods for varying $R_0$. Since each value of $R_0$ yields a different distribution of outbreak sizes, we further examine visually how performance varies as a function of outbreak size.

For each network, we (i)~fix the duration $T_1-T_0$ of the epidemic and assume it is known, (ii)~select a non-zero recovery probability $\mu$, and (iii)~choose three different infection probabilities $\beta$ that result in realistic $R_0$ values. We choose to target $R_0$ values of $1.5$, $2.0$, and $2.5$, which are realistic values for diseases such as Ebola~\cite{muzemboBasicReproductionNumber2024, whoebolaresponseteamEbolaVirusDisease2014} or COVID-19~\cite{locatelliEstimatingBasicReproduction2021, zhangEstimationReproductiveNumber2020, fosterEstimating0Early2022}.\footnote{Note, however, that $R_0$ is not an intrinsic biological constant of a pathogen; it also depends on population density, social behavior, and environmental context~\cite{delamaterComplexityBasicReproduction2019, viceconteCOVID19R0Magic2020}.} Table~\ref{tab:epidemic_params} summarizes the specific choices for each of the eight networks, which are then used to run extensive Monte Carlo SIR simulations (both for generating training data and test data, see Subsections~\ref{sec:training-data} and \ref{sec:test-data}).

Since the Copenhagen network has by far the most contacts per node, we have to choose rather low values of $\beta$ to reach the target values for $R_0$. The same holds, to a lesser extent, for the Boston network. Conversely, for the Escort network, which is quite sparse (0.002), we have to choose a comparatively small $\mu$ together with large values of $\beta$.

\begin{table*}
\centering
\caption{Epidemic simulation parameters per network. Epidemic Duration $T_1 - T_0$ is the length of the simulated epidemic, bounded above by the length of the underlying temporal network. $\beta$ shows the per-contact transmission probabilities chosen to reach the three target reproduction numbers $R_0 \in \{1.5, 2.0, 2.5\}$, respectively. $\mu$ is the per-time-step recovery probability, held fixed across $R_0$ targets. Avg.\ Outbreak Size is the average fraction of nodes infected by the end of the epidemic and Trivial Outbreaks is the percentage of outbreaks of size one, i.e., where the epidemic never spread beyond the source node. All values separated by ``/'' represent the three targets, in order.} 
\label{tab:epidemic_params}
\small
\begin{tabular}{p{3.3cm}ccccc}
\hline
\textbf{Network}
  & \textbf{\begin{tabular}[b]{@{}c@{}}Epidemic Duration\\$T_1 - T_0$\end{tabular}}
  & \textbf{$\beta$}
  & \textbf{$\mu$}
  & \textbf{\begin{tabular}[b]{@{}c@{}}Avg.\ Outbreak\\Size (\%)\end{tabular}}
  & \textbf{\begin{tabular}[b]{@{}c@{}}Trivial\\Outbreaks (\%)\end{tabular}} \\ \hline
Baboon Troop & 25 days & 0.07 / 0.10 / 0.14 & 0.0002 & 21.2 / 33.7 / 51.2 & 20.5 / 11.5 / 5.4 \\[3pt]
Village in Malawi & 12 days & 0.09 / 0.22 / 0.49 & 0.010 & 9.4 / 28.4 / 56.7 & 24.2 / 13.4 / 7.4 \\[3pt]
Boston Students & 21 days & 0.013 / 0.020 / 0.035 & 0.001 & 8.9 / 19.5 / 39.9 & 39.3 / 31.3 / 24.2 \\[3pt]
Office in France & 10 days & 0.12 / 0.17 / 0.25 & 0.010 & 9.6 / 21.7 / 42.7 & 36.6 / 28.0 / 22.6 \\[3pt]
New York Check-ins & 4.1 months & 0.19 / 0.26 / 0.33 & 0.005 & 1.2 / 2.3 / 3.9 & 38.6 / 31.9 / 27.3 \\[3pt]
Copenhagen Students & 12 days & 0.009 / 0.012 / 0.017 & 0.010 & 8.0 / 23.9 / 41.1 & 46.9 / 41.0 / 35.9 \\[3pt]
Tokyo Check-ins & 4.1 months & 0.24 / 0.33 / 0.44 & 0.005 & 1.3 / 3.1 / 6.5 & 39.3 / 32.1 / 24.2 \\[3pt]
Brazilian Escort & 12 months & 0.35 / 0.51 / 0.70 & 0.001 & 2.2 / 6.2 / 13.4 & 36.8 / 27.7 / 19.0 \\
\hline
\end{tabular}
\end{table*}

\subsubsection{Training Data}
\label{sec:training-data}
The Monte Carlo-based methods MCMF and SME, as well as BN (and its ablations SMLP and SGNN), require training data, i.e., simulated snapshots $\mathbf{x}^{c,k}$ for $c \in \mathcal{C}$ and $k = 1, \dots, K$.\footnote{Note that the terminology ``training'' is adopted from machine learning, but we apply it here to MCMF and SME as well, although in the context of statistics it is usually called ``estimation''.} For each node, we use Monte Carlo SIR simulations to generate exactly $K=100$ such snapshots. The deep learning approaches, BN, SMLP and SGNN, further split this data 80/20 into training and validation sets for early stopping.

\subsubsection{Test Data}
\label{sec:test-data}
With the same Monte Carlo process, we independently simulate a separate set of snapshots $\mathbf{x}$ with known sources $s$ as test data. For each network, we aim for roughly 8,000 to 12,000 snapshots in total, which keeps confidence intervals reasonably narrow. For each network, the number of snapshots is a multiple of the number of nodes $|V|$, so that each node serves as the source $s$ equally often. See Section~\ref{sec:evaluation} for the metrics used to evaluate the methods on the test data.

\subsubsection{Parameter Choice}
We aim to keep method parameters constant across networks and follow the settings of the original papers wherever possible. For SME, we use a convergence tolerance of 0.05 for choosing $a$, following the supplementary material of Antulov-Fantulin et al.~\cite{antulov-fantulinIdentificationPatientZero2015a}. For BP, we adopt Braunstein and Ingrosso's~\cite{braunsteinInferenceCausalityEpidemics2016} original choices: an initial source probability of 0.01 per node, a maximum of 50 iterations, and a convergence threshold of 0.001.

For the deep learning approaches, we aim to keep hyperparameters constant across architectures (SMLP, SGNN, BN) to ensure a fair comparison. Ru et al.~\cite{ruInferringPatientZero2023} report a batch size of 128, five layers, a learning rate of 0.001, and the Adam optimizer; we adopt these choices for all three architectures, except that for BN we reduce the batch size on the denser contact networks (details below). Ru et al.'s choices are, however, incomplete: their paper does not report hidden dimension and weight decay. We fill these in with Sterchi et al.'s~\cite{sterchiGraphNeuralNetworks2025} recommendations (a hidden dimension of 128 and a weight decay of 0.0005), and use them for all three architectures alike. Batch normalization, skip connections, and dropout are likewise recommended by Sterchi et al., so we adopt them for SMLP and SGNN, but not for BN, since doing so would not be faithful to Ru et al.'s original architecture. Training stops early once 10 epochs have passed without significant improvement, and we report results from the epoch with the best validation performance rather than the final one.

The previously mentioned exception to keeping hyperparameters constant across networks concerns the denser contact networks (Baboons, Boston, and Copenhagen). With the default batch size of 128, the training loss of BN failed to decrease steadily on these networks. We found that reducing the batch size to 8 (Baboons, Boston) or 4 (Copenhagen), while keeping the learning rate constant, resolved this issue (see Figure~\ref{fig:btnn_learning_curves_appendix} in the Supplementary Material for the corresponding learning curves). Leaving it unresolved would have meant comparing against an unfairly weakened version of BN on these networks.

\subsubsection{Technical Setup}
We run all experiments on a Linux server with an NVIDIA GeForce RTX 5080 (16 GB VRAM) and an AMD Ryzen Threadripper PRO 5965WX (24 cores, 48 threads). We track all runs, as well as data artifacts, using Weights \& Biases.\footnote{L. Biewald, ``Experiment Tracking with Weights and Biases,'' 2020. Software available from wandb.com. \url{https://www.wandb.com/}} For reproducibility, the code, including all parameters and hyperparameters, is available online.\footnote{Available at \href{https://zenodo.org/records/22913467?preview=1&token=eyJhbGciOiJIUzUxMiJ9.eyJpZCI6IjFlYjdiZjRiLTQwOGEtNGMzZS1hNjM0LWE4MDUzNTIzOGNjMiIsImRhdGEiOnt9LCJyYW5kb20iOiIwZTJmNjJlODk1YzA4MGQwNDJjOTU2Zjc3YWM4ZGJjNSJ9.7bm7eCrsuDwpwmYXkt_RzAiDUv_8SnNzwsqWW2Fu4eyh5U0eIfbjHZeYutaS-tkPclBRpol4mSUxDzrWzmovtQ}{zenodo.org/records/22913467}.}

\subsection{Complementary Experiments: Setup}
\label{sec:expsetup_complementary}

We complement the main experiment with three further experiments: the \emph{learning curve} experiment (performance as a function of the amount of training data per node $K$, for methods that rely on training data), the \emph{partiality} experiment (testing how well methods perform when only a percentage of the epidemic states is observed instead of the full snapshot), and the \emph{misspecification} experiment (how methods react to misspecified parameters, such as epidemic duration, infection probability, and recovery probability), each described in more detail below. Whenever possible, they mirror the setup of the main experiment, e.g., using the same amount of test and training data (except for the learning curve experiment). Since these experiments require many runs per method and network, we limit them to one epidemic configuration with $R_0 \approx 2.0$ and four of the eight networks: Baboons, Malawi, Boston, and Office. On the larger networks, a single run of the most expensive methods already takes up to multiple days (e.g., 69 hours for BP on Copenhagen to process the full test data), which makes repeated runs computationally impractical.

\subsubsection{Learning Curves}
The fixed choice of $K=100$ training snapshots per node raises the question: how much performance gain can be achieved with ten or a hundred times as much training data? In fact, Antulov-Fantulin et al.~\cite{antulov-fantulinIdentificationPatientZero2015a} use much more training data for their SME estimator. Thus, for the learning curve experiment, we vary the number of training snapshots per node $K$ logarithmically from 5 to 10,000 (5, 10, 20, 50, \dots, 5,000, 10,000). For small amounts of training data, there is considerable variance in the outcome, so we repeat each run with 10 different seeds. We evaluate the Monte Carlo-based methods (MCMF, SME) and the deep learning approaches (SMLP, SGNN, BN). The other methods that do not require training data are excluded from this experiment.

\subsubsection{Partial Observations}
All experiments up to this point assumed that the full snapshot is available for inference. However, in reality, usually only a fraction of epidemic states is observed. Thus, in this experiment we investigate how performance varies with the percentage $p$ of node states that are observed in a snapshot, from $p=100\%$ down to $p=10\%$ in steps of 10 percentage points. The observed subset is chosen uniformly at random from all nodes $V$. We further ensure that the observed nodes always include at least one infected or recovered node; otherwise, no information on the outbreak would be reported at all.

For all methods, a partial snapshot enlarges the candidate set $\mathcal{C}$: besides the infected and recovered nodes, unobserved nodes are also valid candidates if they can topologically reach all (observed) infected and recovered nodes. Apart from this, most methods have a natural adaptation to handle partial observations. However, there are some subtleties to be considered:
\begin{itemize}
    \item For IBA, DMP, and MCMF, the factorized likelihood is evaluated on the observed subset of the snapshot $\mathbf{x}|_{\{v \,:\, x_v \neq U\}}$. For SME, the Jaccard similarity between the partially observed snapshot and the training snapshots is likewise computed only over the observed nodes ${\{v \,:\, x_v \neq U\}}$, resulting in fewer state comparisons.
    \item For the deep learning approaches (SMLP, SGNN, BN), we additionally mask the training snapshots at the same $p$, but choosing the observed subset randomly. The alternative would be to train the neural networks on snapshots with the \textit{exact} same observed subset as the observed snapshot; however, since that requires separate training for each snapshot in the test data, this is infeasible for a simulation study. As a consequence, our results tend to underestimate the potential of neural networks in practical settings where training on outbreaks with a single fixed observed subset is, in fact, realistic.
\end{itemize}
In contrast to the other methods, ST and BP handle unobserved states naturally by design.

\subsubsection{Parameter Misspecifications}
In practical settings, the exact epidemic parameters, such as the duration of the epidemic and the transmission and recovery probabilities, are usually unknown and would have to be estimated from data, possibly from partial observations \cite{pengInferringStochasticDynamics2026}. So far, we have assumed that all methods know the true parameters; here we test their performance when one of $\beta$, $\mu$, or $T_0$ is misspecified while the other two remain at their true values.

The parameters for generating the test data remain exactly the same as in the other experiments (that is, $\beta$, $\mu$, and $T_0$ chosen to give an $R_0$ of 2.0, see Table~\ref{tab:epidemic_params}), but the methods are given one incorrect parameter. In particular, methods that rely on training data generate it with this misspecified value. For each parameter $\beta$, $\mu$, or $T_0$, and each network, we choose 5 offsets of the form $[-\delta, -\delta/2, 0, \delta/2, \delta]$, where 0 corresponds to the correctly specified model. $\delta$ is chosen depending on the parameter varied and the network as follows:
\begin{itemize}
\item $\beta$: $\delta$ is chosen such that $\beta - \delta$ and $\beta + \delta$ yield an $R_0$ of roughly $1.9$ and $2.1$, respectively\footnote{\label{fn:r0_band}Specifically, $R_0 \in [1.88, 1.92]$ and $R_0 \in [2.08, 2.12]$.}.
\item $\mu$: $\delta$ is chosen such that $\mu + \delta$ and $\mu - \delta$ yield an $R_0$ of roughly $1.9$ and $2.1$, respectively\footref{fn:r0_band}.
\item $T_0$: $\delta$ corresponds to 8--10\% of the true epidemic duration. Since $T_0$ determines the time window of the network, an offset can add or remove nodes, static edges, and contacts. To keep the set of candidate sources fixed across offsets, we treat the node set $V$ of the correctly specified setting as known, so that only contacts and static edges can change.
\end{itemize}

\subsection{Evaluation Metrics}
\label{sec:evaluation}

We primarily evaluate the performance of methods using the \emph{top-$k$ accuracy}, i.e., whether the true source ranks among the top $k$ candidates,
\begin{equation}
    \text{top-}k(\hat{r}, s) = \mathbf{1}_{\{\hat{r}(s) \leq k\}},
\end{equation}
where $\hat{r}(s) \in \{1, \dots, |\mathcal{C}|\}$ denotes the rank of the true source $s$ with respect to $\hat{p}_{\mathbf{x}}$ (or $\hat{f}$ for score-based methods), with ties broken at random. For the uniform baseline, $\text{top-}k$ is $1$ with probability $\frac{k}{\max(k, |\mathcal{C}|)}$, and $0$ otherwise. We empirically approximate the expected value $\mathbb{E}\bigl[\mathbf{1}_{\{\hat{r}(s) \leq k\}}\bigr]$ by averaging across the test data.

For methods returning a probability distribution (all methods except TJC and ST), we additionally report the \emph{Brier score}. For a single observed snapshot $\mathbf{x}$ with true source $s$, it is defined as
\begin{equation}
    \text{Br}(\hat{p}_{\mathbf{x}}, s)
    = \sum_{c \in V} \left(\hat{p}_{\mathbf{x}}(c) - \mathbf{1}_{\{s = c\}}\right)^2,
\end{equation}
and we report the average of $\text{Br}(\hat{p}_{\mathbf{x}}, s)$ across the test data. The Brier score takes values in the interval $[0, 2]$, and lower values indicate better performance. The uniform baseline achieves a score of exactly $\text{Br} = 1 - \frac{1}{|\mathcal{C}|}$, which is roughly $1$ for a large candidate set $\mathcal{C}$. The Brier score's sensitivity to how probability mass is distributed is both an advantage and a downside. On the one hand, it heavily penalizes overconfident distributions, i.e., distributions that concentrate most of the mass on a few wrong nodes. On the other hand, it rewards uniform-like distributions outside the true node, so that the uniform baseline itself already achieves a fairly good Brier score.

Both the $\text{top-}k$ and the Brier score are \emph{proper scoring rules}, but they are not graph-aware: they cannot distinguish whether probability mass has been assigned to a node close to the source or far away. The error distance would be a graph-aware alternative, but it comes with several downsides: it only considers the mode of the distribution (the node with the highest predicted probability) rather than the whole distribution, it is \textit{not} a proper scoring rule, and it is not clearly defined in the temporal case. We therefore do not include the error distance as an evaluation metric. Graph-aware scoring rules that are also strictly proper, such as the resistance and the diffusion score, have so far only been developed for static networks~\cite{hilfikerStrictlyProperScoring2026}.

\section{Results}
\label{sec:results}

\begin{table*}[t]
\centering
\caption{Top-5 accuracy in percent (mean $\pm$ 95\% Wald CI) across networks (primary rows), target $R_0$ (sub-rows), and methods (columns). The best-performing method per row is highlighted in bold and shaded.}
\label{tab:main_results} 
\scriptsize
\setlength{\tabcolsep}{5pt}
\resizebox{\textwidth}{!}{%
\begin{tabular}{ll ccccccccccc}
\hline
\textbf{Network} & \textbf{$R_0$} & \textbf{UB} & \textbf{TJC} & \textbf{ST} & \textbf{IBA} & \textbf{DMP} & \textbf{MCMF} & \textbf{SME} & \textbf{SMLP} & \textbf{SGNN} & \textbf{BN} & \textbf{BP} \\ \hline
\multirow{3}{*}{Baboons}
  & 1.5 & 91.5 \textcolor{gray}{$\pm$ 0.6} & 92.4 \textcolor{gray}{$\pm$ 0.5} & 93.5 \textcolor{gray}{$\pm$ 0.5} & 95.2 \textcolor{gray}{$\pm$ 0.4} & 95.1 \textcolor{gray}{$\pm$ 0.4} & 94.3 \textcolor{gray}{$\pm$ 0.5} & 94.2 \textcolor{gray}{$\pm$ 0.5} & 93.3 \textcolor{gray}{$\pm$ 0.5} & 93.9 \textcolor{gray}{$\pm$ 0.5} & 95.1 \textcolor{gray}{$\pm$ 0.4} & {\setlength{\fboxsep}{1.3pt}\colorbox{gray!15}{\hspace{1.5pt}\textbf{96.1 \textcolor{black!65}{$\pm$ 0.4}}\hspace{1.5pt}}} \\
  & 2.0 & 77.5 \textcolor{gray}{$\pm$ 0.8} & 79.6 \textcolor{gray}{$\pm$ 0.8} & 82.3 \textcolor{gray}{$\pm$ 0.8} & 87.1 \textcolor{gray}{$\pm$ 0.7} & 86.4 \textcolor{gray}{$\pm$ 0.7} & 85.4 \textcolor{gray}{$\pm$ 0.7} & 84.4 \textcolor{gray}{$\pm$ 0.7} & 83.3 \textcolor{gray}{$\pm$ 0.7} & 84.7 \textcolor{gray}{$\pm$ 0.7} & 86.3 \textcolor{gray}{$\pm$ 0.7} & {\setlength{\fboxsep}{1.3pt}\colorbox{gray!15}{\hspace{1.5pt}\textbf{88.1 \textcolor{black!65}{$\pm$ 0.7}}\hspace{1.5pt}}} \\
  & 2.5 & 59.1 \textcolor{gray}{$\pm$ 1.0} & 61.5 \textcolor{gray}{$\pm$ 1.0} & 64.6 \textcolor{gray}{$\pm$ 1.0} & 74.2 \textcolor{gray}{$\pm$ 0.9} & 73.8 \textcolor{gray}{$\pm$ 0.9} & 71.9 \textcolor{gray}{$\pm$ 0.9} & 69.3 \textcolor{gray}{$\pm$ 0.9} & 69.5 \textcolor{gray}{$\pm$ 0.9} & 70.5 \textcolor{gray}{$\pm$ 0.9} & 68.0 \textcolor{gray}{$\pm$ 0.9} & {\setlength{\fboxsep}{1.3pt}\colorbox{gray!15}{\hspace{1.5pt}\textbf{75.5 \textcolor{black!65}{$\pm$ 0.9}}\hspace{1.5pt}}} \\
\hline
\multirow{3}{*}{Malawi}
  & 1.5 & 75.8 \textcolor{gray}{$\pm$ 0.9} & 77.1 \textcolor{gray}{$\pm$ 0.9} & 77.8 \textcolor{gray}{$\pm$ 0.9} & 76.9 \textcolor{gray}{$\pm$ 0.9} & 81.3 \textcolor{gray}{$\pm$ 0.8} & 82.0 \textcolor{gray}{$\pm$ 0.8} & 79.7 \textcolor{gray}{$\pm$ 0.9} & 83.7 \textcolor{gray}{$\pm$ 0.8} & 83.6 \textcolor{gray}{$\pm$ 0.8} & {\setlength{\fboxsep}{1.3pt}\colorbox{gray!15}{\hspace{1.5pt}\textbf{85.4 \textcolor{black!65}{$\pm$ 0.8}}\hspace{1.5pt}}} & 83.9 \textcolor{gray}{$\pm$ 0.8} \\
  & 2.0 & 49.8 \textcolor{gray}{$\pm$ 1.1} & 51.5 \textcolor{gray}{$\pm$ 1.1} & 52.2 \textcolor{gray}{$\pm$ 1.1} & 52.8 \textcolor{gray}{$\pm$ 1.1} & 56.3 \textcolor{gray}{$\pm$ 1.1} & 59.5 \textcolor{gray}{$\pm$ 1.1} & 58.3 \textcolor{gray}{$\pm$ 1.1} & 64.4 \textcolor{gray}{$\pm$ 1.0} & 64.7 \textcolor{gray}{$\pm$ 1.0} & {\setlength{\fboxsep}{1.3pt}\colorbox{gray!15}{\hspace{1.5pt}\textbf{67.6 \textcolor{black!65}{$\pm$ 1.0}}\hspace{1.5pt}}} & 64.2 \textcolor{gray}{$\pm$ 1.0} \\
  & 2.5 & 29.1 \textcolor{gray}{$\pm$ 1.0} & 29.9 \textcolor{gray}{$\pm$ 1.0} & 31.4 \textcolor{gray}{$\pm$ 1.0} & 36.4 \textcolor{gray}{$\pm$ 1.0} & 43.0 \textcolor{gray}{$\pm$ 1.1} & 41.1 \textcolor{gray}{$\pm$ 1.1} & 39.1 \textcolor{gray}{$\pm$ 1.1} & 47.2 \textcolor{gray}{$\pm$ 1.1} & 47.5 \textcolor{gray}{$\pm$ 1.1} & {\setlength{\fboxsep}{1.3pt}\colorbox{gray!15}{\hspace{1.5pt}\textbf{49.7 \textcolor{black!65}{$\pm$ 1.1}}\hspace{1.5pt}}} & 42.6 \textcolor{gray}{$\pm$ 1.1} \\
\hline
\multirow{3}{*}{Boston}
  & 1.5 & 75.2 \textcolor{gray}{$\pm$ 0.9} & 76.3 \textcolor{gray}{$\pm$ 0.9} & 77.1 \textcolor{gray}{$\pm$ 0.9} & 84.0 \textcolor{gray}{$\pm$ 0.8} & 83.7 \textcolor{gray}{$\pm$ 0.8} & 82.1 \textcolor{gray}{$\pm$ 0.8} & 80.0 \textcolor{gray}{$\pm$ 0.8} & 81.8 \textcolor{gray}{$\pm$ 0.8} & 83.6 \textcolor{gray}{$\pm$ 0.8} & {\setlength{\fboxsep}{1.3pt}\colorbox{gray!15}{\hspace{1.5pt}\textbf{85.5 \textcolor{black!65}{$\pm$ 0.7}}\hspace{1.5pt}}} & 84.5 \textcolor{gray}{$\pm$ 0.8} \\
  & 2.0 & 57.4 \textcolor{gray}{$\pm$ 1.0} & 57.9 \textcolor{gray}{$\pm$ 1.0} & 59.0 \textcolor{gray}{$\pm$ 1.0} & 69.1 \textcolor{gray}{$\pm$ 1.0} & 71.5 \textcolor{gray}{$\pm$ 0.9} & 68.8 \textcolor{gray}{$\pm$ 1.0} & 64.5 \textcolor{gray}{$\pm$ 1.0} & 68.9 \textcolor{gray}{$\pm$ 1.0} & 72.0 \textcolor{gray}{$\pm$ 0.9} & {\setlength{\fboxsep}{1.3pt}\colorbox{gray!15}{\hspace{1.5pt}\textbf{74.6 \textcolor{black!65}{$\pm$ 0.9}}\hspace{1.5pt}}} & 71.1 \textcolor{gray}{$\pm$ 1.0} \\
  & 2.5 & 40.3 \textcolor{gray}{$\pm$ 1.0} & 40.4 \textcolor{gray}{$\pm$ 1.0} & 42.3 \textcolor{gray}{$\pm$ 1.0} & 51.6 \textcolor{gray}{$\pm$ 1.1} & 55.8 \textcolor{gray}{$\pm$ 1.0} & 52.6 \textcolor{gray}{$\pm$ 1.0} & 49.3 \textcolor{gray}{$\pm$ 1.1} & 54.2 \textcolor{gray}{$\pm$ 1.0} & 57.0 \textcolor{gray}{$\pm$ 1.0} & {\setlength{\fboxsep}{1.3pt}\colorbox{gray!15}{\hspace{1.5pt}\textbf{59.8 \textcolor{black!65}{$\pm$ 1.0}}\hspace{1.5pt}}} & 53.5 \textcolor{gray}{$\pm$ 1.0} \\
\hline
\multirow{3}{*}{Office}
  & 1.5 & 73.6 \textcolor{gray}{$\pm$ 0.9} & 76.7 \textcolor{gray}{$\pm$ 0.9} & 78.9 \textcolor{gray}{$\pm$ 0.8} & 82.8 \textcolor{gray}{$\pm$ 0.8} & 83.9 \textcolor{gray}{$\pm$ 0.8} & 83.1 \textcolor{gray}{$\pm$ 0.8} & 80.6 \textcolor{gray}{$\pm$ 0.8} & 84.4 \textcolor{gray}{$\pm$ 0.7} & 85.3 \textcolor{gray}{$\pm$ 0.7} & {\setlength{\fboxsep}{1.3pt}\colorbox{gray!15}{\hspace{1.5pt}\textbf{89.0 \textcolor{black!65}{$\pm$ 0.6}}\hspace{1.5pt}}} & {\setlength{\fboxsep}{1.3pt}\colorbox{gray!15}{\hspace{1.5pt}\textbf{89.0 \textcolor{black!65}{$\pm$ 0.6}}\hspace{1.5pt}}} \\
  & 2.0 & 54.1 \textcolor{gray}{$\pm$ 1.0} & 55.7 \textcolor{gray}{$\pm$ 1.0} & 60.2 \textcolor{gray}{$\pm$ 1.0} & 66.4 \textcolor{gray}{$\pm$ 1.0} & 68.4 \textcolor{gray}{$\pm$ 1.0} & 67.3 \textcolor{gray}{$\pm$ 1.0} & 62.4 \textcolor{gray}{$\pm$ 1.0} & 70.2 \textcolor{gray}{$\pm$ 0.9} & 70.8 \textcolor{gray}{$\pm$ 0.9} & {\setlength{\fboxsep}{1.3pt}\colorbox{gray!15}{\hspace{1.5pt}\textbf{78.1 \textcolor{black!65}{$\pm$ 0.8}}\hspace{1.5pt}}} & 77.8 \textcolor{gray}{$\pm$ 0.9} \\
  & 2.5 & 40.0 \textcolor{gray}{$\pm$ 1.0} & 40.4 \textcolor{gray}{$\pm$ 1.0} & 44.4 \textcolor{gray}{$\pm$ 1.0} & 54.1 \textcolor{gray}{$\pm$ 1.0} & 57.0 \textcolor{gray}{$\pm$ 1.0} & 53.9 \textcolor{gray}{$\pm$ 1.0} & 47.7 \textcolor{gray}{$\pm$ 1.0} & 56.8 \textcolor{gray}{$\pm$ 1.0} & 58.1 \textcolor{gray}{$\pm$ 1.0} & {\setlength{\fboxsep}{1.3pt}\colorbox{gray!15}{\hspace{1.5pt}\textbf{65.1 \textcolor{black!65}{$\pm$ 1.0}}\hspace{1.5pt}}} & 64.0 \textcolor{gray}{$\pm$ 1.0} \\
\hline
\multirow{3}{*}{New York}
  & 1.5 & 94.0 \textcolor{gray}{$\pm$ 0.5} & 96.7 \textcolor{gray}{$\pm$ 0.3} & 97.8 \textcolor{gray}{$\pm$ 0.3} & 97.6 \textcolor{gray}{$\pm$ 0.3} & 97.5 \textcolor{gray}{$\pm$ 0.3} & 97.4 \textcolor{gray}{$\pm$ 0.3} & 97.2 \textcolor{gray}{$\pm$ 0.3} & 98.0 \textcolor{gray}{$\pm$ 0.3} & 98.0 \textcolor{gray}{$\pm$ 0.3} & 99.5 \textcolor{gray}{$\pm$ 0.1} & {\setlength{\fboxsep}{1.3pt}\colorbox{gray!15}{\hspace{1.5pt}\textbf{99.6 \textcolor{black!65}{$\pm$ 0.1}}\hspace{1.5pt}}} \\
  & 2.0 & 89.9 \textcolor{gray}{$\pm$ 0.6} & 92.0 \textcolor{gray}{$\pm$ 0.5} & 96.0 \textcolor{gray}{$\pm$ 0.4} & 95.8 \textcolor{gray}{$\pm$ 0.4} & 95.6 \textcolor{gray}{$\pm$ 0.4} & 95.4 \textcolor{gray}{$\pm$ 0.4} & 94.6 \textcolor{gray}{$\pm$ 0.4} & 96.8 \textcolor{gray}{$\pm$ 0.3} & 97.9 \textcolor{gray}{$\pm$ 0.3} & 99.0 \textcolor{gray}{$\pm$ 0.2} & {\setlength{\fboxsep}{1.3pt}\colorbox{gray!15}{\hspace{1.5pt}\textbf{99.2 \textcolor{black!65}{$\pm$ 0.2}}\hspace{1.5pt}}} \\
  & 2.5 & 85.6 \textcolor{gray}{$\pm$ 0.7} & 87.5 \textcolor{gray}{$\pm$ 0.6} & 95.0 \textcolor{gray}{$\pm$ 0.4} & 94.9 \textcolor{gray}{$\pm$ 0.4} & 94.2 \textcolor{gray}{$\pm$ 0.5} & 93.7 \textcolor{gray}{$\pm$ 0.5} & 91.9 \textcolor{gray}{$\pm$ 0.5} & 96.0 \textcolor{gray}{$\pm$ 0.4} & 96.4 \textcolor{gray}{$\pm$ 0.4} & 98.7 \textcolor{gray}{$\pm$ 0.2} & {\setlength{\fboxsep}{1.3pt}\colorbox{gray!15}{\hspace{1.5pt}\textbf{99.1 \textcolor{black!65}{$\pm$ 0.2}}\hspace{1.5pt}}} \\
\hline
\multirow{3}{*}{Copenhagen}
  & 1.5 & 64.1 \textcolor{gray}{$\pm$ 0.9} & 64.1 \textcolor{gray}{$\pm$ 0.9} & 64.1 \textcolor{gray}{$\pm$ 0.9} & 69.3 \textcolor{gray}{$\pm$ 0.9} & 69.7 \textcolor{gray}{$\pm$ 0.9} & 68.2 \textcolor{gray}{$\pm$ 0.9} & 65.7 \textcolor{gray}{$\pm$ 0.9} & 69.4 \textcolor{gray}{$\pm$ 0.9} & {\setlength{\fboxsep}{1.3pt}\colorbox{gray!15}{\hspace{1.5pt}\textbf{72.8 \textcolor{black!65}{$\pm$ 0.9}}\hspace{1.5pt}}} & 71.1 \textcolor{gray}{$\pm$ 0.9} & 70.9 \textcolor{gray}{$\pm$ 0.9} \\
  & 2.0 & 51.4 \textcolor{gray}{$\pm$ 1.0} & 51.3 \textcolor{gray}{$\pm$ 1.0} & 51.5 \textcolor{gray}{$\pm$ 1.0} & 55.7 \textcolor{gray}{$\pm$ 1.0} & 56.1 \textcolor{gray}{$\pm$ 1.0} & 53.7 \textcolor{gray}{$\pm$ 1.0} & 52.2 \textcolor{gray}{$\pm$ 1.0} & 55.6 \textcolor{gray}{$\pm$ 1.0} & 58.5 \textcolor{gray}{$\pm$ 1.0} & {\setlength{\fboxsep}{1.3pt}\colorbox{gray!15}{\hspace{1.5pt}\textbf{60.8 \textcolor{black!65}{$\pm$ 1.0}}\hspace{1.5pt}}} & 57.2 \textcolor{gray}{$\pm$ 1.0} \\
  & 2.5 & 42.1 \textcolor{gray}{$\pm$ 1.0} & 42.2 \textcolor{gray}{$\pm$ 1.0} & 42.2 \textcolor{gray}{$\pm$ 1.0} & 46.2 \textcolor{gray}{$\pm$ 1.0} & 46.6 \textcolor{gray}{$\pm$ 1.0} & 42.9 \textcolor{gray}{$\pm$ 1.0} & 42.7 \textcolor{gray}{$\pm$ 1.0} & 44.8 \textcolor{gray}{$\pm$ 1.0} & 46.1 \textcolor{gray}{$\pm$ 1.0} & {\setlength{\fboxsep}{1.3pt}\colorbox{gray!15}{\hspace{1.5pt}\textbf{47.6 \textcolor{black!65}{$\pm$ 1.0}}\hspace{1.5pt}}} & 45.4 \textcolor{gray}{$\pm$ 1.0} \\
\hline
\multirow{3}{*}{Tokyo}
  & 1.5 & 87.6 \textcolor{gray}{$\pm$ 0.7} & 91.0 \textcolor{gray}{$\pm$ 0.6} & 93.7 \textcolor{gray}{$\pm$ 0.5} & 94.4 \textcolor{gray}{$\pm$ 0.5} & 94.2 \textcolor{gray}{$\pm$ 0.5} & 93.4 \textcolor{gray}{$\pm$ 0.5} & 92.5 \textcolor{gray}{$\pm$ 0.5} & 94.4 \textcolor{gray}{$\pm$ 0.5} & 95.7 \textcolor{gray}{$\pm$ 0.4} & 97.9 \textcolor{gray}{$\pm$ 0.3} & {\setlength{\fboxsep}{1.3pt}\colorbox{gray!15}{\hspace{1.5pt}\textbf{98.6 \textcolor{black!65}{$\pm$ 0.2}}\hspace{1.5pt}}} \\
  & 2.0 & 82.0 \textcolor{gray}{$\pm$ 0.8} & 84.4 \textcolor{gray}{$\pm$ 0.8} & 91.2 \textcolor{gray}{$\pm$ 0.6} & 92.0 \textcolor{gray}{$\pm$ 0.6} & 91.8 \textcolor{gray}{$\pm$ 0.6} & 90.3 \textcolor{gray}{$\pm$ 0.6} & 88.0 \textcolor{gray}{$\pm$ 0.7} & 93.0 \textcolor{gray}{$\pm$ 0.5} & 94.5 \textcolor{gray}{$\pm$ 0.5} & 97.6 \textcolor{gray}{$\pm$ 0.3} & {\setlength{\fboxsep}{1.3pt}\colorbox{gray!15}{\hspace{1.5pt}\textbf{98.4 \textcolor{black!65}{$\pm$ 0.3}}\hspace{1.5pt}}} \\
  & 2.5 & 76.1 \textcolor{gray}{$\pm$ 0.9} & 77.2 \textcolor{gray}{$\pm$ 0.9} & 90.1 \textcolor{gray}{$\pm$ 0.6} & 92.1 \textcolor{gray}{$\pm$ 0.6} & 91.4 \textcolor{gray}{$\pm$ 0.6} & 88.5 \textcolor{gray}{$\pm$ 0.7} & 87.3 \textcolor{gray}{$\pm$ 0.7} & 91.9 \textcolor{gray}{$\pm$ 0.6} & 94.0 \textcolor{gray}{$\pm$ 0.5} & 97.5 \textcolor{gray}{$\pm$ 0.3} & {\setlength{\fboxsep}{1.3pt}\colorbox{gray!15}{\hspace{1.5pt}\textbf{98.1 \textcolor{black!65}{$\pm$ 0.3}}\hspace{1.5pt}}} \\
\hline
\multirow{3}{*}{Escort}
  & 1.5 & 82.6 \textcolor{gray}{$\pm$ 0.7} & 84.7 \textcolor{gray}{$\pm$ 0.7} & 91.1 \textcolor{gray}{$\pm$ 0.5} & 92.3 \textcolor{gray}{$\pm$ 0.5} & 91.2 \textcolor{gray}{$\pm$ 0.5} & 89.9 \textcolor{gray}{$\pm$ 0.6} & 89.1 \textcolor{gray}{$\pm$ 0.6} & 88.2 \textcolor{gray}{$\pm$ 0.6} & 93.5 \textcolor{gray}{$\pm$ 0.5} & 97.7 \textcolor{gray}{$\pm$ 0.3} & {\setlength{\fboxsep}{1.3pt}\colorbox{gray!15}{\hspace{1.5pt}\textbf{99.0 \textcolor{black!65}{$\pm$ 0.2}}\hspace{1.5pt}}} \\
  & 2.0 & 74.0 \textcolor{gray}{$\pm$ 0.8} & 74.5 \textcolor{gray}{$\pm$ 0.8} & 89.3 \textcolor{gray}{$\pm$ 0.6} & 91.8 \textcolor{gray}{$\pm$ 0.5} & 89.3 \textcolor{gray}{$\pm$ 0.6} & 86.0 \textcolor{gray}{$\pm$ 0.7} & 86.5 \textcolor{gray}{$\pm$ 0.6} & 86.3 \textcolor{gray}{$\pm$ 0.7} & 91.3 \textcolor{gray}{$\pm$ 0.5} & 96.6 \textcolor{gray}{$\pm$ 0.3} & {\setlength{\fboxsep}{1.3pt}\colorbox{gray!15}{\hspace{1.5pt}\textbf{98.8 \textcolor{black!65}{$\pm$ 0.2}}\hspace{1.5pt}}} \\
  & 2.5 & 69.4 \textcolor{gray}{$\pm$ 0.9} & 67.9 \textcolor{gray}{$\pm$ 0.9} & 91.8 \textcolor{gray}{$\pm$ 0.5} & 94.7 \textcolor{gray}{$\pm$ 0.4} & 93.1 \textcolor{gray}{$\pm$ 0.5} & 86.6 \textcolor{gray}{$\pm$ 0.6} & 89.9 \textcolor{gray}{$\pm$ 0.6} & 86.6 \textcolor{gray}{$\pm$ 0.6} & 91.1 \textcolor{gray}{$\pm$ 0.5} & 96.7 \textcolor{gray}{$\pm$ 0.3} & {\setlength{\fboxsep}{1.3pt}\colorbox{gray!15}{\hspace{1.5pt}\textbf{99.2 \textcolor{black!65}{$\pm$ 0.2}}\hspace{1.5pt}}} \\
\hline
\end{tabular}}
\end{table*}

\begin{figure*}[t]
    \centering
    \renewcommand{\arraystretch}{0.5}
    \setlength{\tabcolsep}{3pt}
    \begin{tabular}{r ccc}
        &
        \footnotesize\textbf{Top-5} &
        \footnotesize\textbf{Top-1} &
        \footnotesize\textbf{Brier Score} \\[4pt]

        \adjustbox{valign=c}{\rotatebox{90}{\scriptsize\textbf{Baboons}}} &
        \adjustbox{valign=c}{\includegraphics[width=0.30\textwidth]{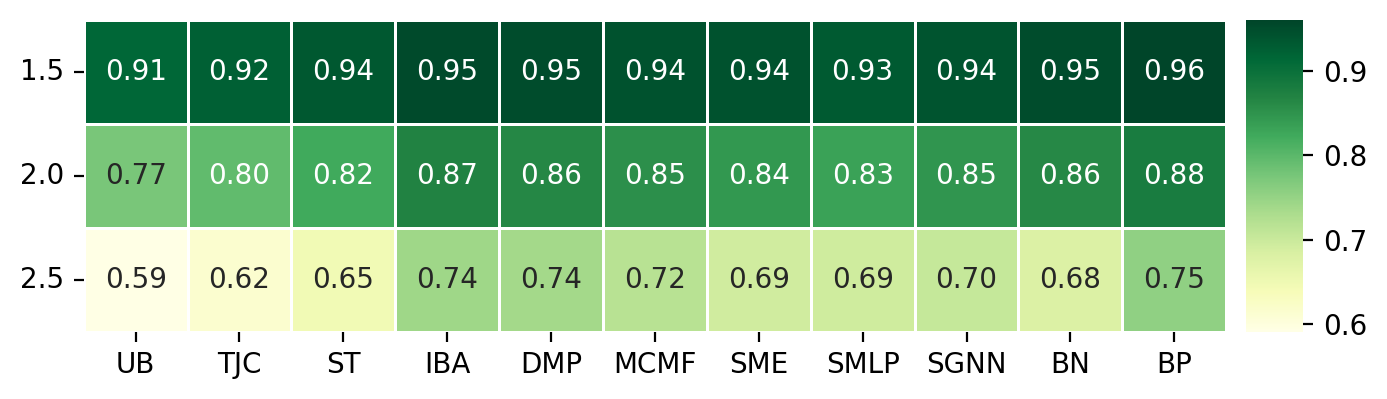}} &
        \adjustbox{valign=c}{\includegraphics[width=0.30\textwidth]{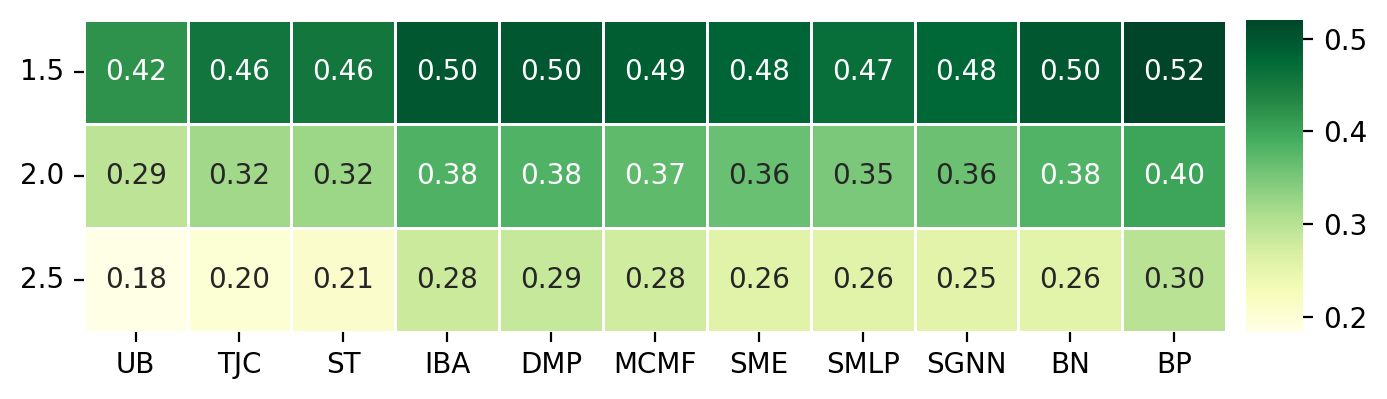}} &
        \adjustbox{valign=c}{\includegraphics[width=0.30\textwidth]{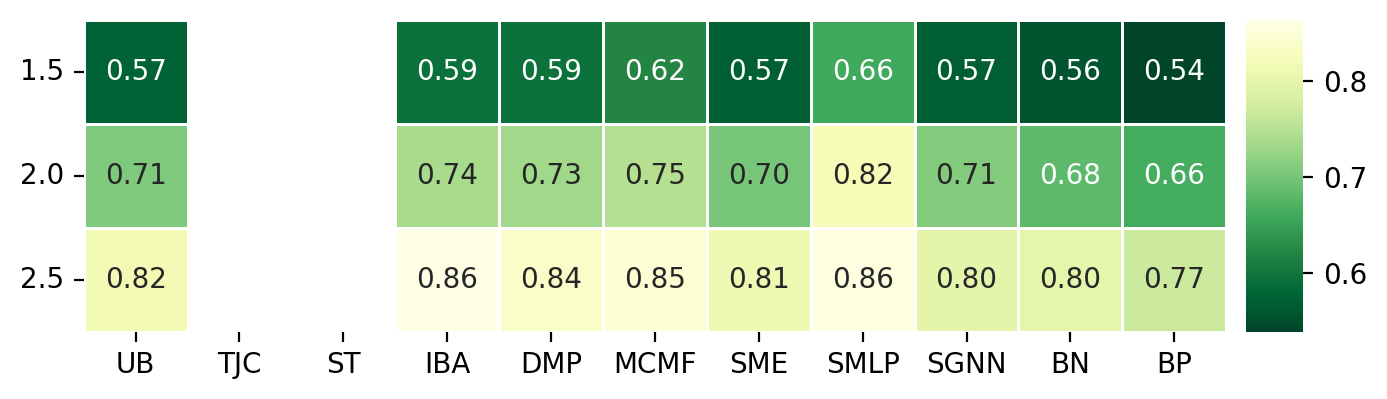}} \\[2pt]

        \adjustbox{valign=c}{\rotatebox{90}{\scriptsize\textbf{Malawi}}} &
        \adjustbox{valign=c}{\includegraphics[width=0.30\textwidth]{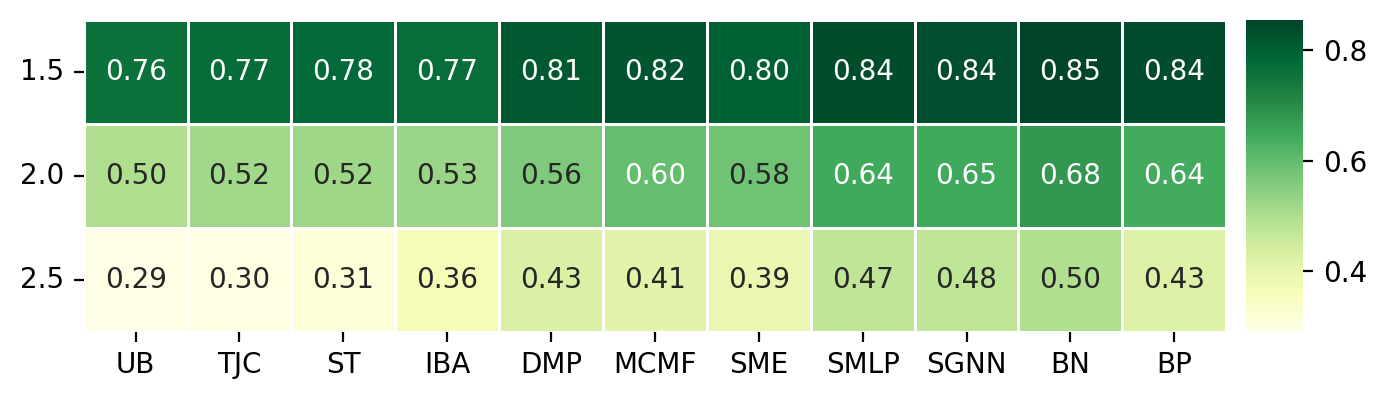}} &
        \adjustbox{valign=c}{\includegraphics[width=0.30\textwidth]{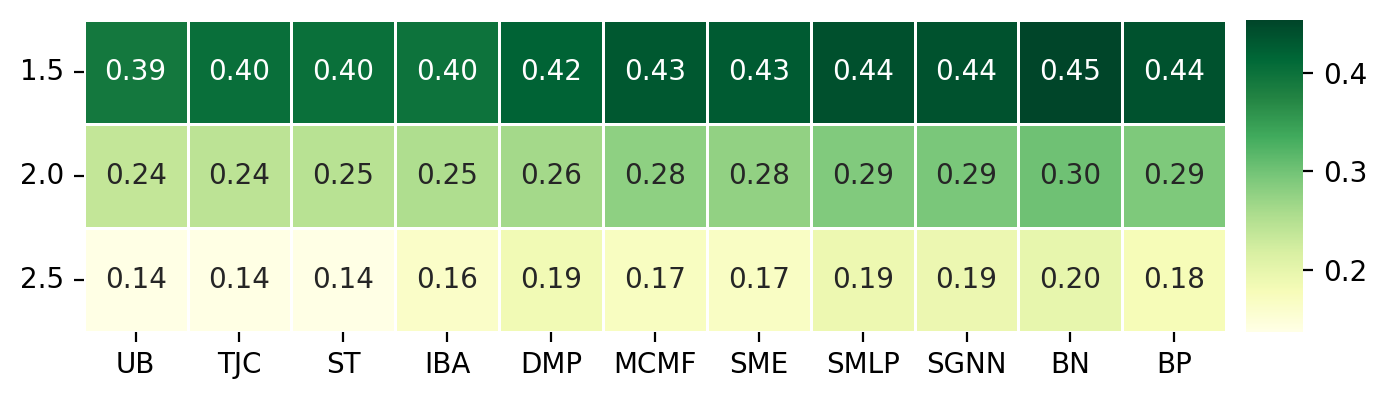}} &
        \adjustbox{valign=c}{\includegraphics[width=0.30\textwidth]{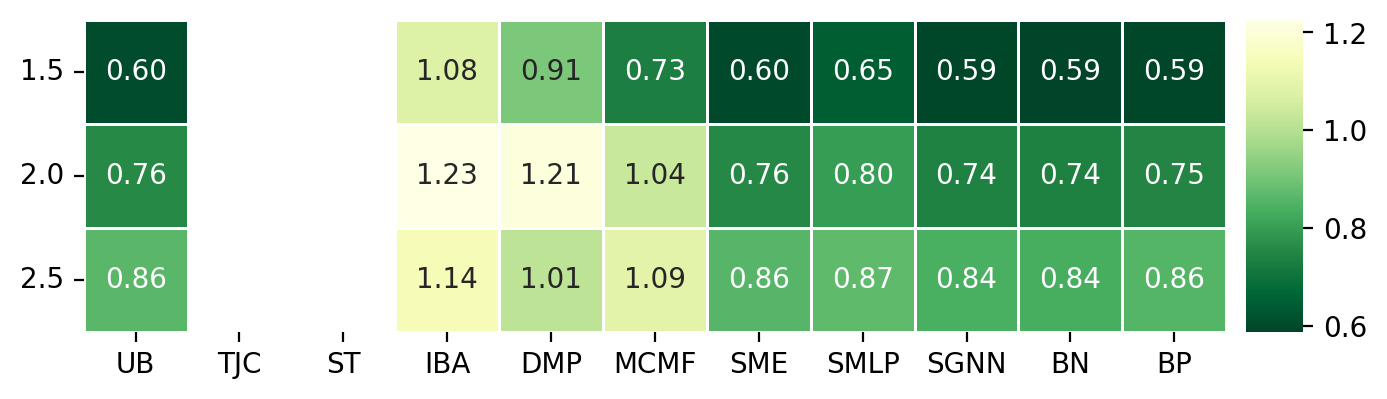}} \\[2pt]

        \adjustbox{valign=c}{\rotatebox{90}{\scriptsize\textbf{Boston}}} &
        \adjustbox{valign=c}{\includegraphics[width=0.30\textwidth]{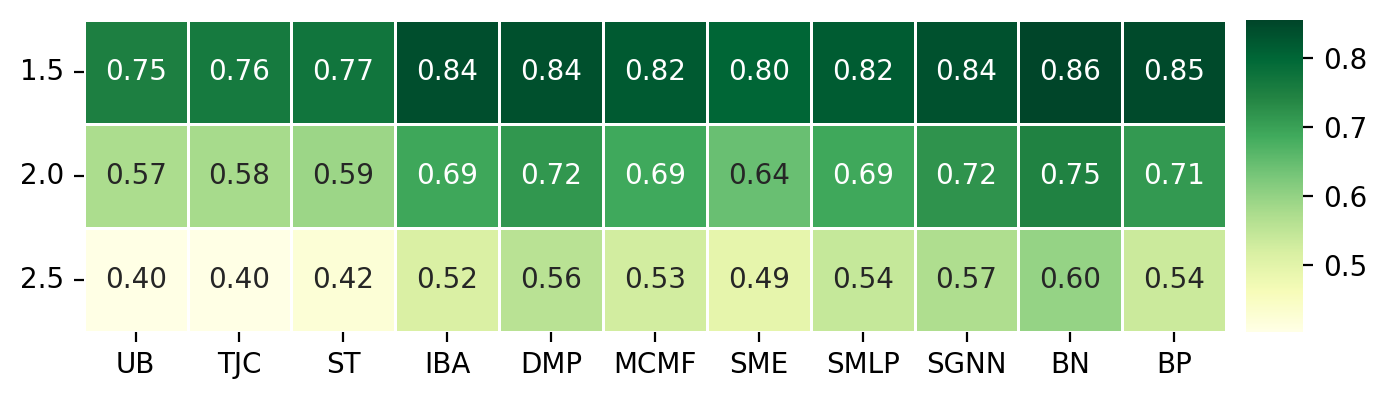}} &
        \adjustbox{valign=c}{\includegraphics[width=0.30\textwidth]{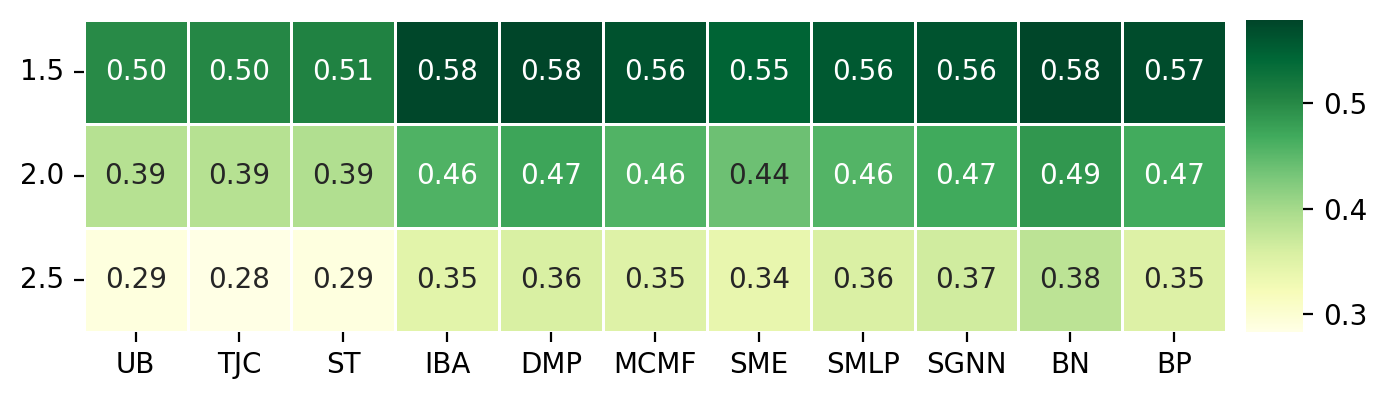}} &
        \adjustbox{valign=c}{\includegraphics[width=0.30\textwidth]{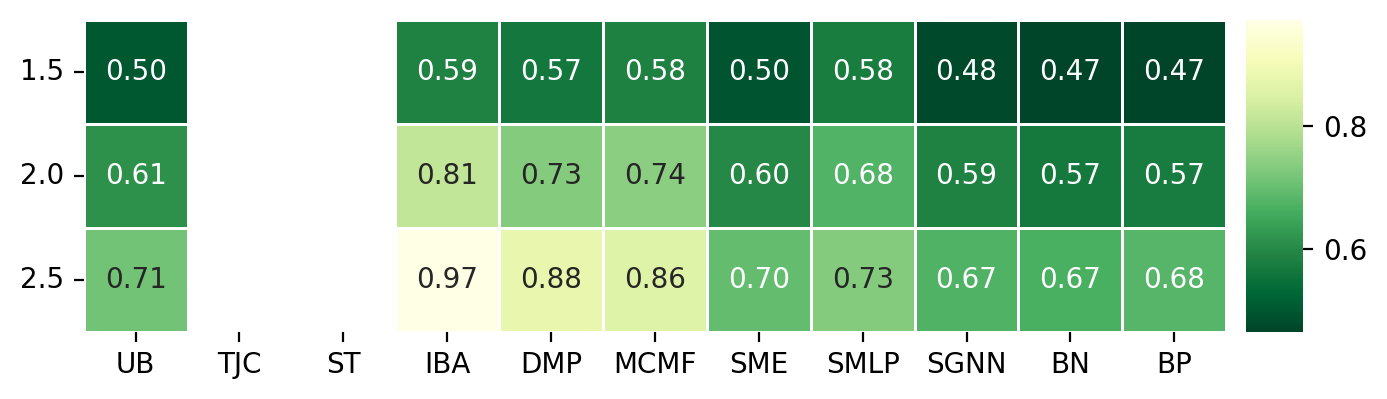}} \\[2pt]

        \adjustbox{valign=c}{\rotatebox{90}{\scriptsize\textbf{Office}}} &
        \adjustbox{valign=c}{\includegraphics[width=0.30\textwidth]{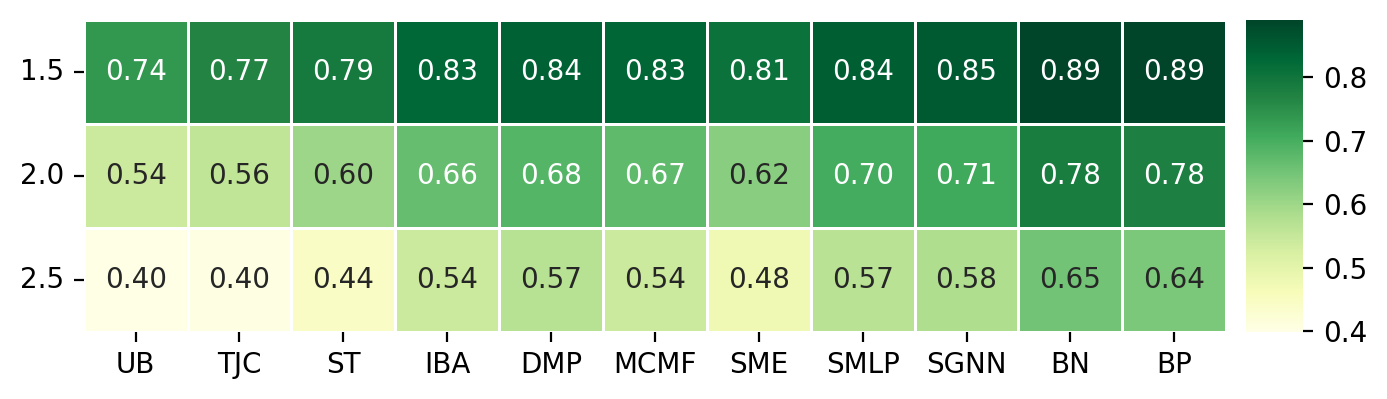}} &
        \adjustbox{valign=c}{\includegraphics[width=0.30\textwidth]{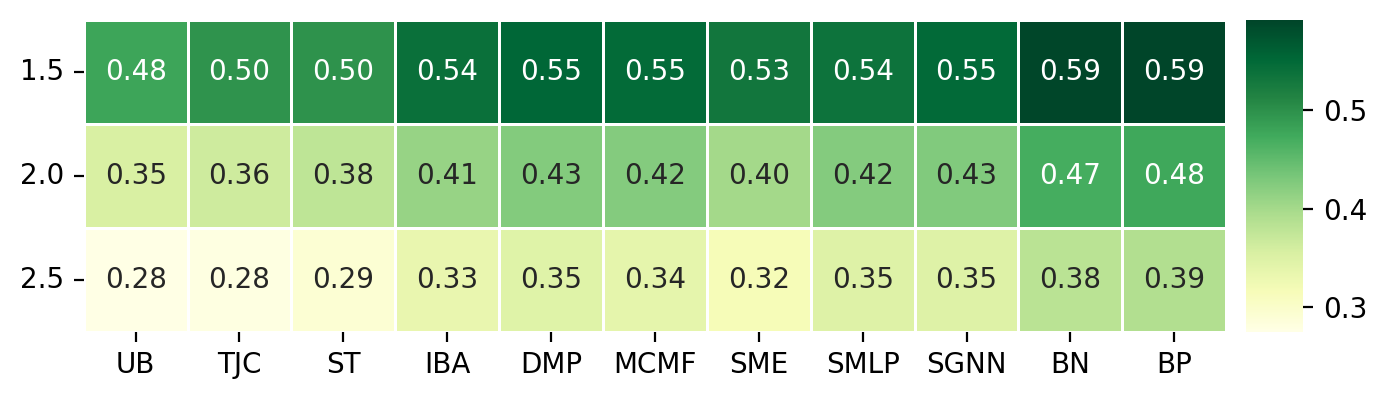}} &
        \adjustbox{valign=c}{\includegraphics[width=0.30\textwidth]{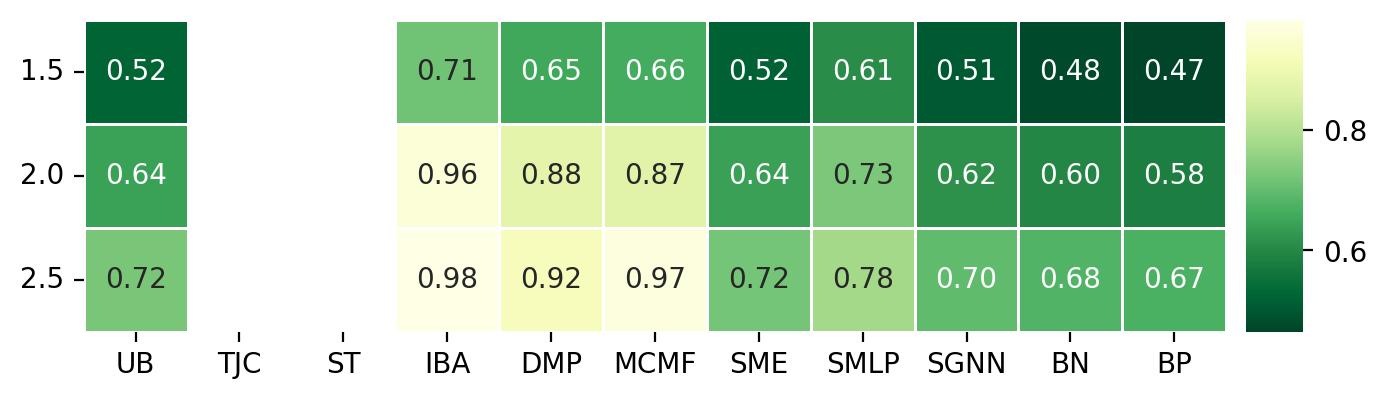}} \\[2pt]

        \adjustbox{valign=c}{\rotatebox{90}{\scriptsize\textbf{New Y.}}} &
        \adjustbox{valign=c}{\includegraphics[width=0.30\textwidth]{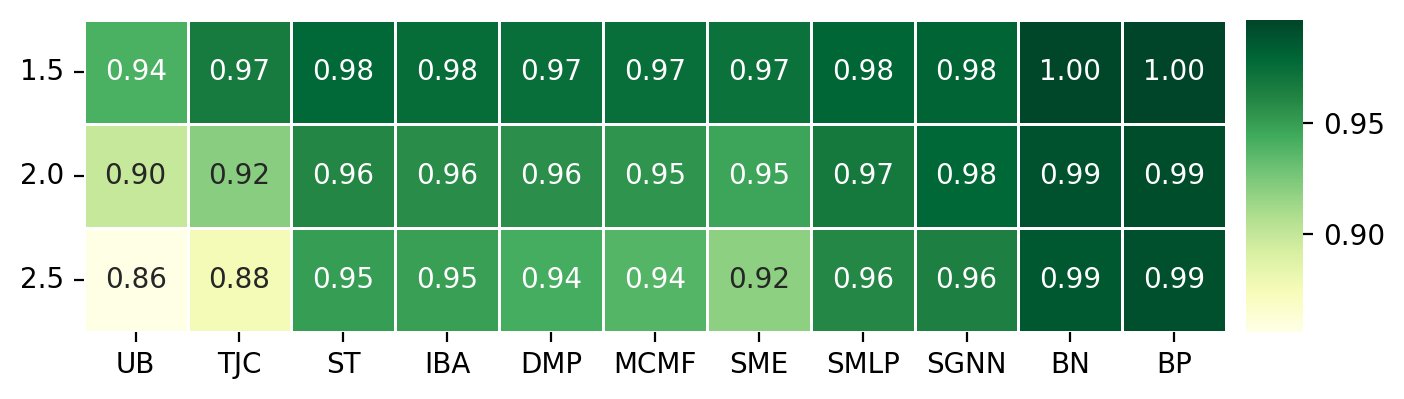}} &
        \adjustbox{valign=c}{\includegraphics[width=0.30\textwidth]{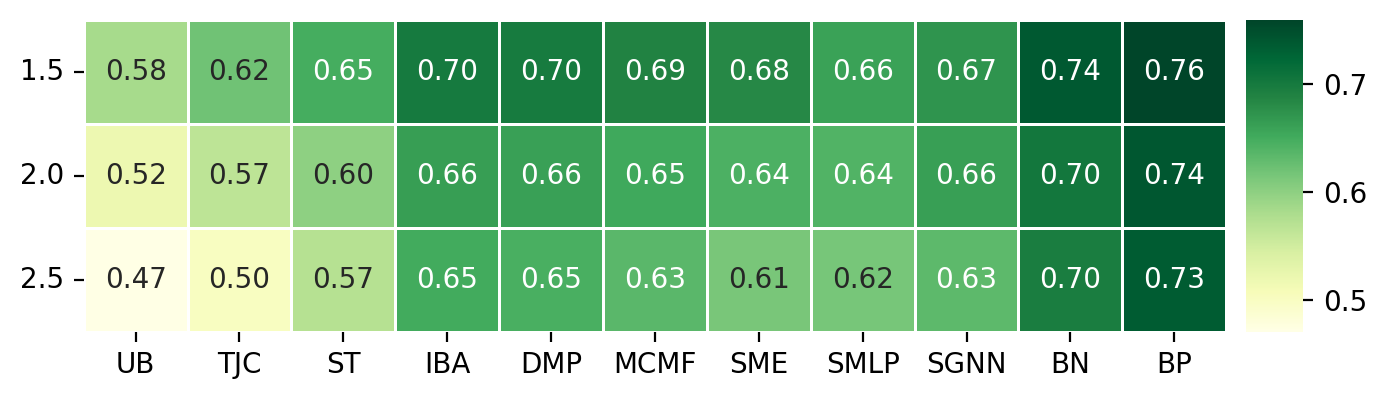}} &
        \adjustbox{valign=c}{\includegraphics[width=0.30\textwidth]{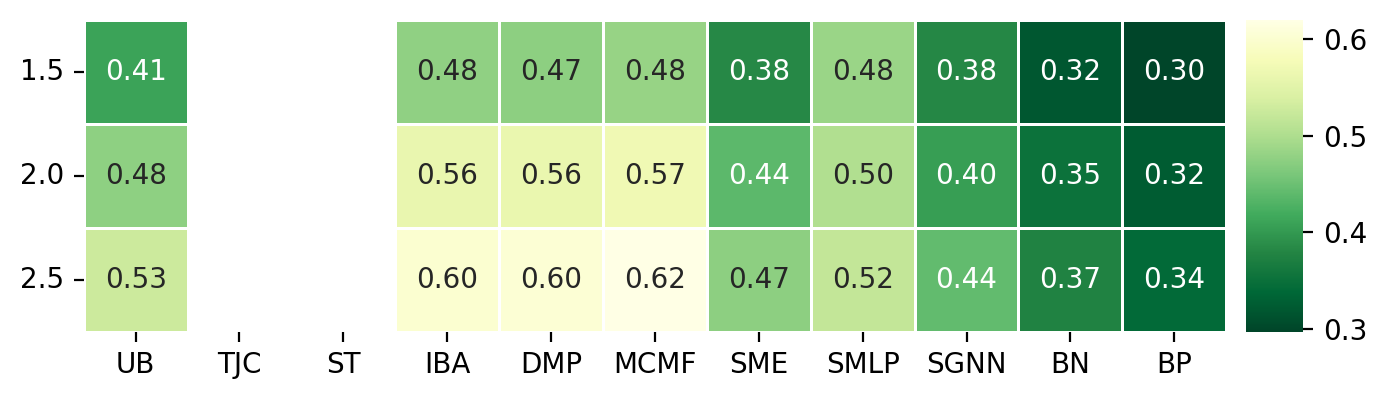}} \\[2pt]

        \adjustbox{valign=c}{\rotatebox{90}{\scriptsize\textbf{Copenh.}}} &
        \adjustbox{valign=c}{\includegraphics[width=0.30\textwidth]{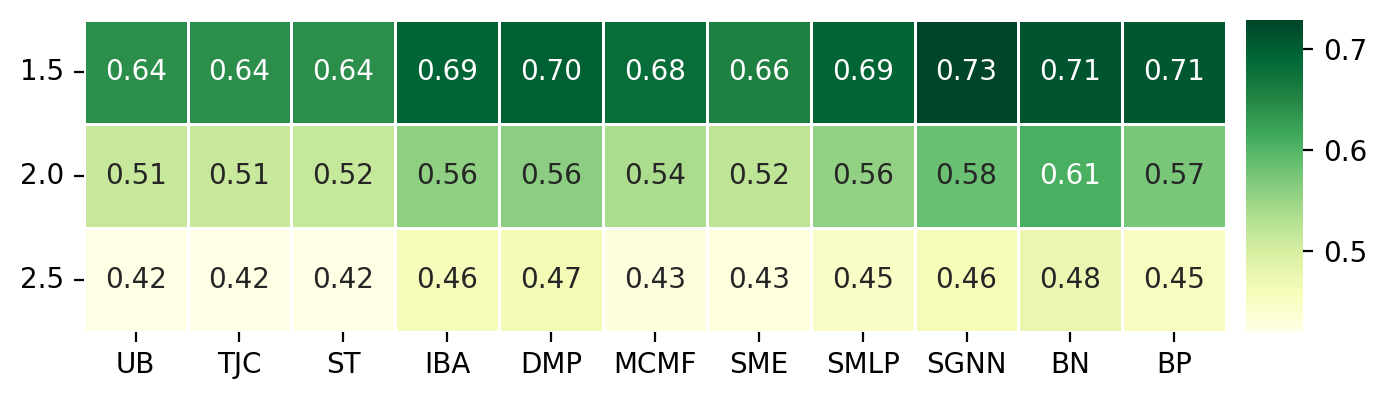}} &
        \adjustbox{valign=c}{\includegraphics[width=0.30\textwidth]{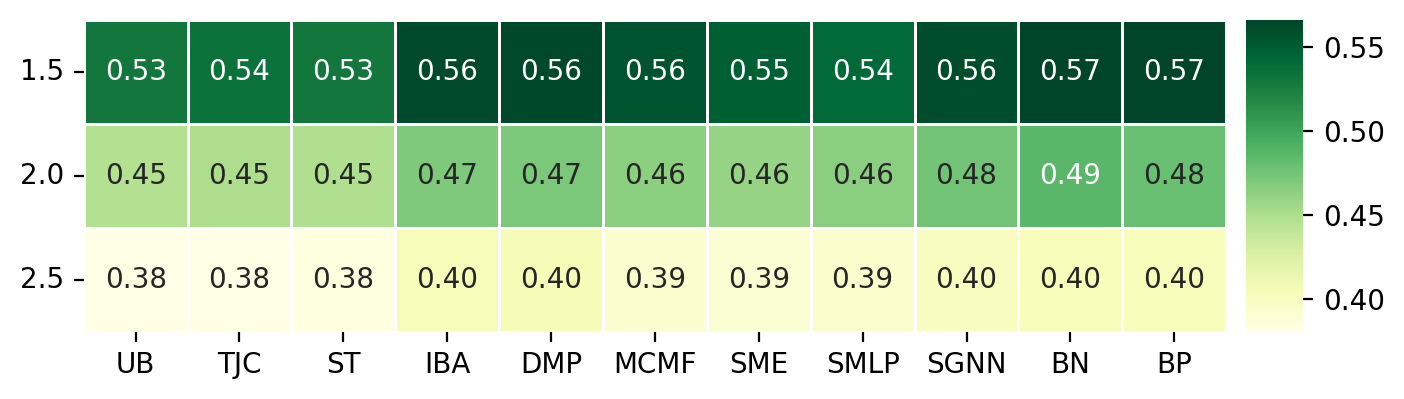}} &
        \adjustbox{valign=c}{\includegraphics[width=0.30\textwidth]{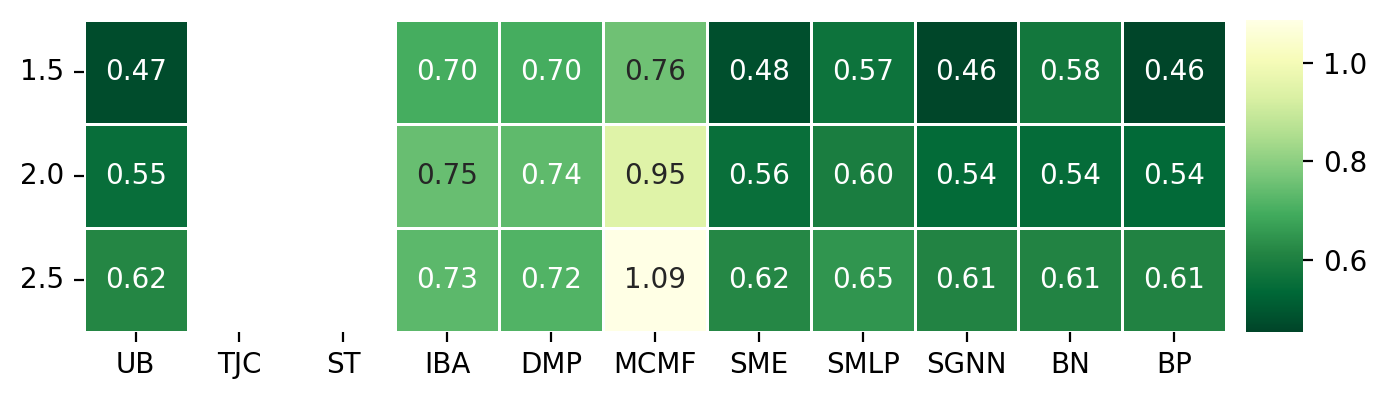}} \\[2pt]

        \adjustbox{valign=c}{\rotatebox{90}{\scriptsize\textbf{Tokyo}}} &
        \adjustbox{valign=c}{\includegraphics[width=0.30\textwidth]{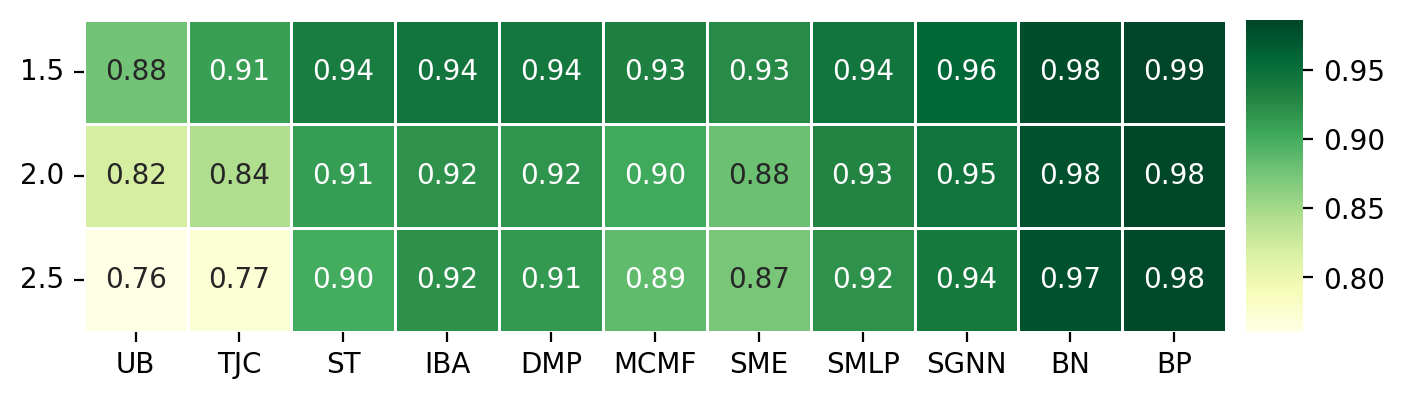}} &
        \adjustbox{valign=c}{\includegraphics[width=0.30\textwidth]{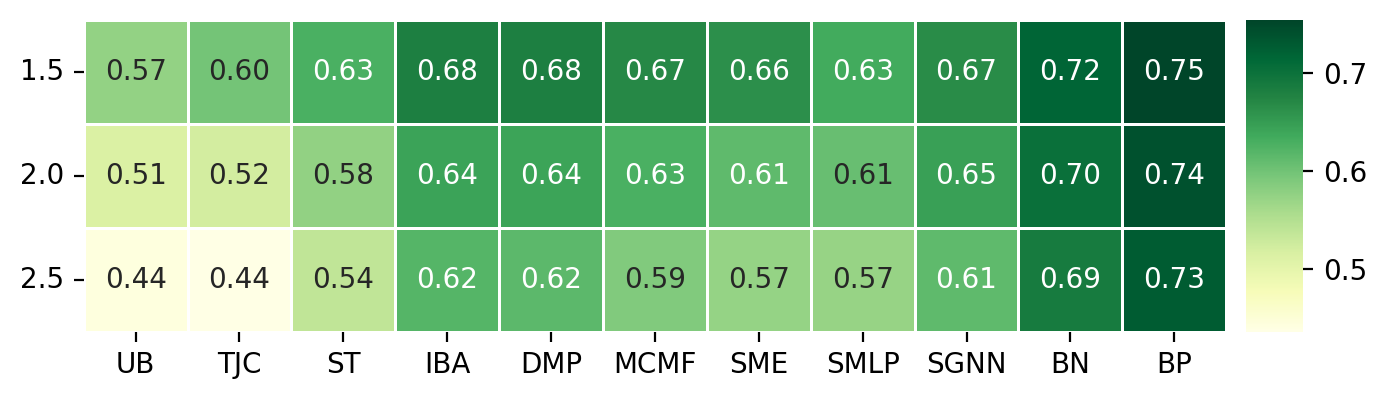}} &
        \adjustbox{valign=c}{\includegraphics[width=0.30\textwidth]{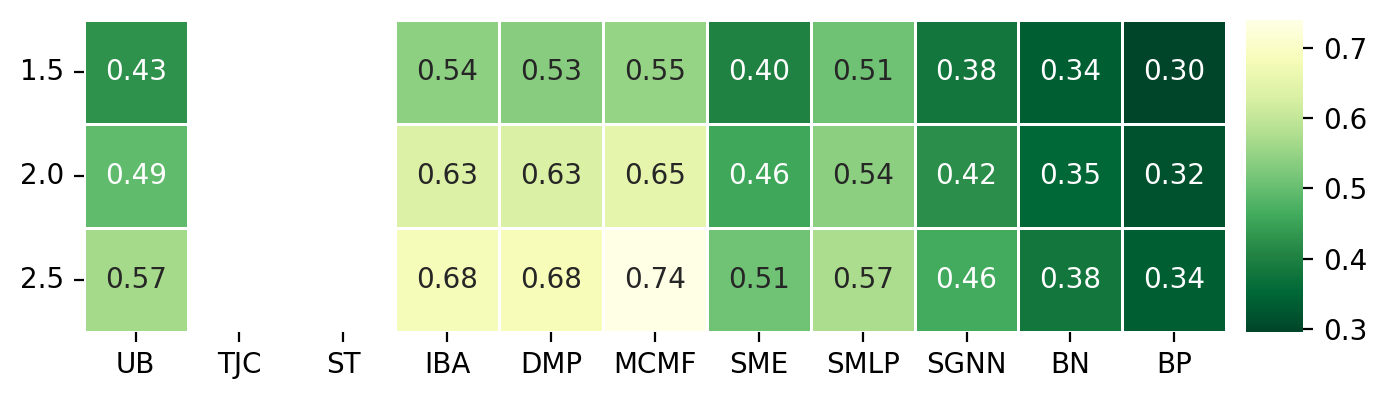}} \\[2pt]

        \adjustbox{valign=c}{\rotatebox{90}{\scriptsize\textbf{Escort}}} &
        \adjustbox{valign=c}{\includegraphics[width=0.30\textwidth]{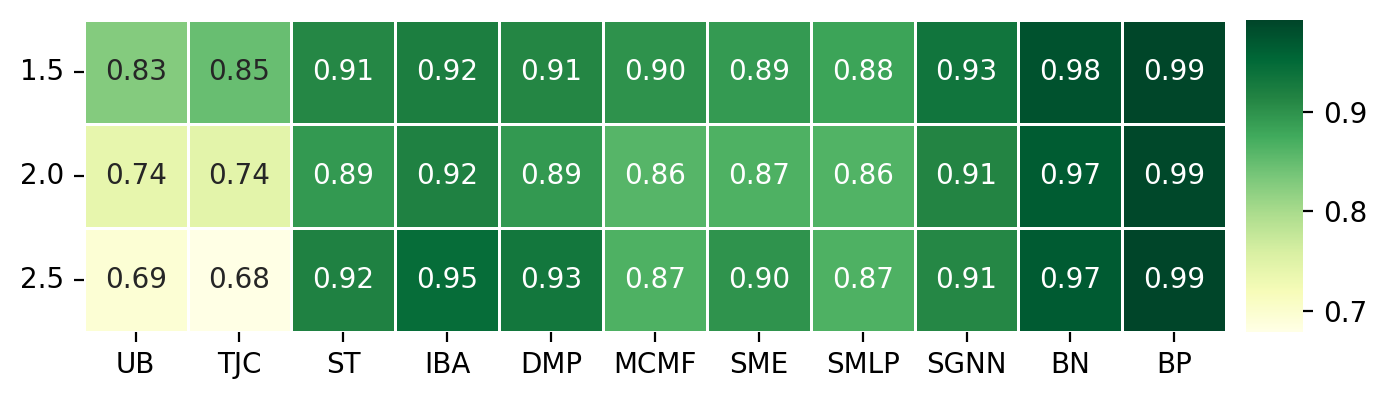}} &
        \adjustbox{valign=c}{\includegraphics[width=0.30\textwidth]{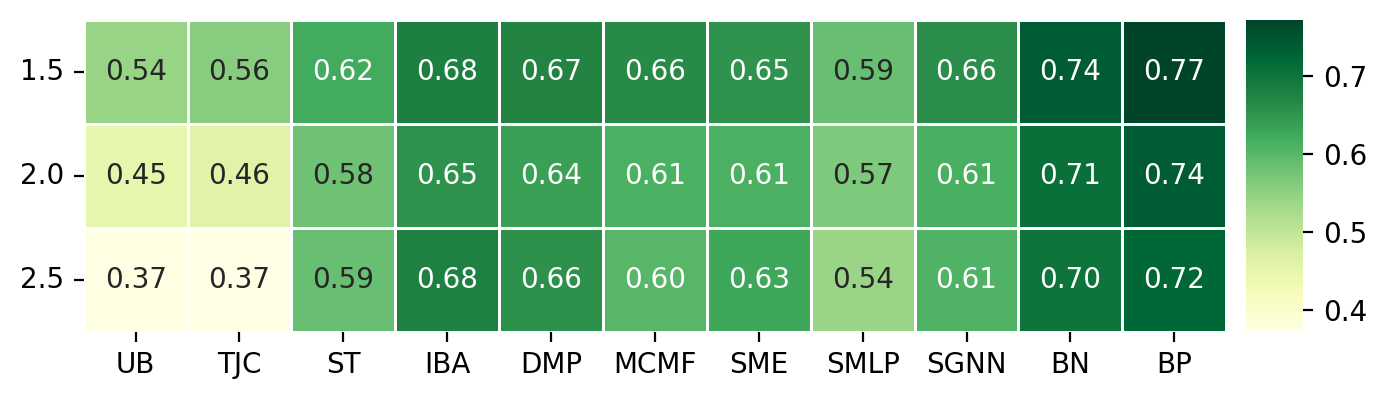}} &
        \adjustbox{valign=c}{\includegraphics[width=0.30\textwidth]{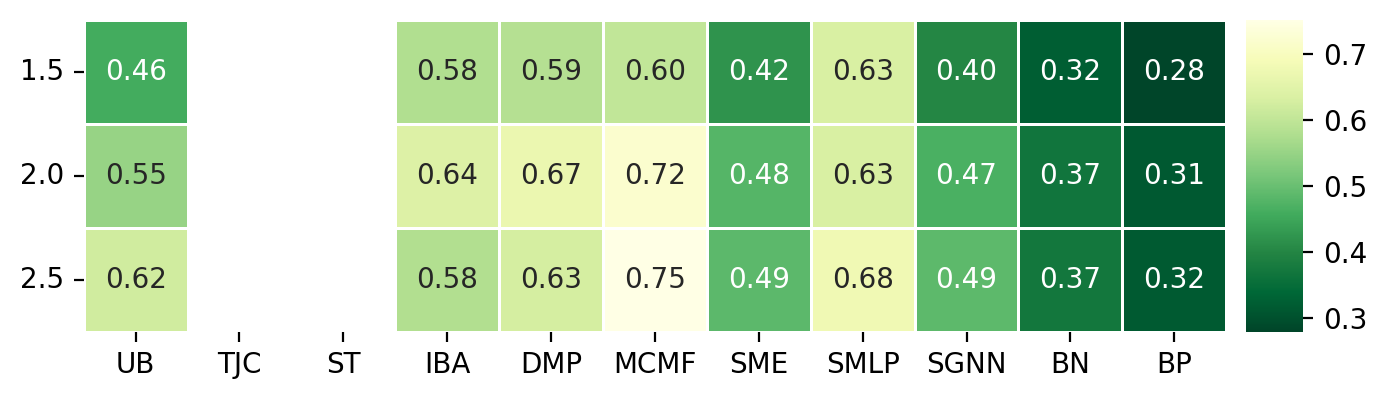}} \\

    \end{tabular}
    \caption{Evaluation heatmaps (top-5 accuracy, top-1 accuracy, and Brier score) across contact networks and methods. A darker color indicates better performance. The Brier score is not applicable to TJC and ST, since they do not output a probability distribution.}
    \label{fig:heatmaps}
\end{figure*}

\begin{figure*}[t]
    \centering
    \renewcommand{\arraystretch}{0.5}
    \setlength{\tabcolsep}{3pt}
    \begin{tabular}{r ccc c}
        &
        \footnotesize\textbf{$R_0 = 1.5$} &
        \footnotesize\textbf{$R_0 = 2.0$} &
        \footnotesize\textbf{$R_0 = 2.5$} &
        \\[4pt]

        \adjustbox{valign=c}{\rotatebox{90}{\footnotesize\textbf{Office}}} &
        \adjustbox{valign=c}{\includegraphics[width=0.28\textwidth]{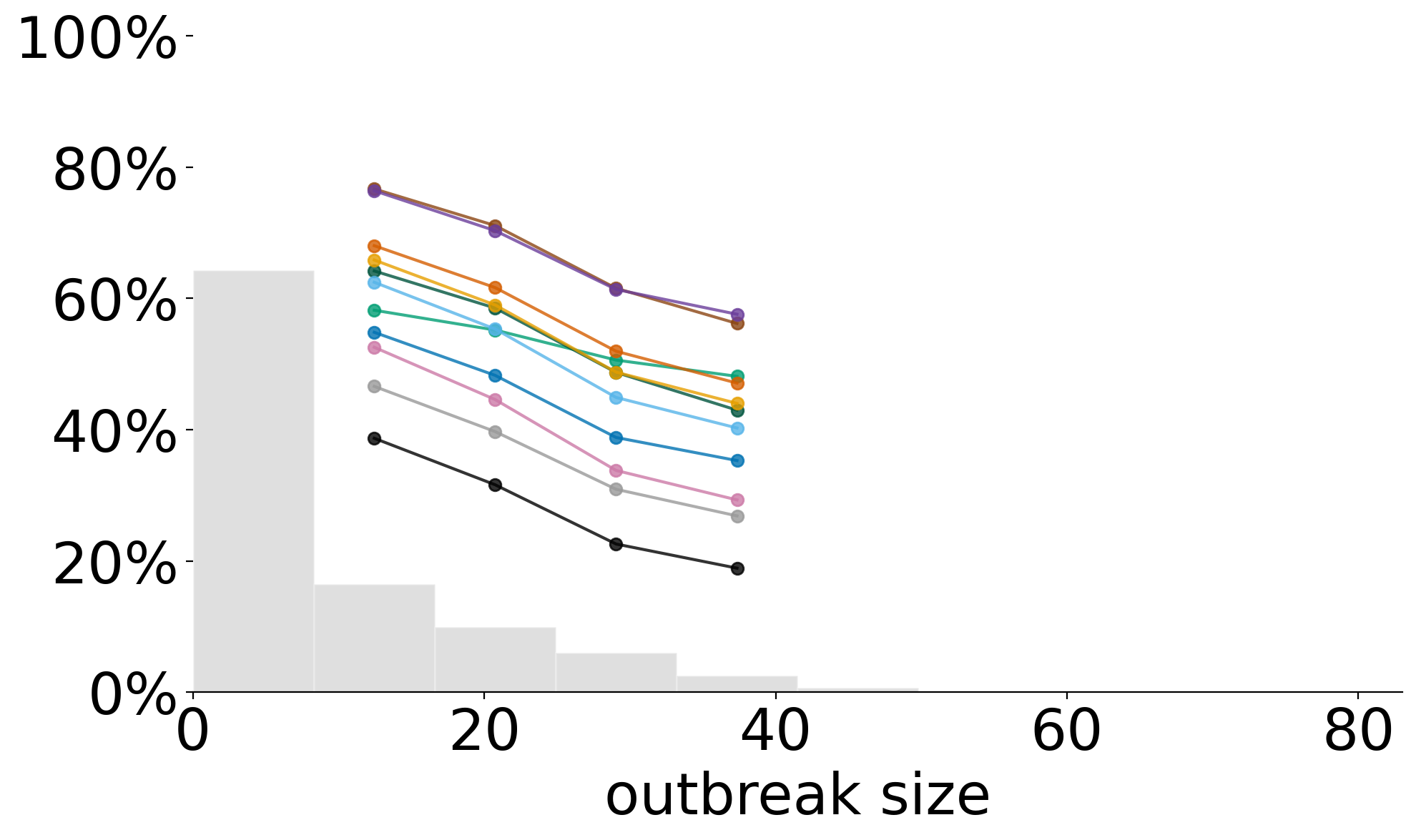}} &
        \adjustbox{valign=c}{\includegraphics[width=0.28\textwidth]{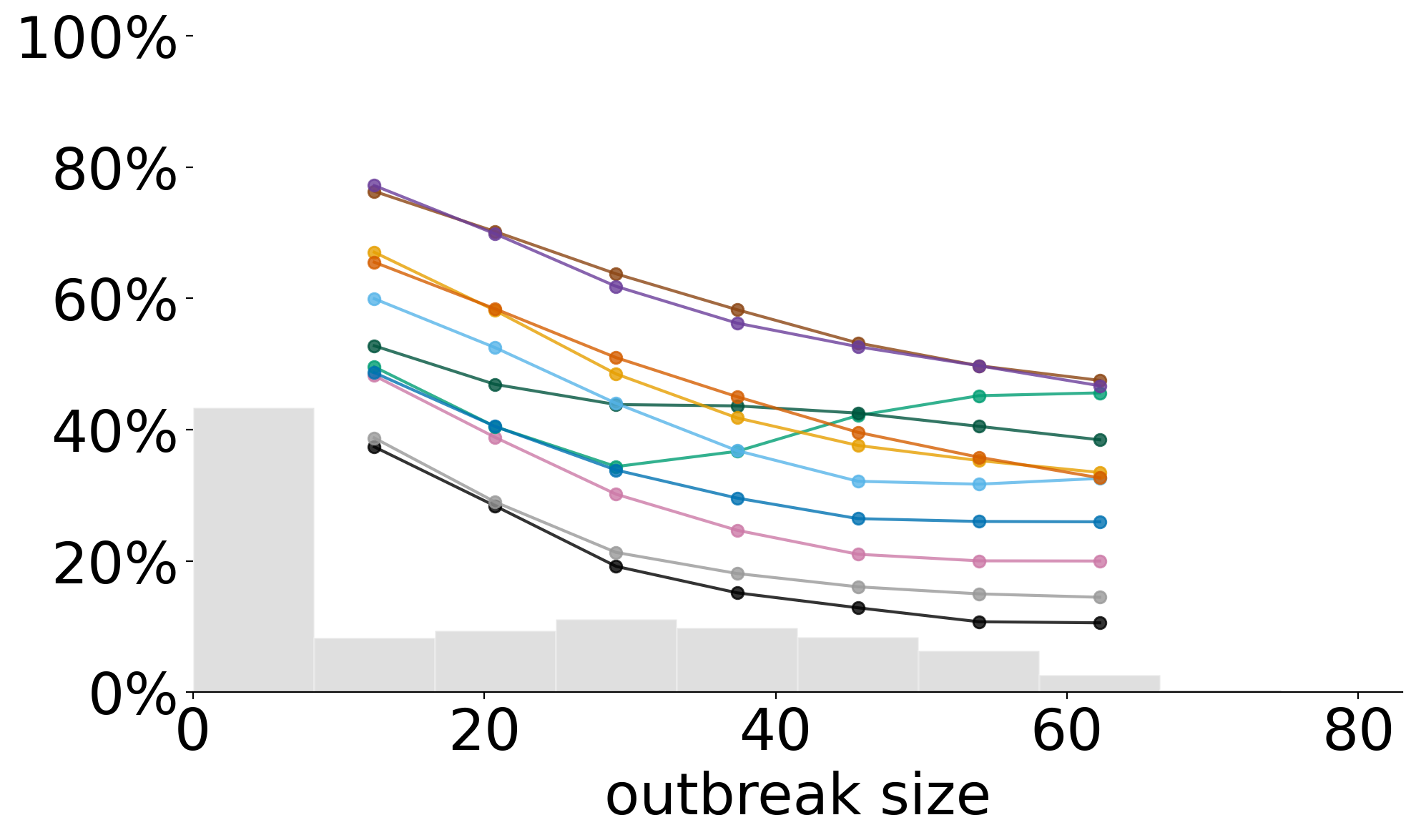}} &
        \adjustbox{valign=c}{\includegraphics[width=0.28\textwidth]{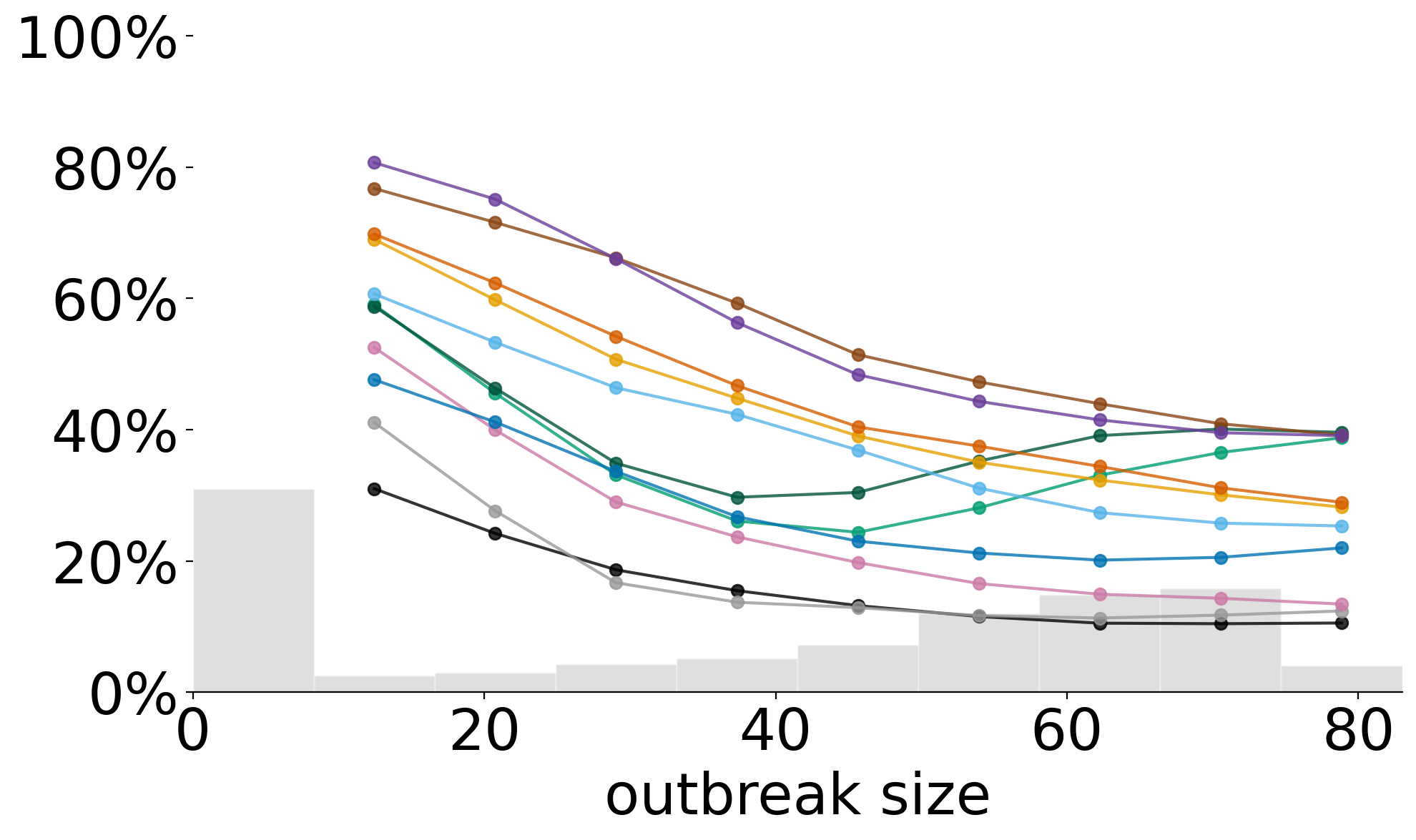}} &
        \raisebox{-0.5\height}[0pt][0pt]{\includegraphics[width=0.0889\textwidth]{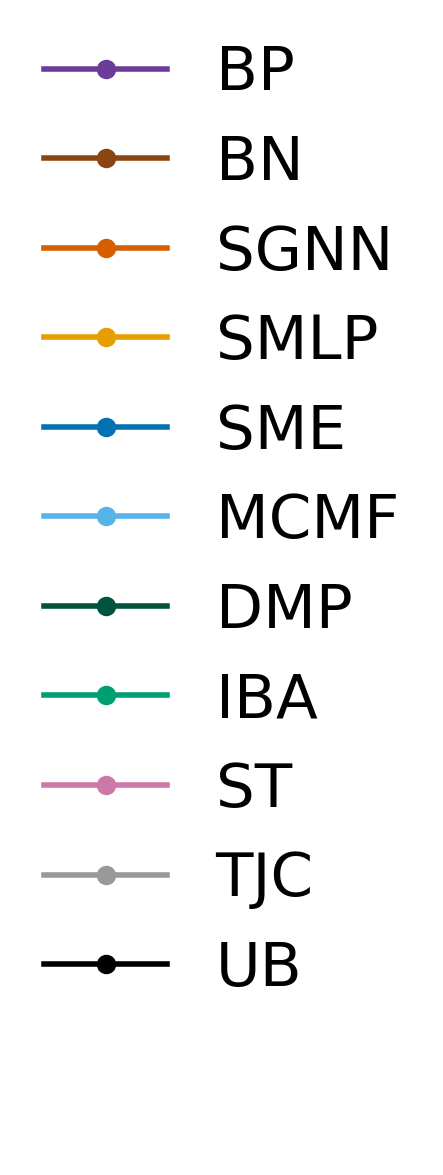}} \\[2pt]

        \adjustbox{valign=c}{\rotatebox{90}{\footnotesize\textbf{Escort}}} &
        \adjustbox{valign=c}{\includegraphics[width=0.28\textwidth]{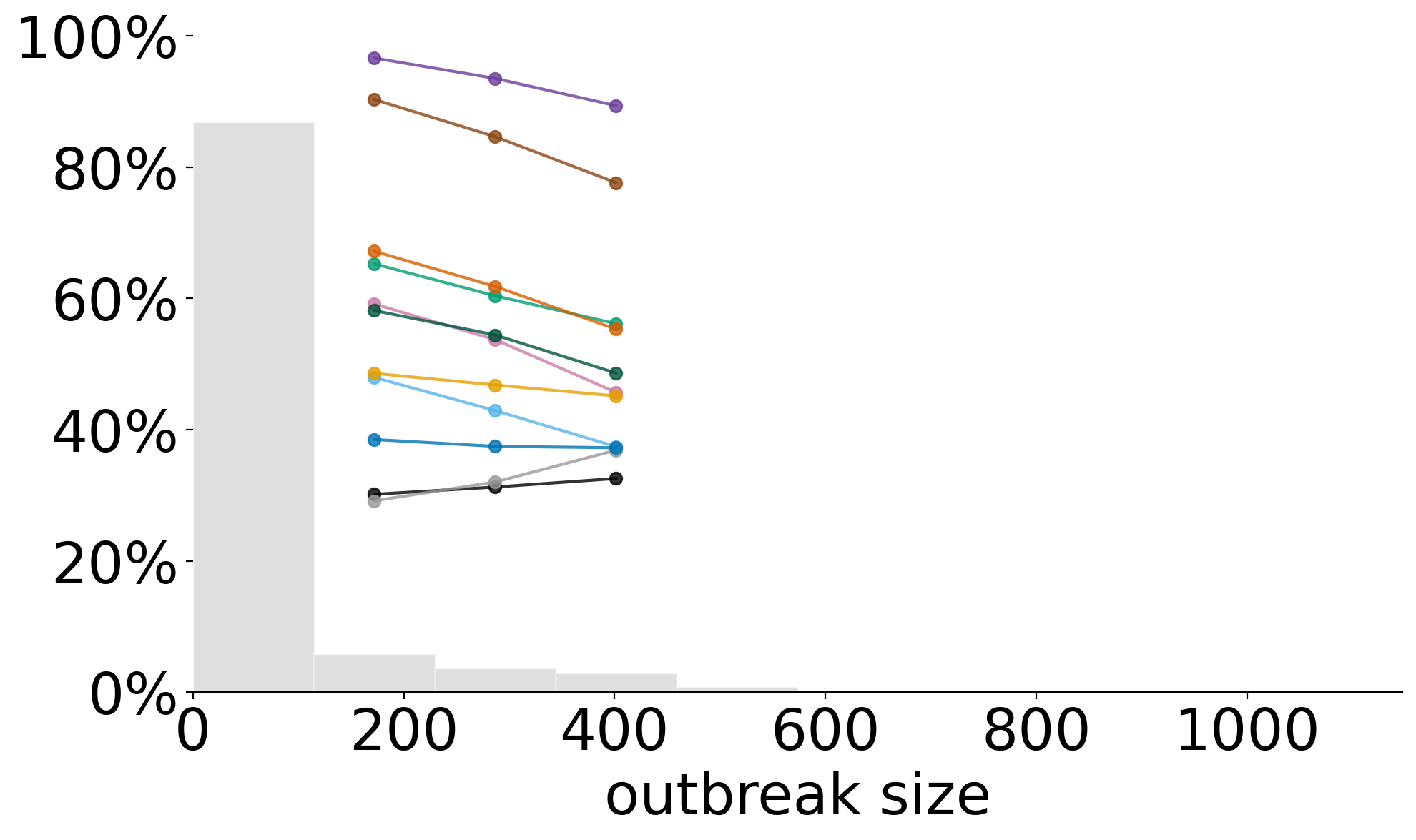}} &
        \adjustbox{valign=c}{\includegraphics[width=0.28\textwidth]{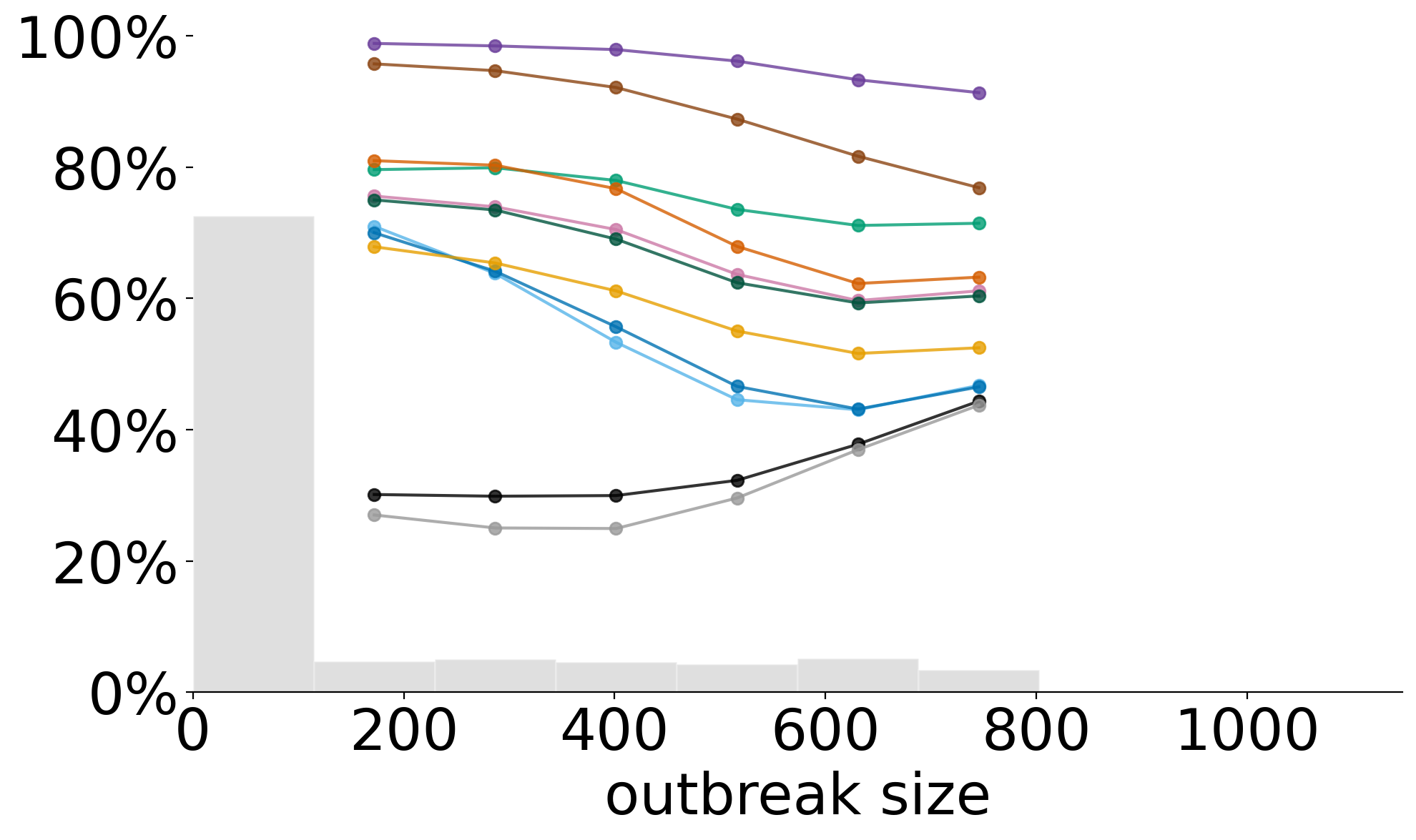}} &
        \adjustbox{valign=c}{\includegraphics[width=0.28\textwidth]{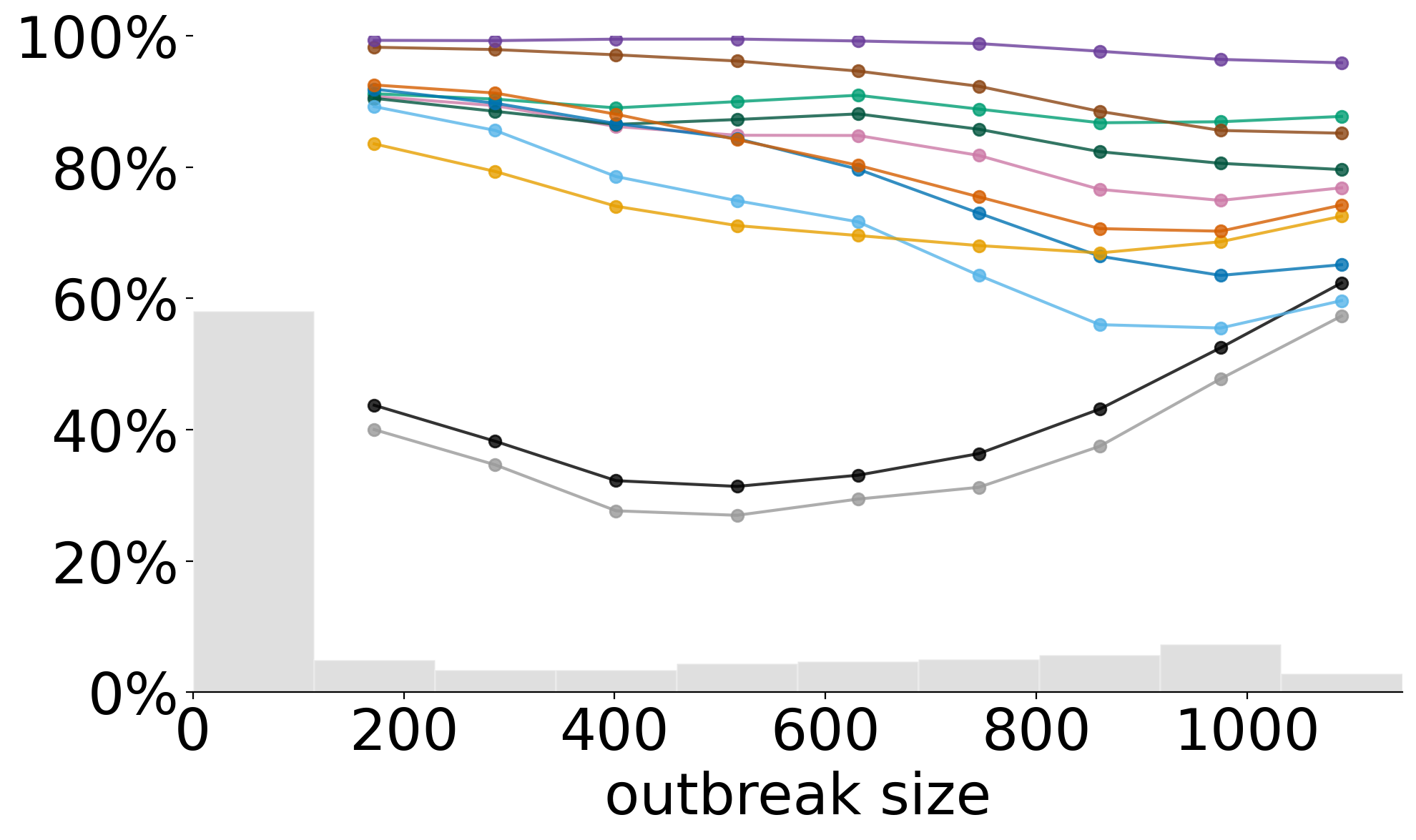}} &
         \\

    \end{tabular}
    \caption{Top-5 accuracy vs.\ outbreak size on the Office and Escort networks and for the three values of $R_0$. The test data is grouped into 10 bins of equal width in outbreak size. The gray histogram in the background shows the distribution of outbreak sizes. Lines are smoothed with a three-bin weighted moving average with weights $(\tfrac{1}{4}, \tfrac{1}{2}, \tfrac{1}{4})$.}
    \label{fig:ranks_vs_outbreak_size}
\end{figure*}

The following four subsections present the results of our main experiment and the three complementary experiments.

\subsection{Main Experiment}
\label{sec:main_exp}

Table~\ref{tab:main_results} summarizes the results of our main experiment, in terms of top-$5$ accuracy. Figure~\ref{fig:heatmaps} complements the table with heatmaps that also show top-$1$ accuracy and the Brier score. Figure~\ref{fig:ranks_vs_outbreak_size} shows the top-$5$ accuracy as a function of outbreak size for two selected networks, namely Office and Escort; the corresponding plots for the other networks are reported in the Supplementary Material, Figure~\ref{fig:ranks_vs_outbreak_size_appendix}. Below, we discuss the main findings, starting with the most important ones.

\paragraph{\textit{BN and BP are the strongest methods}}
The most striking result to emerge from the data is that BN or BP ranks first on every network and for every $R_0$, with a single exception (SGNN on Copenhagen, $R_0=1.5$). BP performs best on Baboons, New York, Tokyo, and Escort, whereas BN performs best on Malawi, Boston, Office, and Copenhagen ($R_0=2.0$ and $2.5$). There is a tentative pattern: BP tends to win on more tree-like networks, and BN on more cyclic ones (measured in terms of the static GCC, see Table~\ref{tab:networks}; the tiny Baboons network is the exception). With only eight networks, however, this observation should be treated as preliminary.

\paragraph{\textit{Factorized-likelihood methods are overconfident}}
The performance of the factorized-likelihood methods IBA, DMP, and MCMF falls clearly behind the two leading methods. The factorized likelihood produces overconfident source distributions that concentrate most of their probability mass on one or a few nodes (Figure~\ref{fig:source_probs_appendix} in the Supplementary Material). This is reflected in particularly poor Brier scores, which, in some cases, are even worse than those of the uniform baseline (Figure~\ref{fig:heatmaps}). The SMLP is similarly overconfident, albeit for a different reason, most likely because it disregards the graph structure.

\paragraph{\textit{Leakiness works in favor of IBA and DMP}}
IBA and DMP are known to be ``leaky'': unlike MCMF, which approximates the true average of the individual node-state probabilities, they systematically overestimate the I/R probabilities (Figure~\ref{fig:state_density_appendix} in the Supplementary Material). Although this biases their node-state estimates, it appears to benefit source detection, as IBA and DMP perform slightly better than MCMF on most networks. We suspect that this also explains why, on the Office and Escort networks, the performance of IBA and DMP declines less steeply with outbreak size---or in some cases even increases (Figure~\ref{fig:ranks_vs_outbreak_size}). In short, overestimated I/R probabilities seem to better match large outbreaks. However, the effect does not appear on all networks (Figure~\ref{fig:ranks_vs_outbreak_size_appendix} in the Supplementary Material).

\paragraph{\textit{SME is not competitive on full snapshots}}
By design, SME avoids the overconfidence of the factorized-likelihood methods and achieves comparatively good Brier scores (Figure~\ref{fig:heatmaps}). In terms of top-$1$ and top-$5$ accuracy, however, SME is not competitive: it performs worse than MCMF on almost all networks, with the exception of Escort for $R_0 \geq 2.0$. As shown in Section~\ref{sec:results_partial}, this picture changes when only a partial snapshot is observed.

\paragraph{\textit{The SMLP is surprisingly strong on some networks}}
On some networks, the SMLP is nearly as accurate as the SGNN, even though it has no access to the static projection of the network. On Baboons, Malawi, Office, and New York, the two differ by less than 1.5 percentage points in top-$5$ accuracy. However, on others, most notably on Escort, the SGNN clearly outperforms the SMLP.

\paragraph{\textit{Graph-geometric methods outperform UB}}
While not among the most competitive methods, ST outperforms the two baselines (UB and TJC), most notably on the large and rather sparse Tokyo and Escort networks. TJC, in turn, outperforms UB on most networks. The exceptions are Copenhagen, where the two are essentially tied, and Escort, where TJC falls behind UB at higher $R_0$.

\paragraph{\textit{High baseline accuracy is an artifact of small outbreaks}}
The accuracy of UB may seem high at first. This is because many outbreaks are trivial or small: for $R_0=1.5$, between 20 and 47\% of outbreaks, depending on the network, never spread beyond the source (Table~\ref{tab:epidemic_params}), and 57--74\% infect at most five nodes. For these outbreaks, the top-$5$ accuracy is 100\% by definition, regardless of the method. Interestingly, UB and TJC sometimes perform better on large outbreaks than on medium-sized ones, for example on Escort (Figure~\ref{fig:ranks_vs_outbreak_size}). This is because large outbreaks are reachable from fewer nodes, which reduces the candidate set $\mathcal{C}$. The effect is particularly pronounced in sparse networks.

\subsection{Learning Curves}
\label{sec:results_learning_curves}

\begin{figure*}[t]
    \centering
    \setlength{\tabcolsep}{3pt}
    \begin{tabular}{ccccc}
        \footnotesize\textbf{Baboons} & \footnotesize\textbf{Malawi} & \footnotesize\textbf{Boston} & \footnotesize\textbf{Office} & \\[2pt]
        \raisebox{-0.5\height}{\includegraphics[width=0.21\textwidth]{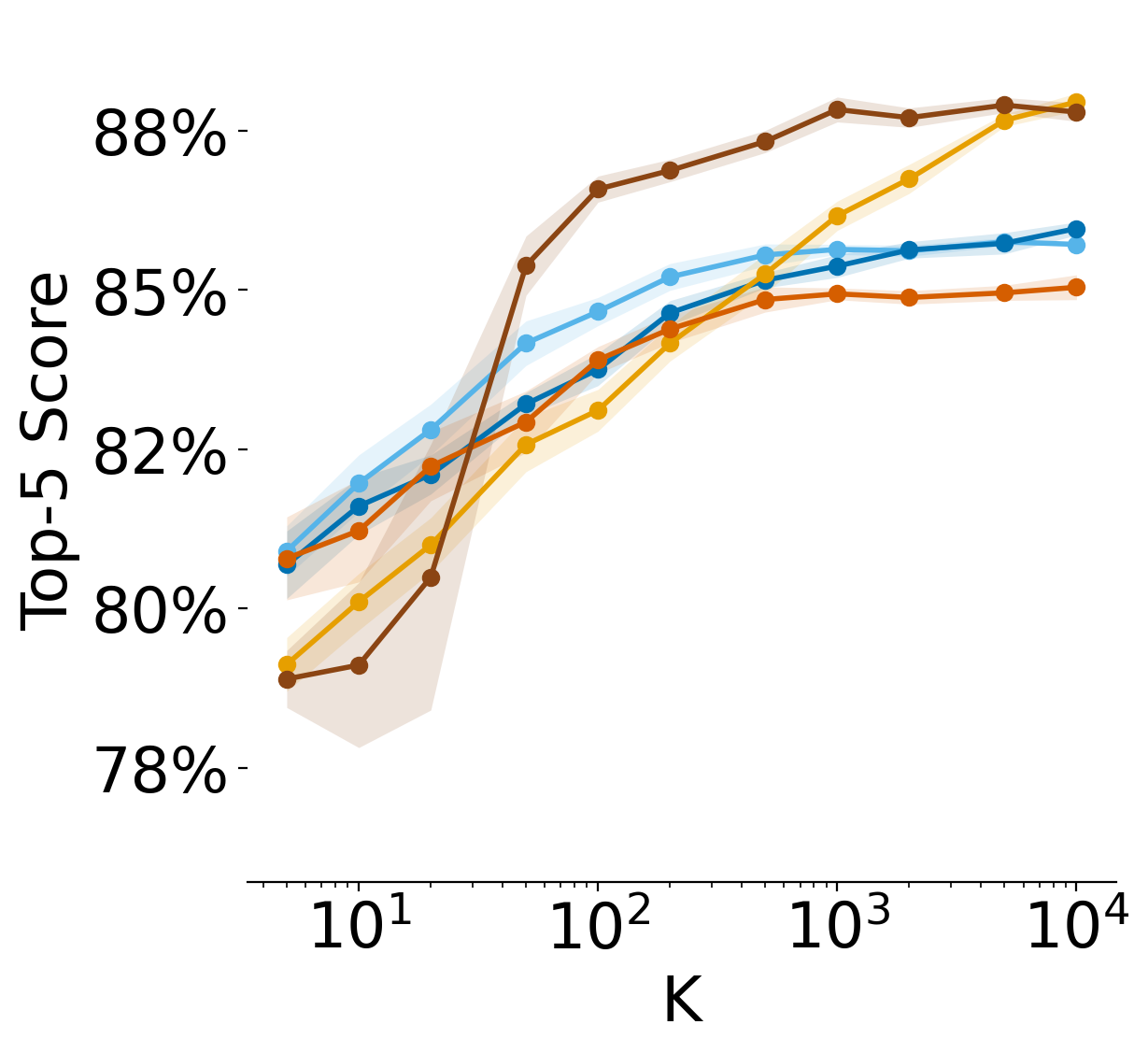}} &
        \raisebox{-0.5\height}{\includegraphics[width=0.21\textwidth]{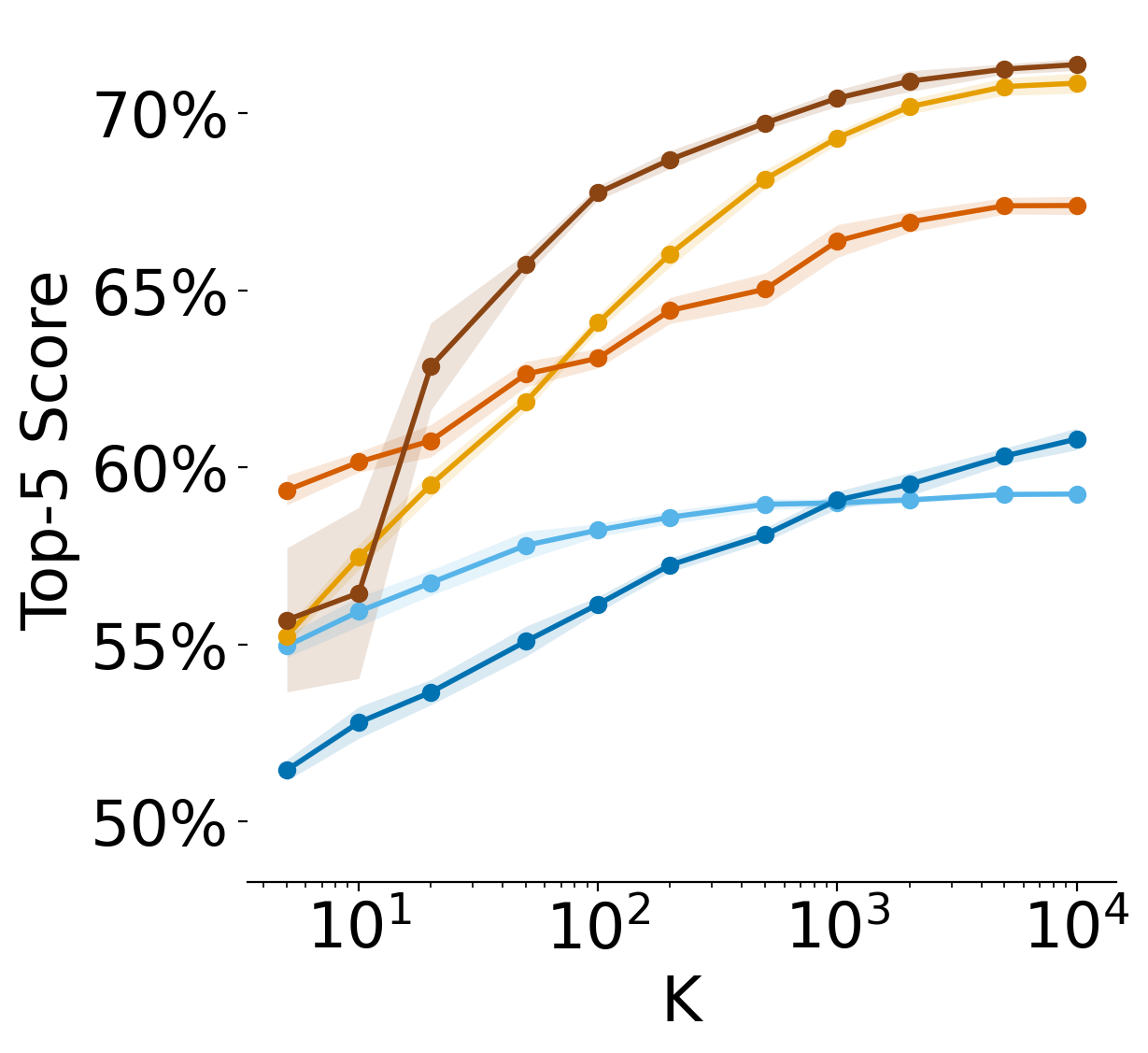}} &
        \raisebox{-0.5\height}{\includegraphics[width=0.21\textwidth]{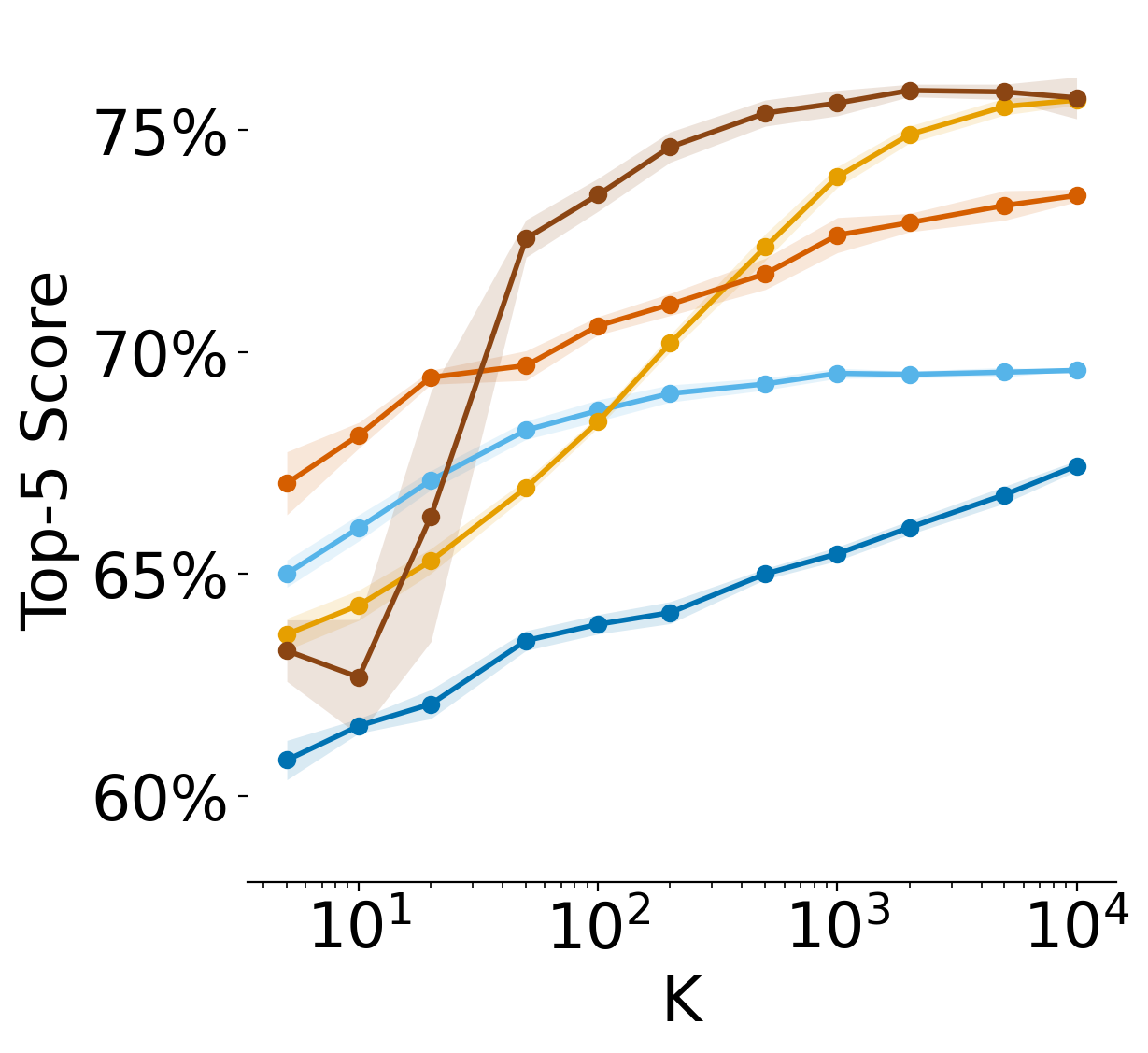}} &
        \raisebox{-0.5\height}{\includegraphics[width=0.21\textwidth]{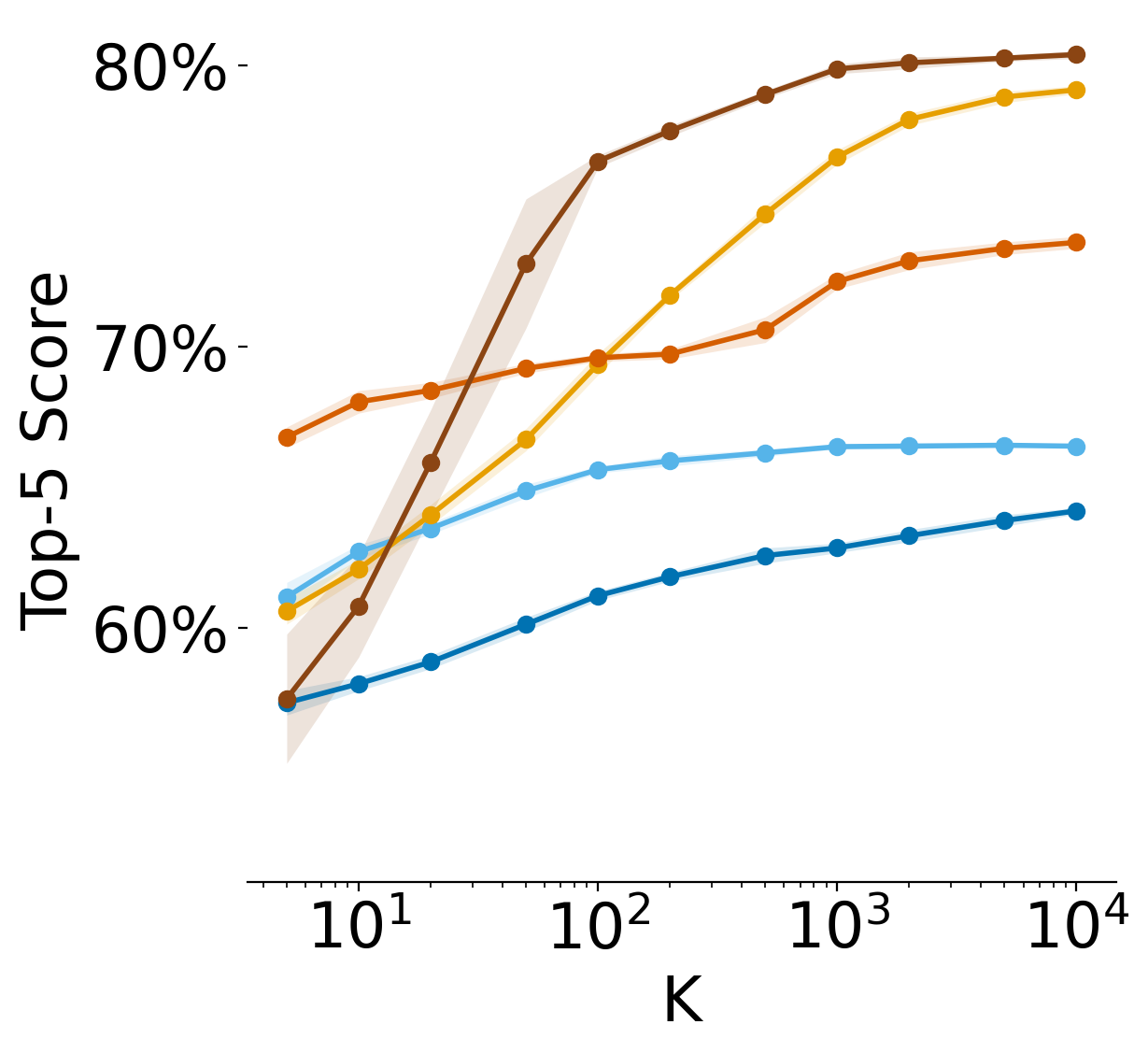}} &
        \raisebox{-0.5\height}[0pt][0pt]{\includegraphics[width=0.0889\textwidth]{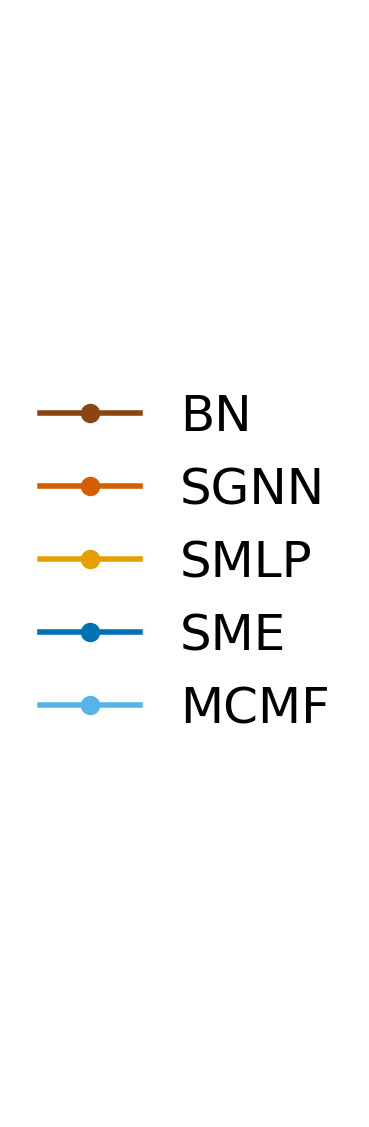}} \\
    \end{tabular}
    \caption{Top-5 accuracy vs.\ number of training snapshots per node $K$ (log-scale), for a selection of four networks. Only methods that rely on training data can be shown (MCMF, SME, SMLP, SGNN, BN). Since training outcomes vary considerably for small amounts of training data, in particular for BN, each run is repeated with 10 different seeds, and shaded bands denote the 95\% CI across these repeats. Unlike the Wald CIs in Table~\ref{tab:main_results}, these bands thus capture variability across training runs rather than across the test data.}
    \label{fig:learning_curves}
\end{figure*}

Figure~\ref{fig:learning_curves} shows the learning curves in terms of top-5 accuracy. The most noteworthy finding to emerge is that the deep learning approaches improve much more with additional training data than MCMF and SME. The quick saturation of MCMF makes sense: for a large enough $K$, the individual node-state probabilities are already well approximated, and there is little room for further improvement beyond that point, since the systematic error induced by the factorized likelihood is unaffected. SME also improves rather slowly. Its soft margin function is meant to reduce the number of simulations needed by Direct Monte Carlo, which grows exponentially in $|V|$ (there are $3^{|V|}$ possible snapshot configurations in the SIR model). Nevertheless, its learning curve remains rather flat.

The SGNN and SMLP show an interesting pattern: SGNN starts out stronger, but at some point SMLP overtakes it as more training data becomes available. We speculate that SGNN's early advantage results from having access to the entire static network. As the amount of training data grows, however, the SMLP is able to \textit{implicitly} learn the network topology, while the static topology increasingly acts as a negative inductive bias for the SGNN. In other words, the discarding of the temporal information can be seen as a form of imprecise regularization. This is consistent with the main results (Table~\ref{tab:main_results}), where the SMLP is competitive on some networks, presumably those whose topology it can implicitly learn from the available training data. Moreover, as noted before, the SMLP also suffers from an overconfidence problem, which is often not reflected in the top-$k$ score but manifests in the Brier score (Figure~\ref{fig:heatmaps}).

Finally, BN performs poorly with little training data. We suspect that this is due to its large number of parameters, resulting from its evolving edge features and weight matrices. Its wide CI bands further show that it is not only inaccurate but also unstable with little training data. With more training data, however, it gains the most and surpasses the other deep learning methods.

\subsection{Partial Observations}
\label{sec:results_partial}

\begin{figure*}[t]
    \centering
    \setlength{\tabcolsep}{3pt}
    \begin{tabular}{ccccc}
        \footnotesize\textbf{Baboons} & \footnotesize\textbf{Malawi} & \footnotesize\textbf{Boston} & \footnotesize\textbf{Office} & \\[2pt]
        \raisebox{-0.5\height}{\includegraphics[width=0.21\textwidth]{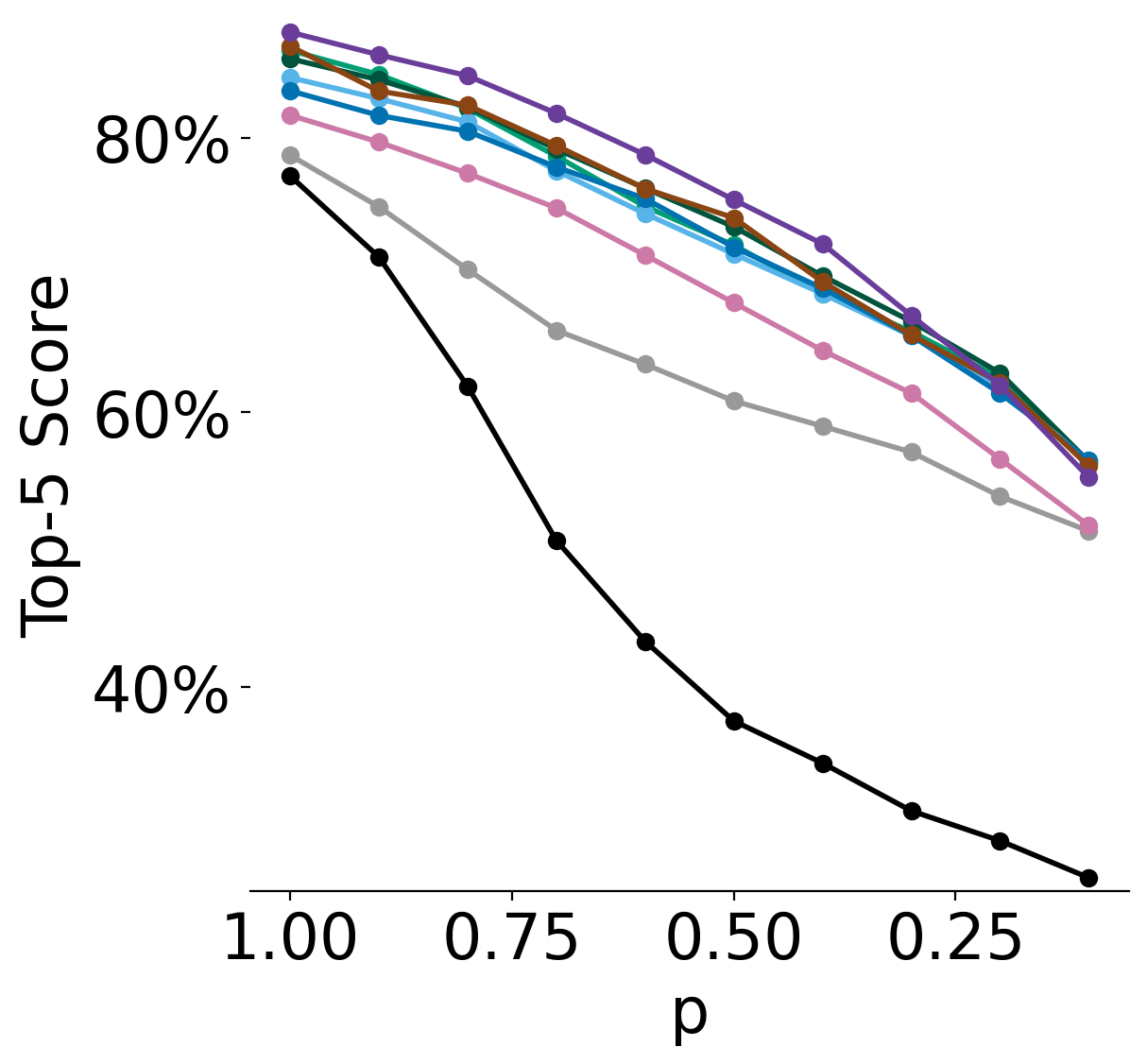}} &
        \raisebox{-0.5\height}{\includegraphics[width=0.21\textwidth]{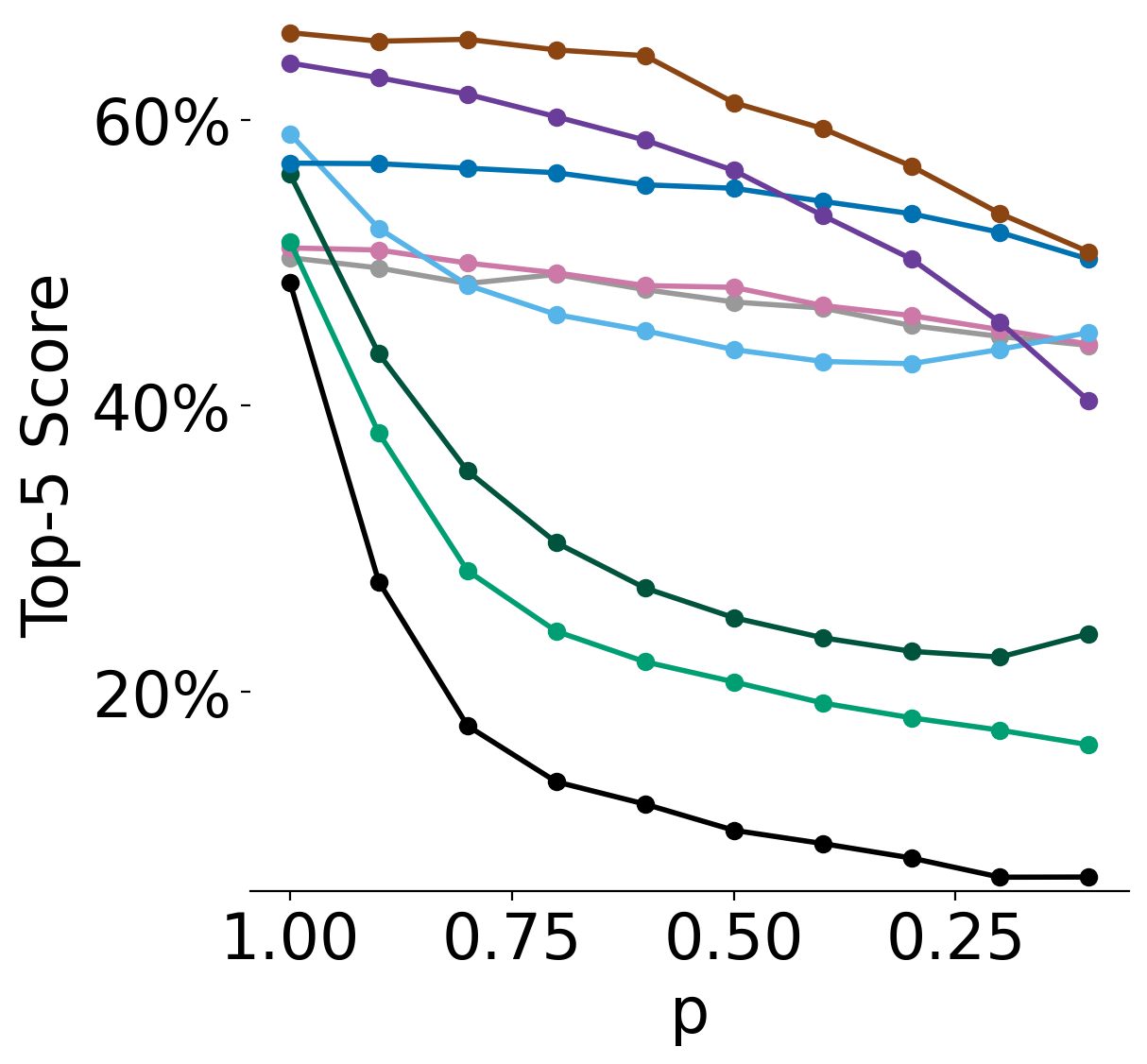}} &
        \raisebox{-0.5\height}{\includegraphics[width=0.21\textwidth]{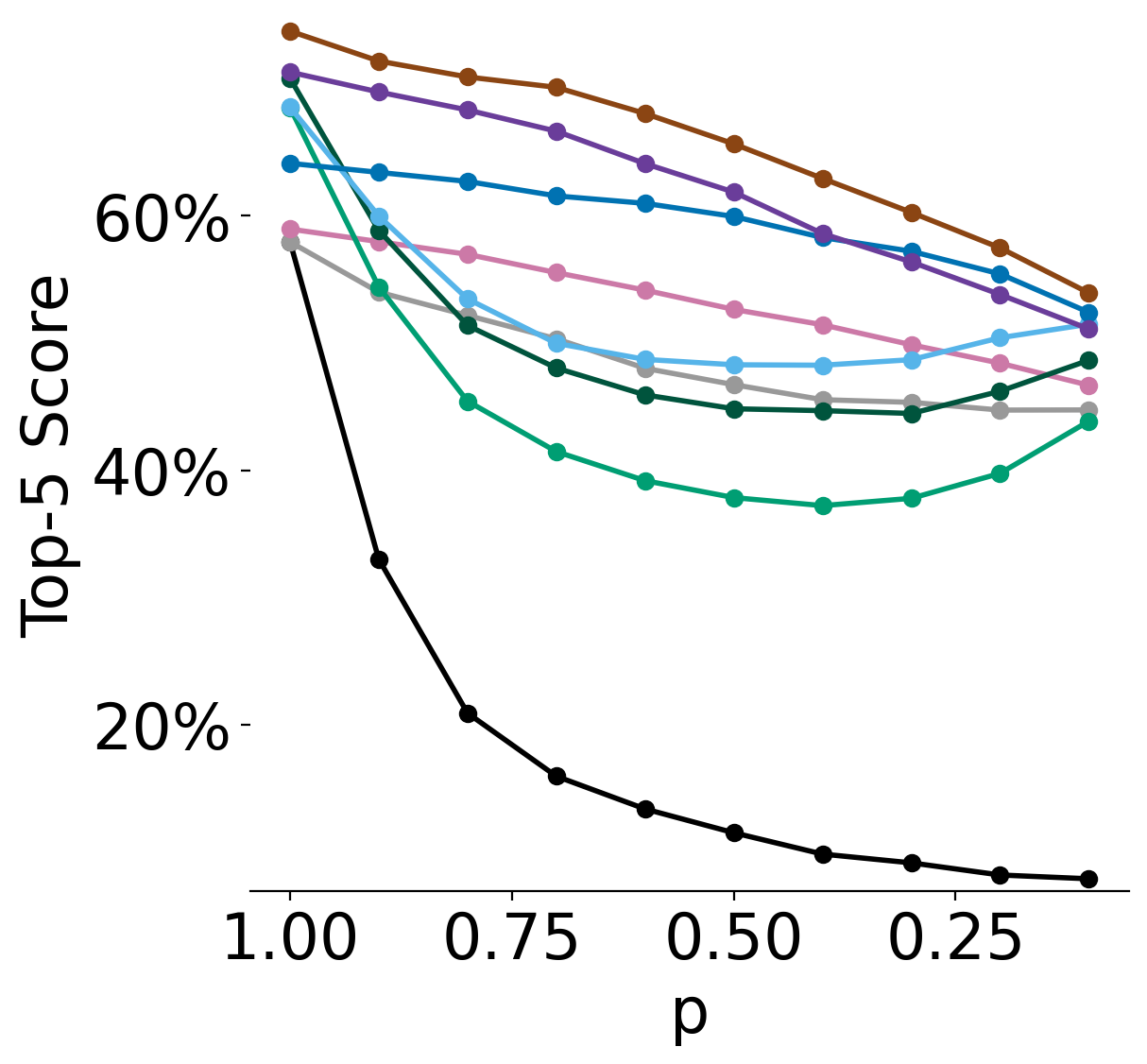}} &
        \raisebox{-0.5\height}{\includegraphics[width=0.21\textwidth]{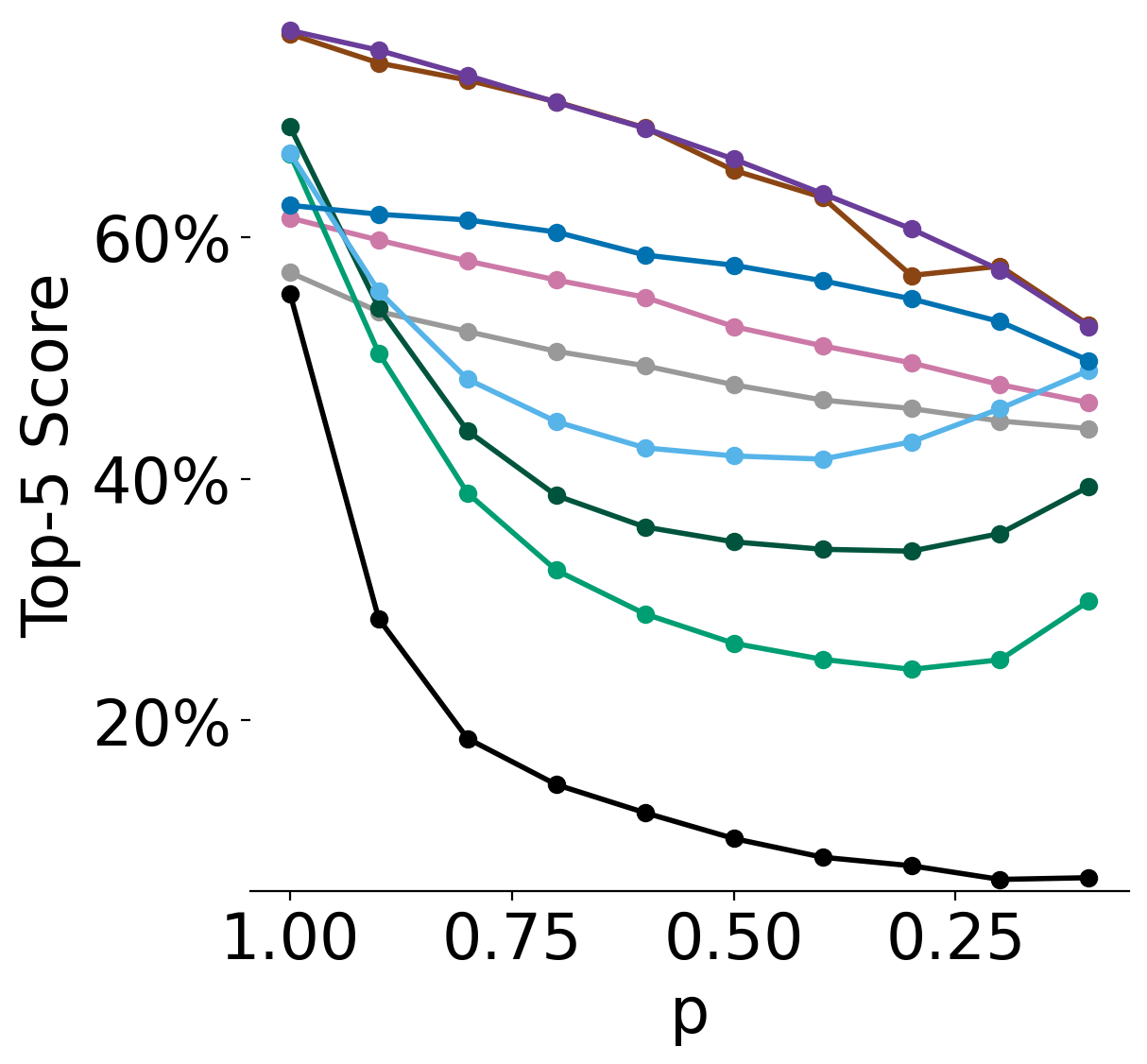}} &
        \raisebox{-0.5\height}[0pt][0pt]{\includegraphics[width=0.0889\textwidth]{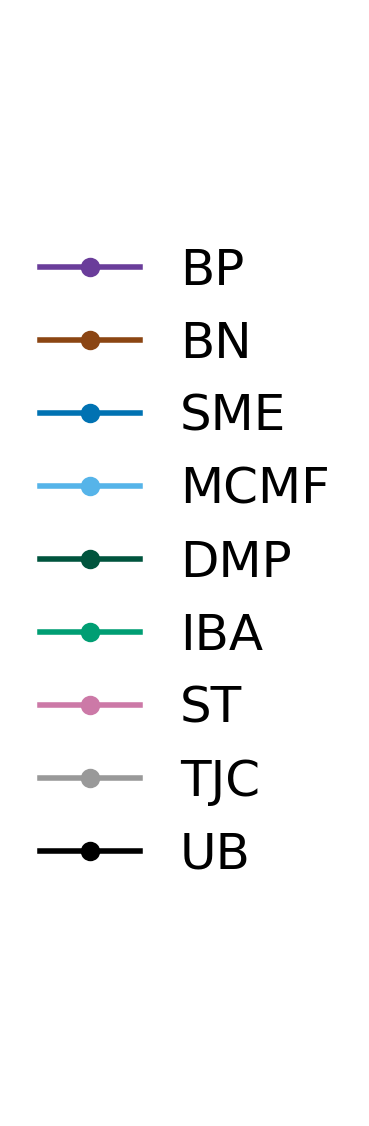}} \\
    \end{tabular}
    \caption{Top-5 accuracy vs.\ fraction of observed nodes $p$, for a selection of four networks. The SMLP and SGNN, which merely serve as benchmarks for BN, as well as uncertainty bands, are omitted for visual clarity.}
    \label{fig:partiality_behaviour}
\end{figure*}

Figure~\ref{fig:partiality_behaviour} shows how source detection performance, measured by top-5 accuracy, depends on the fraction $p$ of the snapshot that is observed. The most notable result is that SME performs very strongly for small $p$, competitive with the best methods, BN and BP. The likely reason is that, as $p$ decreases, the complexity of the overlap between snapshots decreases exponentially and the soft margin increasingly resembles an exact match on the observed subset. In other words, it approaches Direct Monte Carlo matching for a growing share of (partial) observed snapshots $\mathbf{x}$ in the test data. BN and BP are also quite robust overall; only on the Malawi network does BP deteriorate rather quickly with decreasing $p$.

The results further reinforce our doubts about the factorized-likelihood assumption: apart from the Baboons network, IBA, DMP, and MCMF all show a roughly U-shaped pattern. Their performance drops sharply as $p$ falls below 100\% but recovers somewhat at very small $p$. Interestingly, MCMF shows a less pronounced U-shaped pattern and performs better than IBA and DMP under partial observation. TJC and ST, by contrast, are more robust: their performance deteriorates more slowly than that of the factorized-likelihood-based methods.

\subsection{Parameter Misspecifications}
\label{sec:results_misspec}

\begin{figure*}[tp]
    \centering
    \renewcommand{\arraystretch}{0.5}
    \setlength{\tabcolsep}{3pt}
    \begin{tabular}{r ccc c}

        \adjustbox{valign=c,raise=0.03\linewidth}{\rotatebox{90}{\footnotesize\textbf{Baboons}}} &
        \adjustbox{valign=c}{\includegraphics[width=0.23\linewidth]{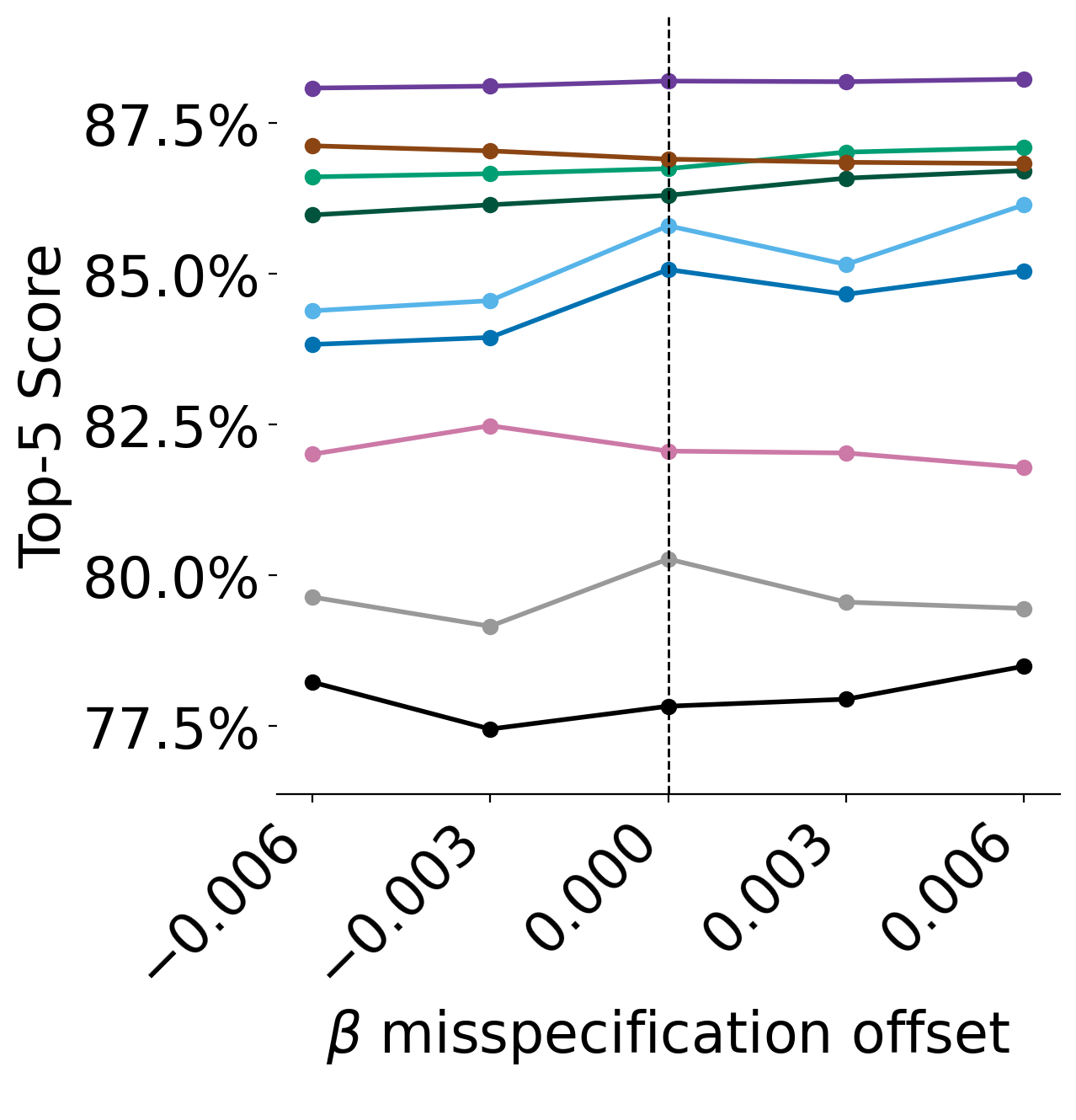}} &
        \adjustbox{valign=c}{\includegraphics[width=0.23\linewidth]{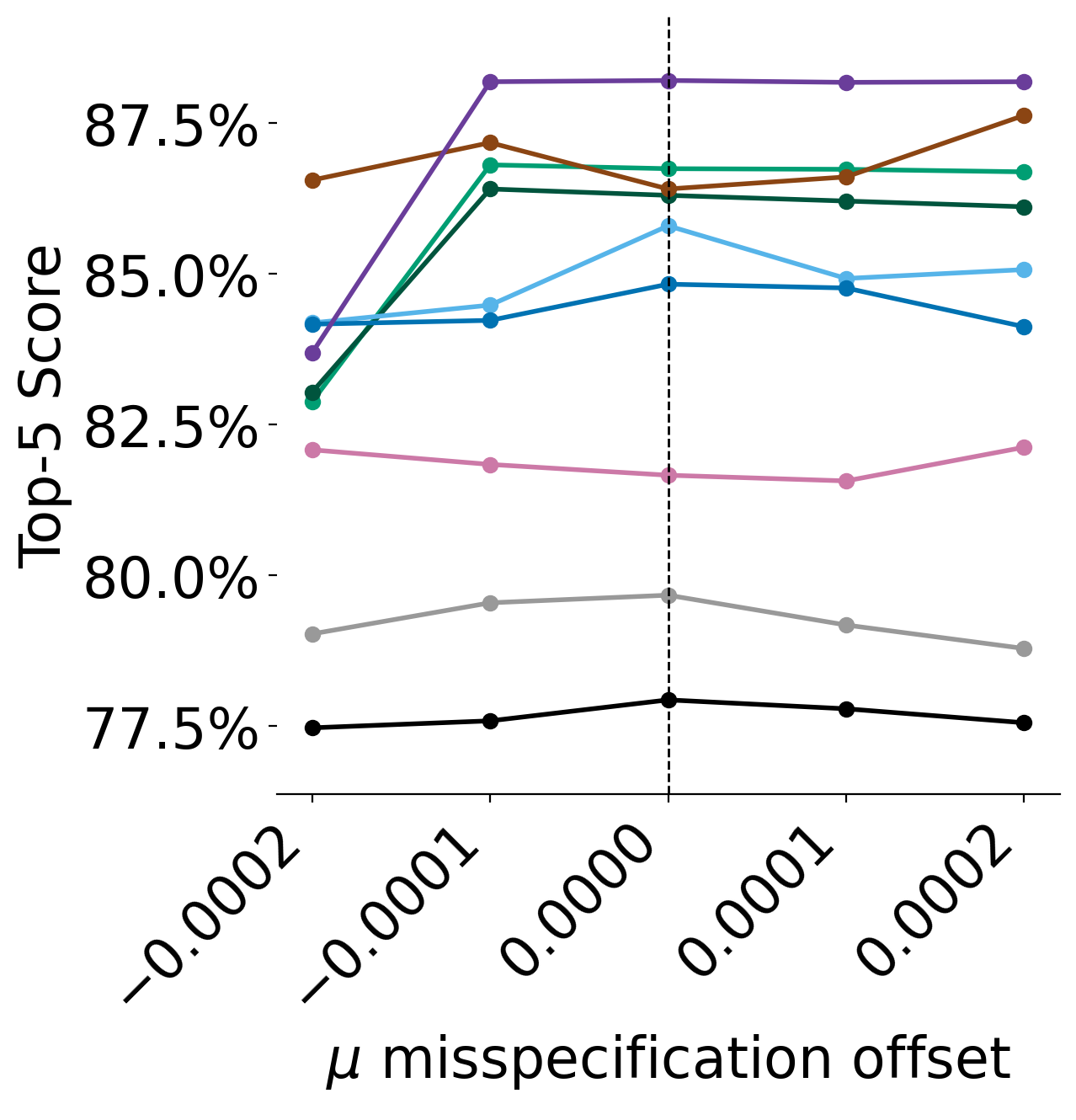}} &
        \adjustbox{valign=c}{\includegraphics[width=0.23\linewidth]{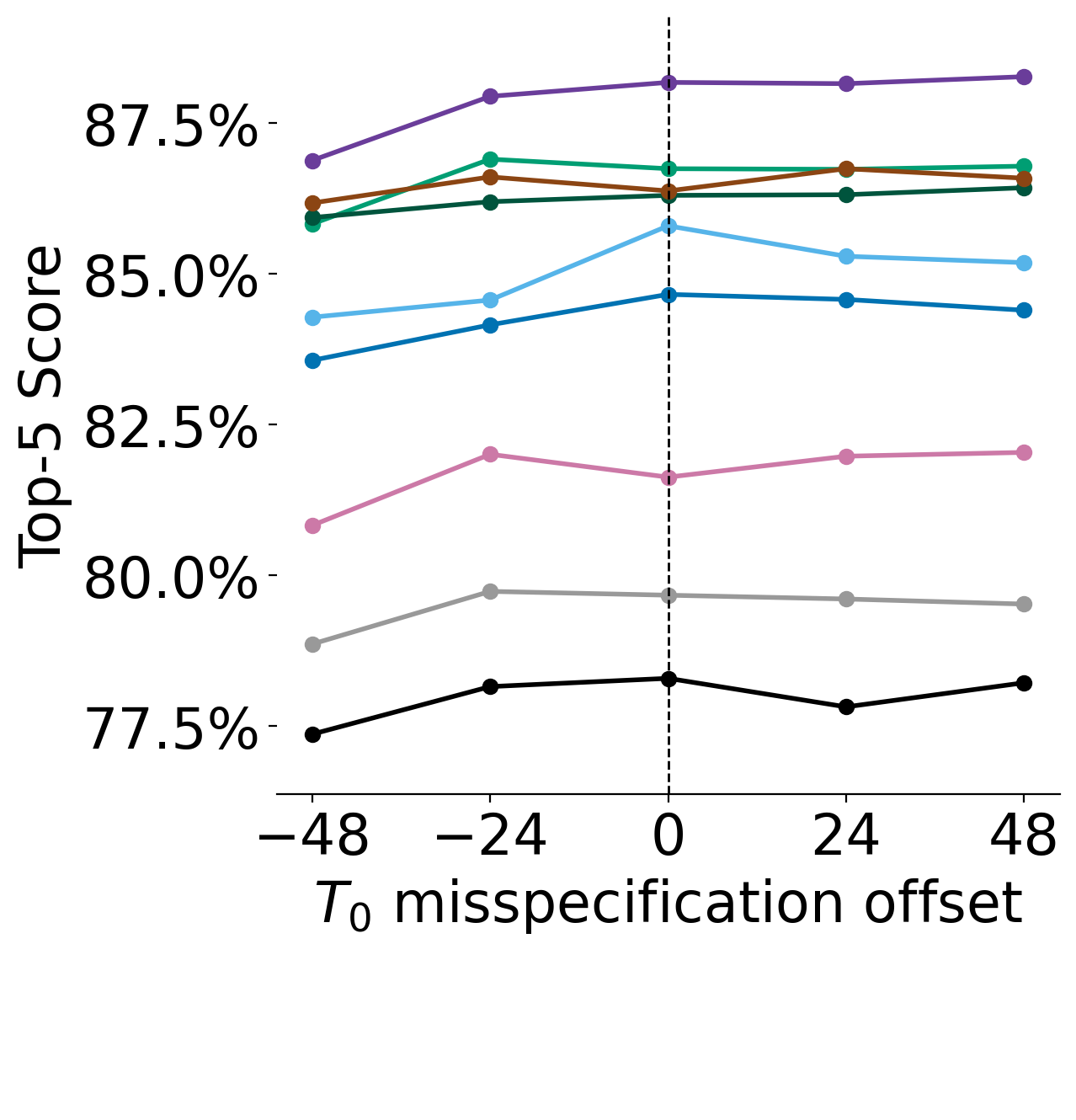}} &
        \raisebox{-0.2007\linewidth}[0pt][0pt]{\includegraphics[width=0.0889\linewidth]{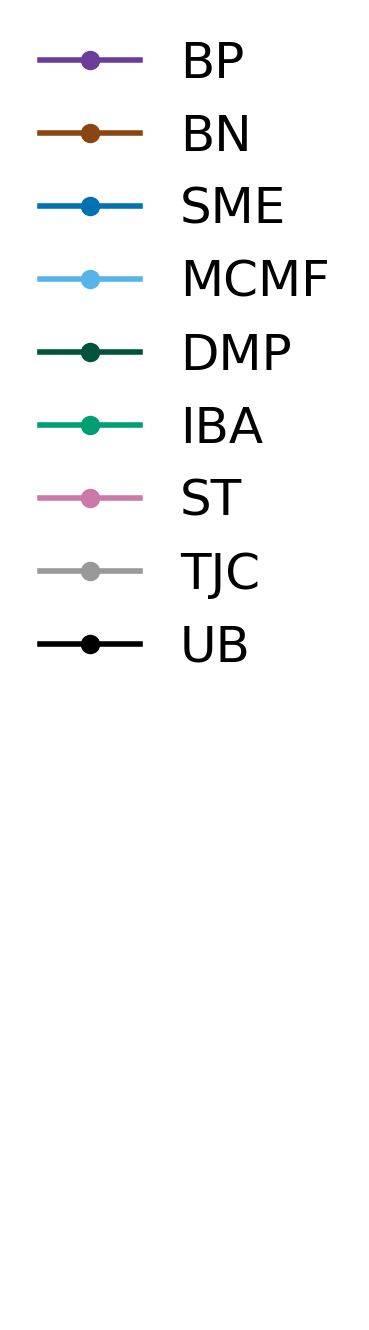}} \\[2pt]

        \adjustbox{valign=c,raise=0.03\linewidth}{\rotatebox{90}{\footnotesize\textbf{Malawi}}} &
        \adjustbox{valign=c}{\includegraphics[width=0.23\linewidth]{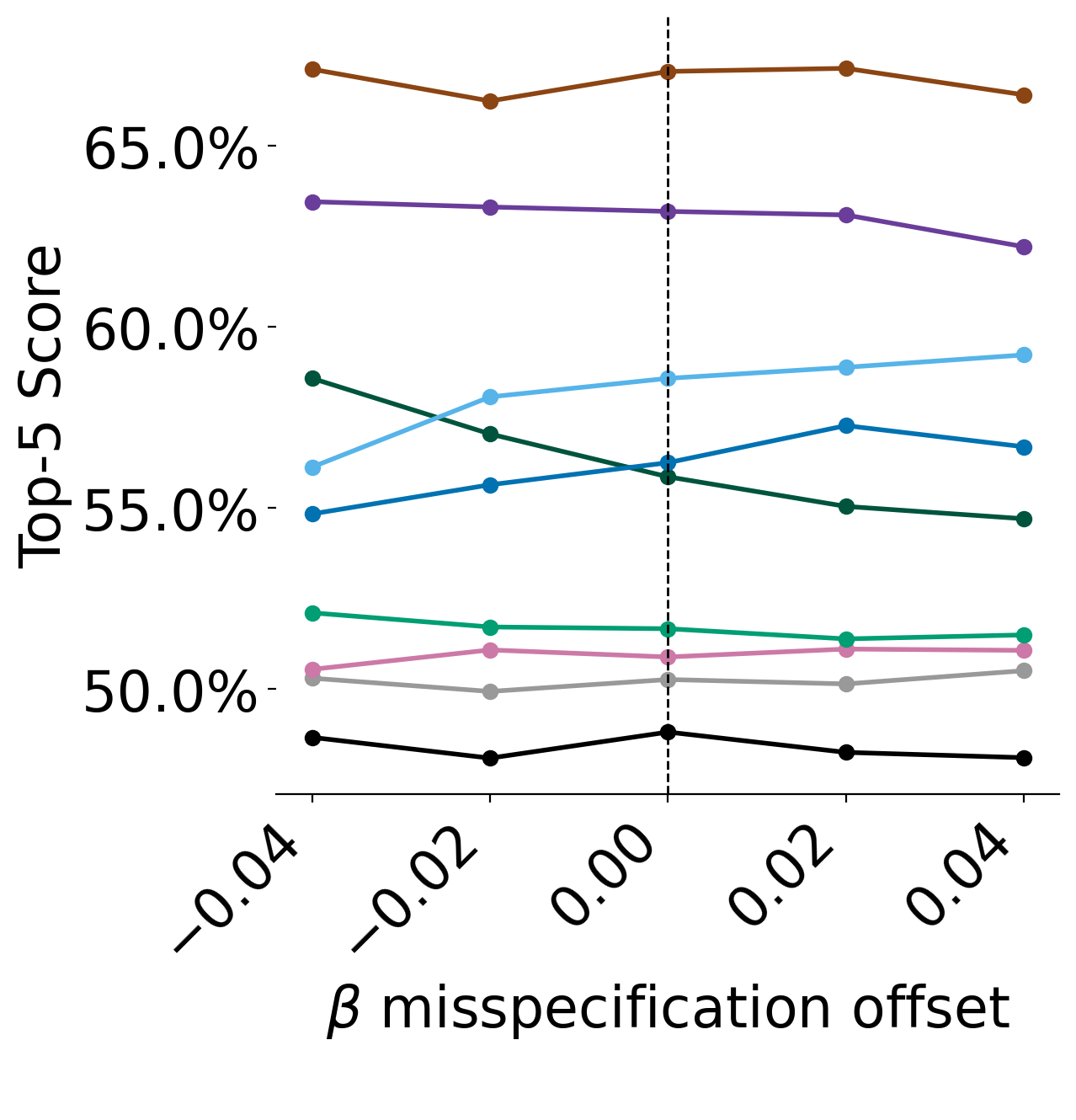}} &
        \adjustbox{valign=c}{\includegraphics[width=0.23\linewidth]{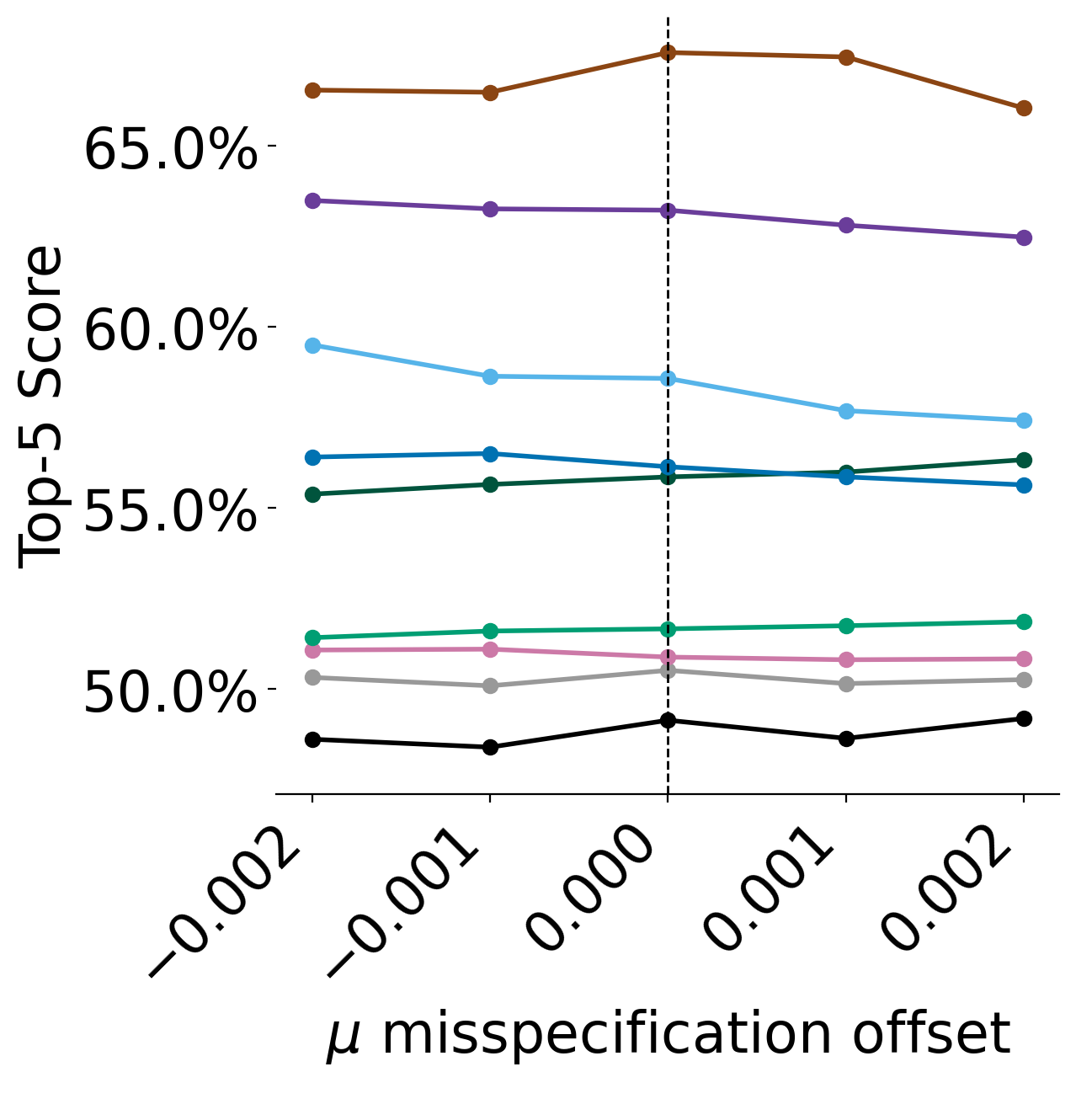}} &
        \adjustbox{valign=c}{\includegraphics[width=0.23\linewidth]{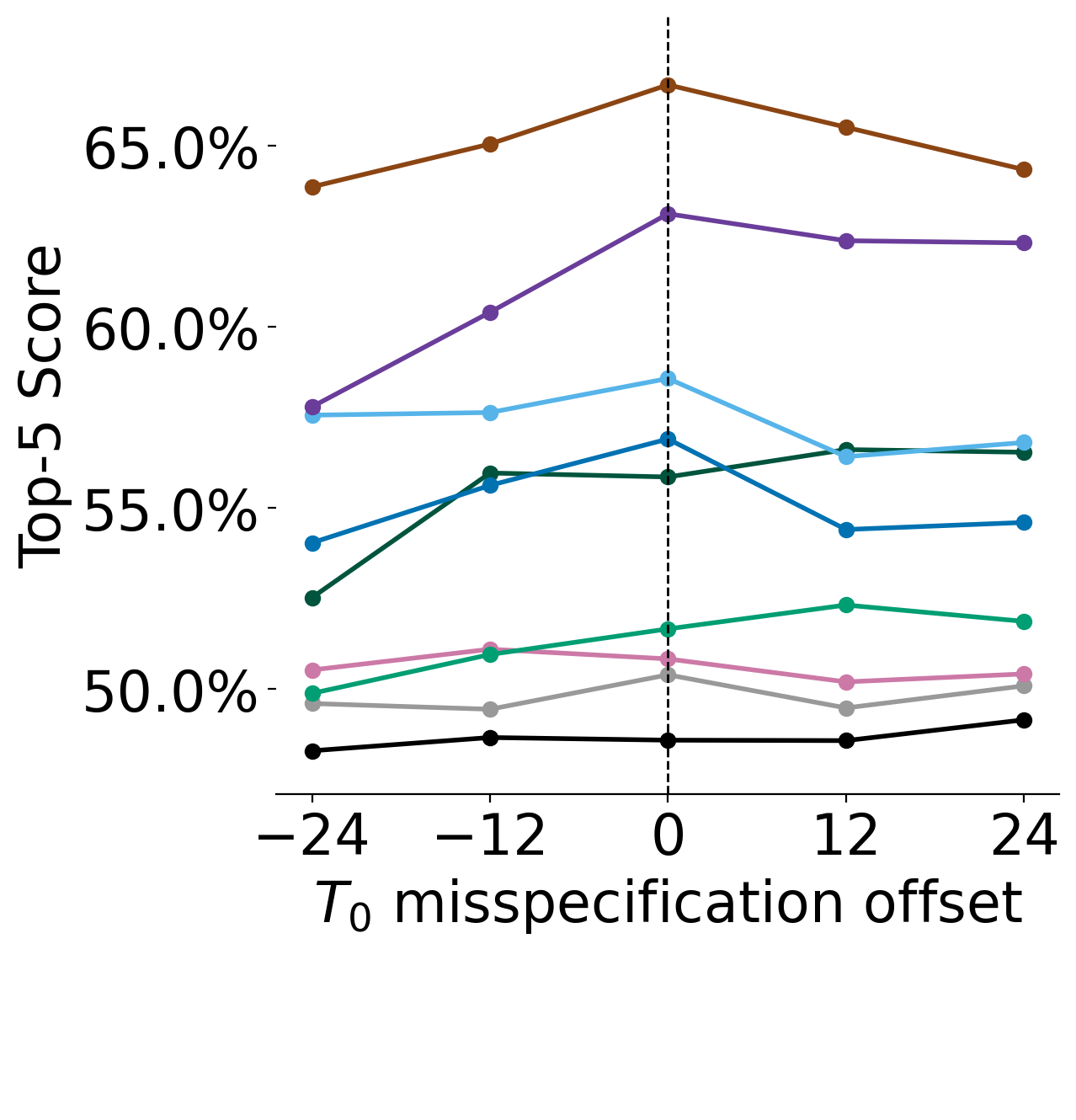}} &
         \\[2pt]

        \adjustbox{valign=c,raise=0.03\linewidth}{\rotatebox{90}{\footnotesize\textbf{Boston}}} &
        \adjustbox{valign=c}{\includegraphics[width=0.23\linewidth]{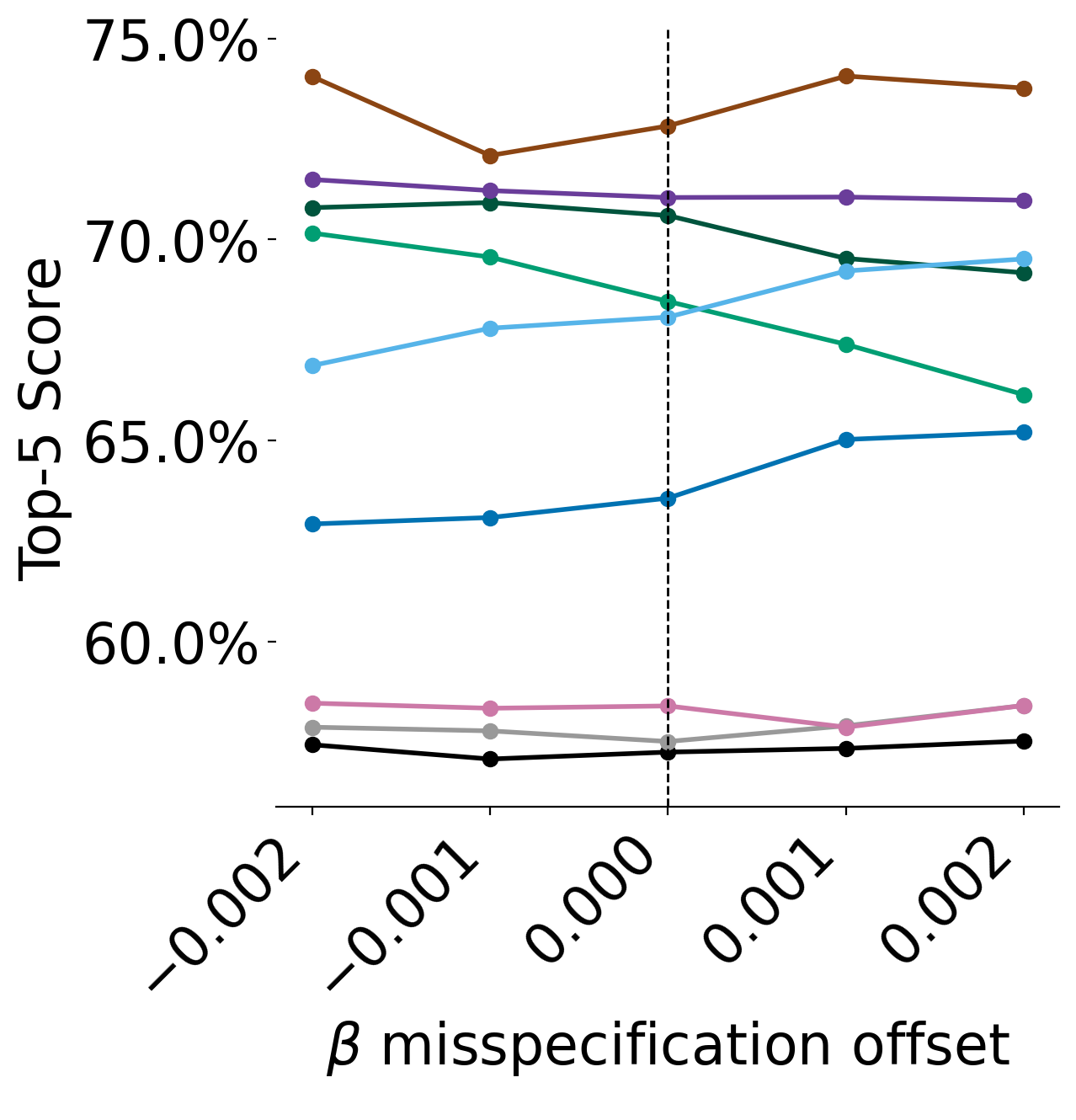}} &
        \adjustbox{valign=c}{\includegraphics[width=0.23\linewidth]{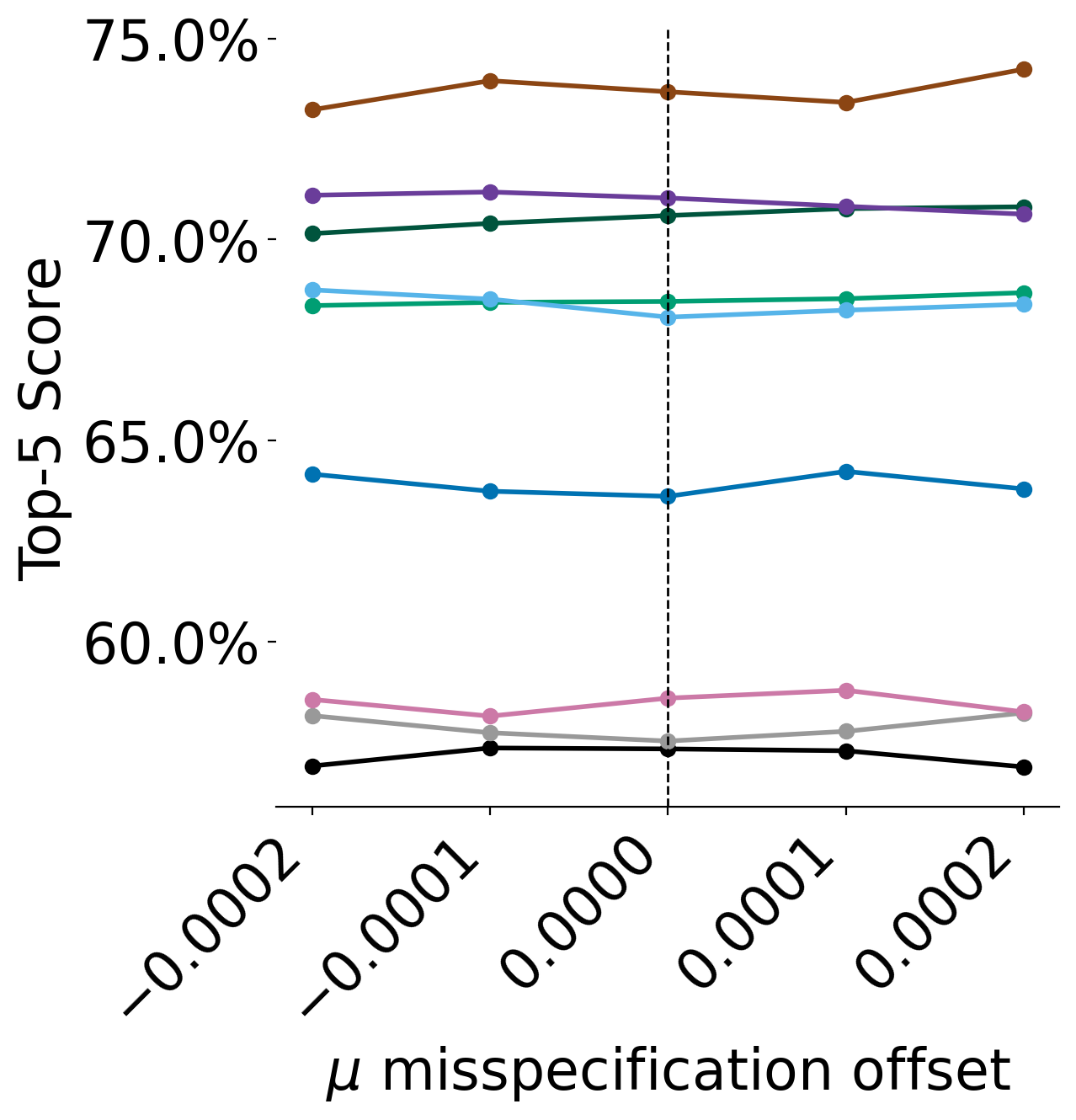}} &
        \adjustbox{valign=c}{\includegraphics[width=0.23\linewidth]{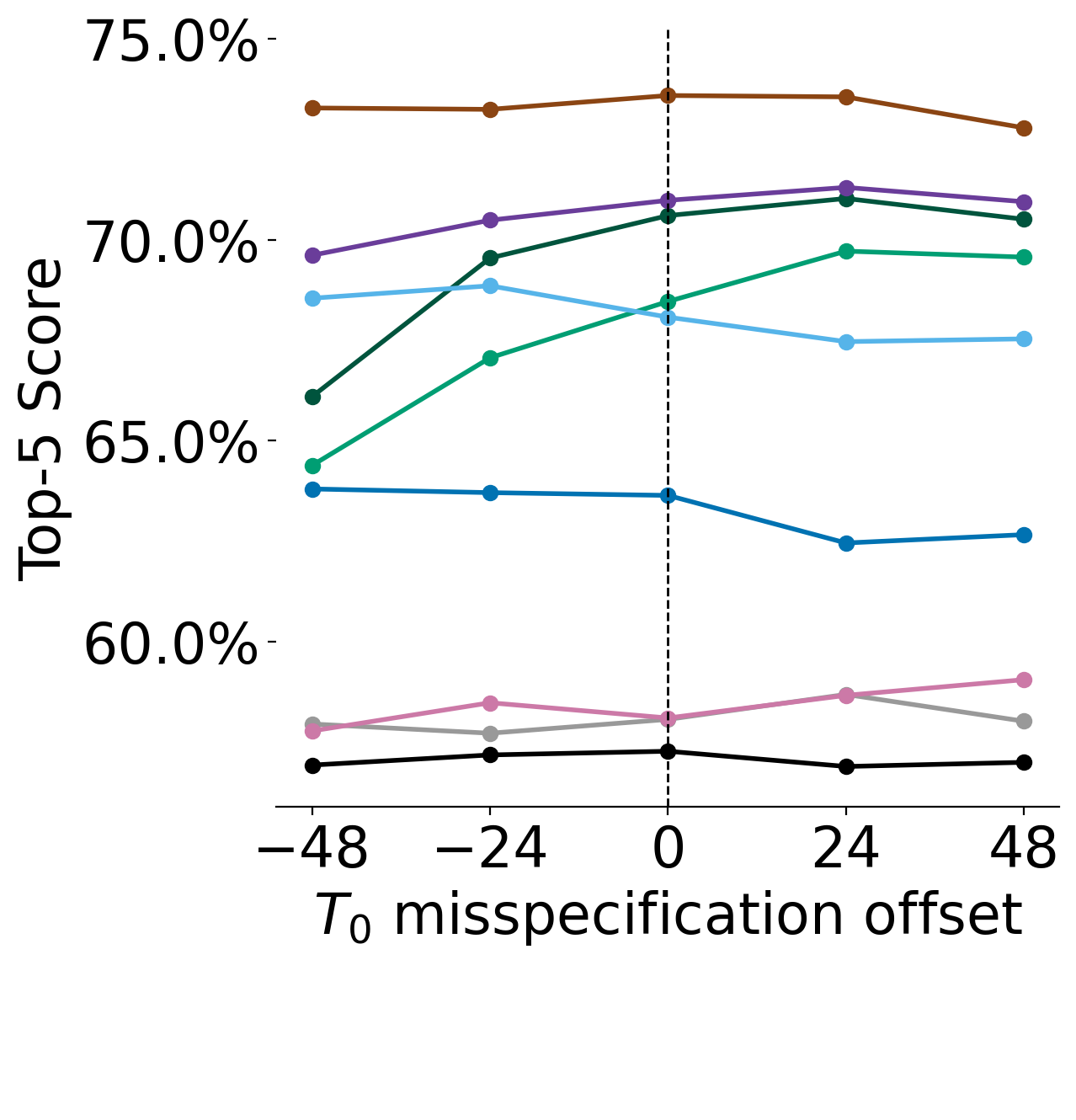}} &
         \\[2pt]

        \adjustbox{valign=c,raise=0.03\linewidth}{\rotatebox{90}{\footnotesize\textbf{Office}}} &
        \adjustbox{valign=c}{\includegraphics[width=0.23\linewidth]{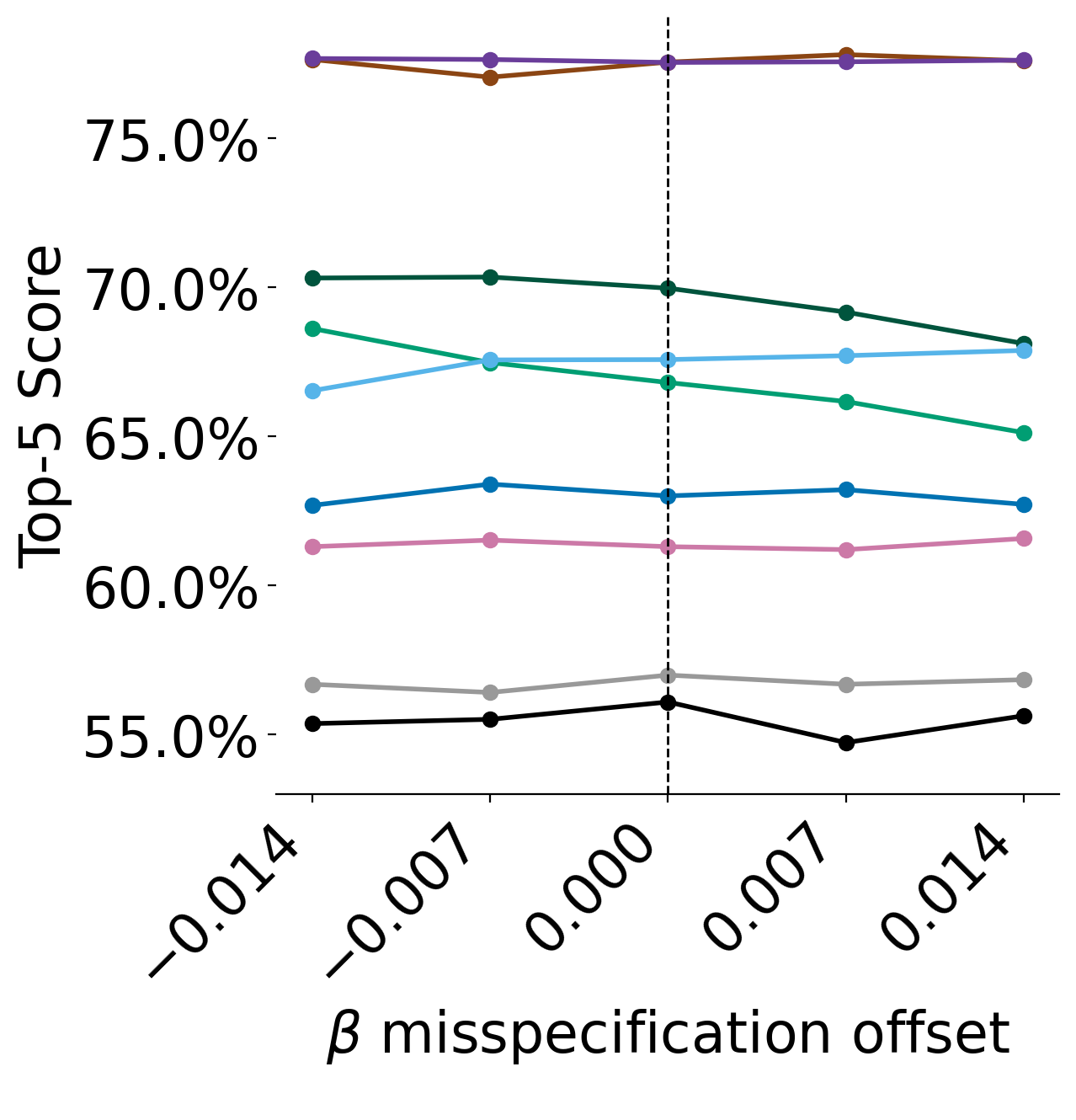}} &
        \adjustbox{valign=c}{\includegraphics[width=0.23\linewidth]{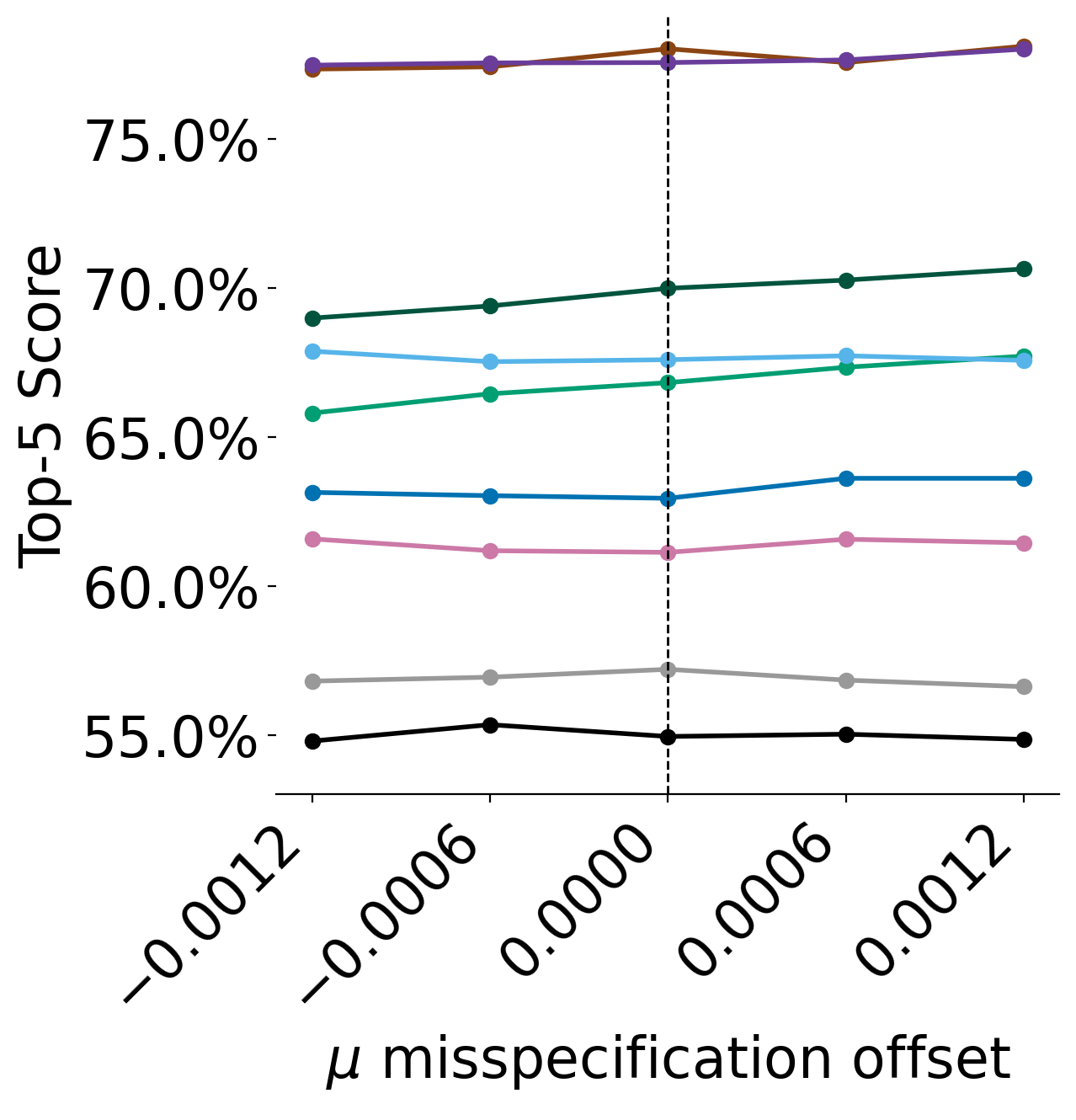}} &
        \adjustbox{valign=c}{\includegraphics[width=0.23\linewidth]{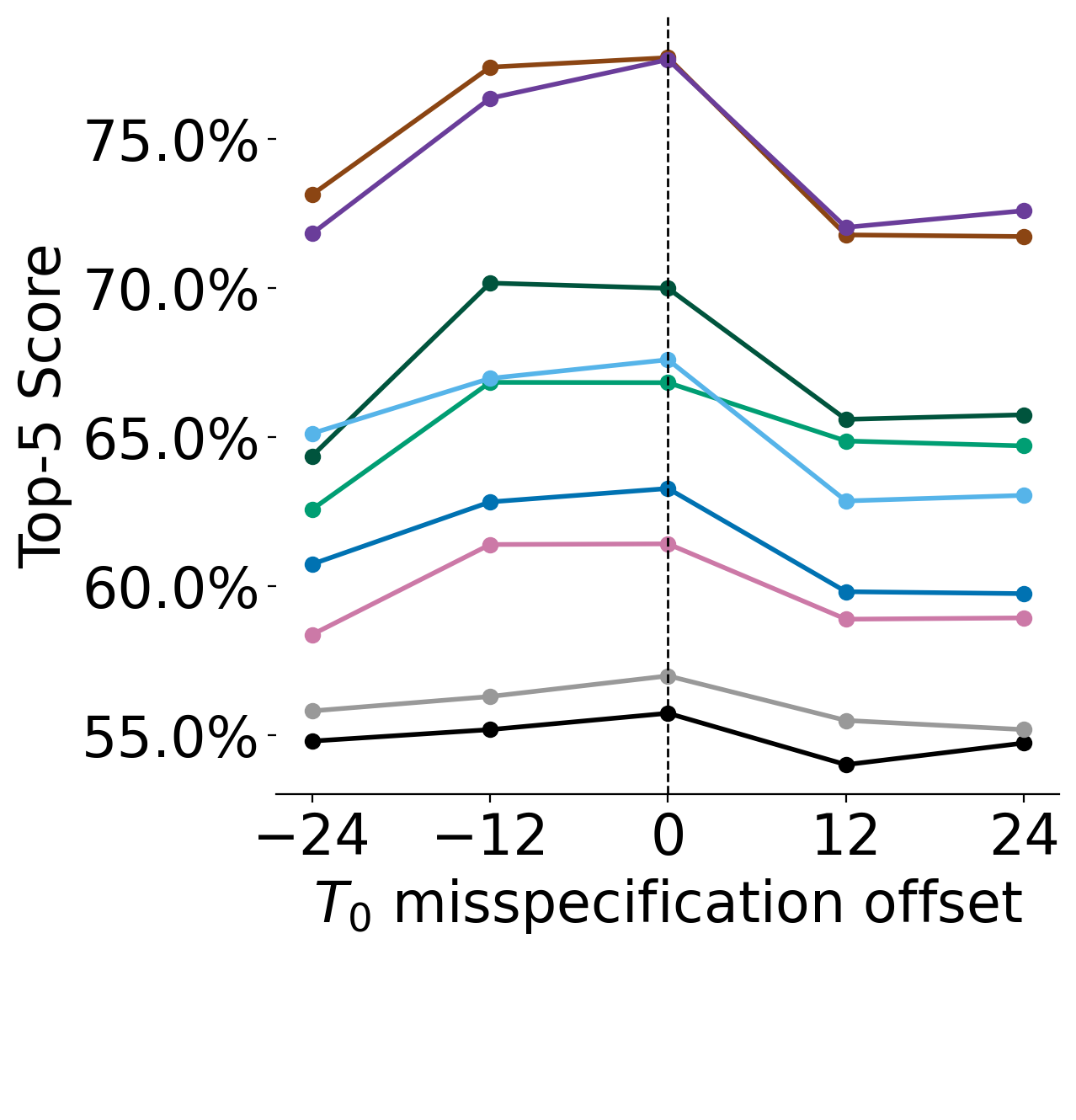}} &
         \\

    \end{tabular}
    \caption{Top-5 accuracy vs.\ misspecification of $\beta$, $\mu$, and $T_0$, for a selection of four networks. The SMLP and SGNN, which merely serve as benchmarks for BN, as well as uncertainty bands, are omitted for visual clarity. The largest offset for $\beta$ and $\mu$ is chosen such that the resulting $R_0$ is roughly $1.9$ or $2.1$, depending on the sign of the offset; the $T_0$ offsets are chosen relative to each network's epidemic duration.}
    \label{fig:misspec_behaviour}
\end{figure*}

Figure~\ref{fig:misspec_behaviour} shows the results of the misspecification experiment in terms of top-5 accuracy. The main takeaway is that the ranking of methods remains largely stable even under misspecification of $\beta$, $\mu$, and $T_0$. In fact, a misspecified BN or BP often still outperforms well-specified competitors.

The largest sensitivity results from a misspecified epidemic start $T_0$ on the Malawi and Office networks: it is the only setting where BN and BP show a significant performance drop. IBA and DMP show a similar pattern on these networks for a misspecified $T_0$, albeit at a lower level. A further sensitivity appears on the Baboons network for a misspecification of $\mu$ to a lower value. An offset of $-0.0002$ effectively sets $\mu$ to 0 and thus reduces the SIR model to an SI model, negatively impacting the performance of IBA, DMP, and BP, but \emph{not} the other methods. Other than that, no further noteworthy effects stand out.

\section{Discussion}
\label{sec:discussion}

Our comparative study was conducted with a harmonized problem formulation, a shared set of networks, and a shared experimental framework. It directly addresses the issues in prior work that we raised in the Introduction: diverging problem formulations, heterogeneous network choices, and inconsistent baseline selection.

Our central finding is that BP and BN are the strongest methods overall, and we see these two methods as the most promising avenue for future research on source detection for temporal contact networks (see Limitations and Future Work below). SME can be a strong alternative when only a small fraction of node states is observed, but it improves only slowly with more training data. IBA, DMP, and MCMF are systematically flawed because they all rely on the factorized likelihood, which performs poorly and leads to overconfident probability distributions. ST and TJC are non-probabilistic, purely graph-geometric alternatives that offer a solid baseline for source detection.

Our results add a twist to the broader narrative around graph neural networks for source detection: recent literature, including our own prior work, has argued that GNN-based approaches substantially outperform classical methods in source detection~\cite{sterchiGraphNeuralNetworks2025, ruInferringPatientZero2023}. Our results partially confirm this---BN is indeed one of the strongest approaches---but the fact that a classical, training-free method like BP remains highly competitive, and even outperforms BN on some networks (e.g., Escort), complicates this narrative. To our knowledge, no prior GNN-based source detection study has benchmarked against BP. The reported GNN superiority may therefore partly reflect the inconsistent baseline selection discussed in the Introduction.

Comparing the computational complexities of the two best methods (Table~\ref{tab:methods}) is also instructive: by number of operations, BN is clearly the more expensive method. However, this comparison hides an important asymmetry: BN pays this cost once, during training, after which inference on a new observed snapshot is nearly free, whereas BP pays its full complexity on every single query, since it has no notion of a trained, reusable model. Consider, for example, the Copenhagen network with $R_0 = 2.5$: one BP inference took about 25 seconds, whereas training BN took roughly 10 hours. Yet for the roughly 10,000 snapshots in our test data, BP needed almost three days (69 hours) in total to iterate over them, while BN's vectorized inference took only a few minutes. Note that we do not report runtimes systematically, since the implementations differ widely (BN in PyTorch on GPU, BP parallelized on CPU) and are therefore not directly comparable. However, this example suffices for illustrative purposes: BN can become cheaper than BP for repeated inference on the same network, once the training cost is amortized, while BP remains preferable for one-off inference or on networks that keep changing.

Finally, we suspect that the SMLP's strong performance on some networks is the reason why many GNN-based source detection papers do not report an MLP baseline at all, even though it is an obvious benchmark. That a method without any access to the graph performs embarrassingly well on several networks shows that (static) graph-awareness does not automatically pay off. This should not be overstated, however: on other networks, most notably Escort, the SMLP falls clearly behind the SGNN. Furthermore, SMLP suffers from the same overconfidence problem as the factorized-likelihood-based methods.

\subsection*{Limitations and Future Work}

This study has several limitations.

First, we only consider a single observed snapshot. How these methods would need to be adapted to a setting with multiple snapshots over time, as in Sha et al.~\cite{shaSourceDetectionNetworks2021}, remains open. It is likewise unclear how they would extend to a setting where information about individual nodes' states arrives incrementally as an epidemic unfolds, as in real-world surveillance.

Second, scalability to substantially larger networks remains untested: the largest network in our study has 2{,}133 nodes (Escort network), while real-world contact networks relevant to public health, such as city- or country-scale mobility data, can be orders of magnitude larger. We only know the theoretical computational complexity of BP and BN (Table~\ref{tab:methods}), but their practical performance on large networks---say, a hundred thousand nodes or more---remains untested.

Third, and relatedly, due to computational restrictions, the complementary experiments, which often required many repeated runs per method and network, were performed only on the four smaller networks. The evidence from these experiments is therefore weaker than for the main results, which span all eight networks, including the larger ones. For example, SME's strong performance under partial observation (Section~\ref{sec:results_partial}) is only established on these four small networks.

\section{Conclusion}
\label{sec:conclusion}

Identifying the patient zero of an epidemic directly supports public health responses such as outbreak containment and resource allocation. Systematic comparisons of source detection methods exist for static networks, both for classical methods~\cite{jiangIdentifyingPropagationSources2017} and, more recently, for deep learning approaches~\cite{sterchiGraphNeuralNetworks2025}. But no such systematic comparison has been done for epidemic source detection on \textit{temporal} contact networks, even though in most cases temporal contact structure captures epidemic spreading more realistically than a static approximation. This paper closes that gap through a harmonized problem formulation and a shared experimental framework. We systematically compared ST, IBA, DMP, MCMF, SME, BN, and BP under a common set of conditions.

Overall, our results show that a deep learning approach (BN) and (loopy) belief propagation (BP) outperform all other methods across varying network types and epidemic settings. Both yield relatively well-calibrated outputs with high accuracy, and both are fairly robust to parameter misspecification and to settings where only a partial snapshot is observed. Their main difference lies in computational cost: a single BP run is cheaper than training BN, but BN's inference on new snapshots is nearly free. In practice, BP may therefore be preferable for one-off source detection, while BN may be more attractive if source detection is performed repeatedly on the same network. Finally, we see scaling these approaches to substantially larger networks as the most pressing direction for future work.

\section*{Acknowledgments}
This work was supported by the Swiss National Science Foundation (SNSF) [grant number 228527].

\bibliographystyle{IEEEtran}
\bibliography{references}

\clearpage
\onecolumn
\setcounter{section}{0}\setcounter{figure}{0}\setcounter{table}{0}
\renewcommand{\thesection}{\Alph{section}}
\renewcommand{\thefigure}{S\arabic{figure}}
\renewcommand{\thetable}{S\arabic{table}}
\renewcommand{\theHsection}{S\Alph{section}}
\renewcommand{\theHfigure}{S\arabic{figure}}
\renewcommand{\theHtable}{S\arabic{table}}
\begin{center}{\LARGE Supplementary Material}\end{center}
\vspace{1.5em}

\section{Preprocessing of the Foursquare Check-in Data}
\label{sec:foursquare_appendix}

We preprocess the raw Foursquare check-in data into temporal contact networks, a process that depends on several parameters. First, before building contacts, we remove check-ins at venue categories where being at the same venue does not imply physical proximity, mostly large transit hubs and other high-throughput venues: Train Stations, Subways, Light Rails, Airports, Bus Stations, Ferries, Harbors / Marinas, Convention Centers, Electronics Stores, Stadiums, Malls, and Arcades. Second, we define a contact as follows: two users are considered in contact if they checked in at the same venue within 60 minutes of each other (no checkout times are provided). A contact spanning midnight (e.g., 23:45 and 00:15) is kept and recorded on the earlier of the two days. Third, we exclude users with fewer than 5 recorded contacts in a given city, as this is too few to characterize their contact pattern. Applying this pipeline separately to the New York City and Tokyo check-ins, drawn from the same underlying dataset, yields two temporal contact networks, which we refer to as the New York check-ins and Tokyo check-ins networks. Although the full dataset spans almost a year, for the epidemic simulations we only consider the first 140 days, since a substantial number of days later in the dataset (roughly days 140--200) contain irregularities and missing data.

\vspace{1.5em}
\section{Additional Figures}

\begin{figure}[htbp]
    \centering
    \renewcommand{\arraystretch}{0.5}
    \setlength{\tabcolsep}{3pt}
    \begin{tabular}{ccc}
        \footnotesize\textbf{Baboons} &
        \footnotesize\textbf{Boston} &
        \footnotesize\textbf{Copenhagen} \\[4pt]

        \includegraphics[width=0.28\textwidth]{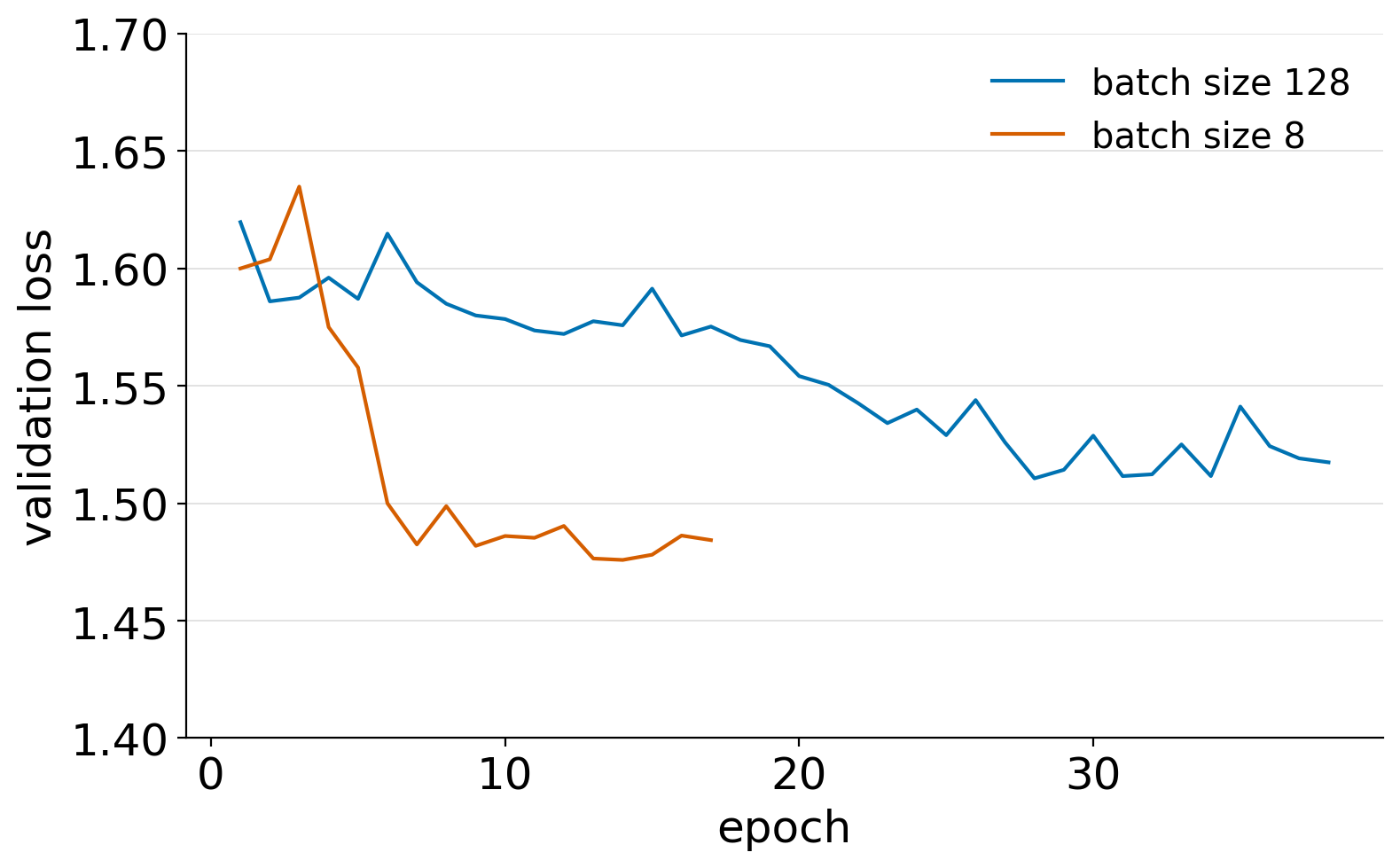} &
        \includegraphics[width=0.28\textwidth]{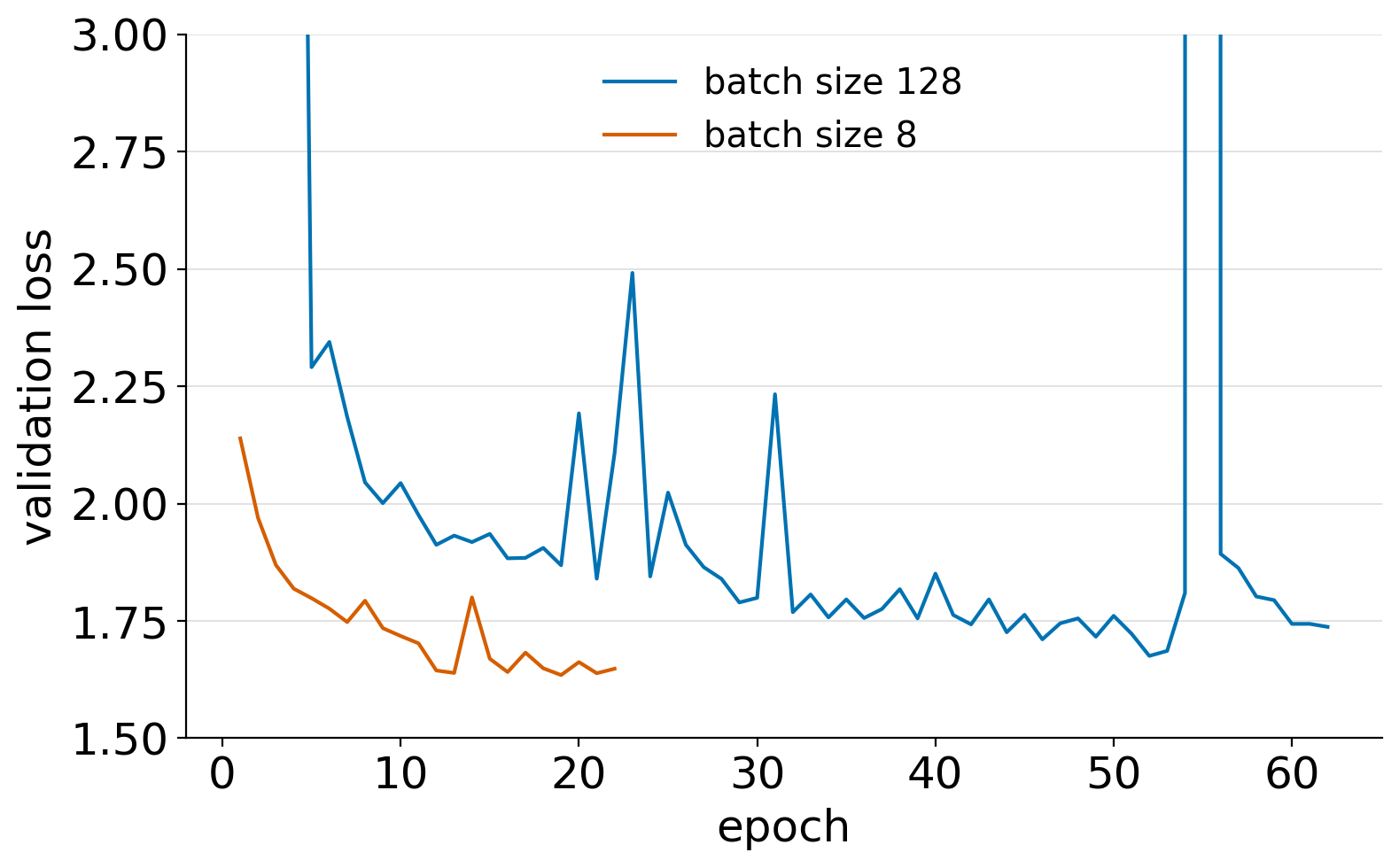} &
        \includegraphics[width=0.28\textwidth]{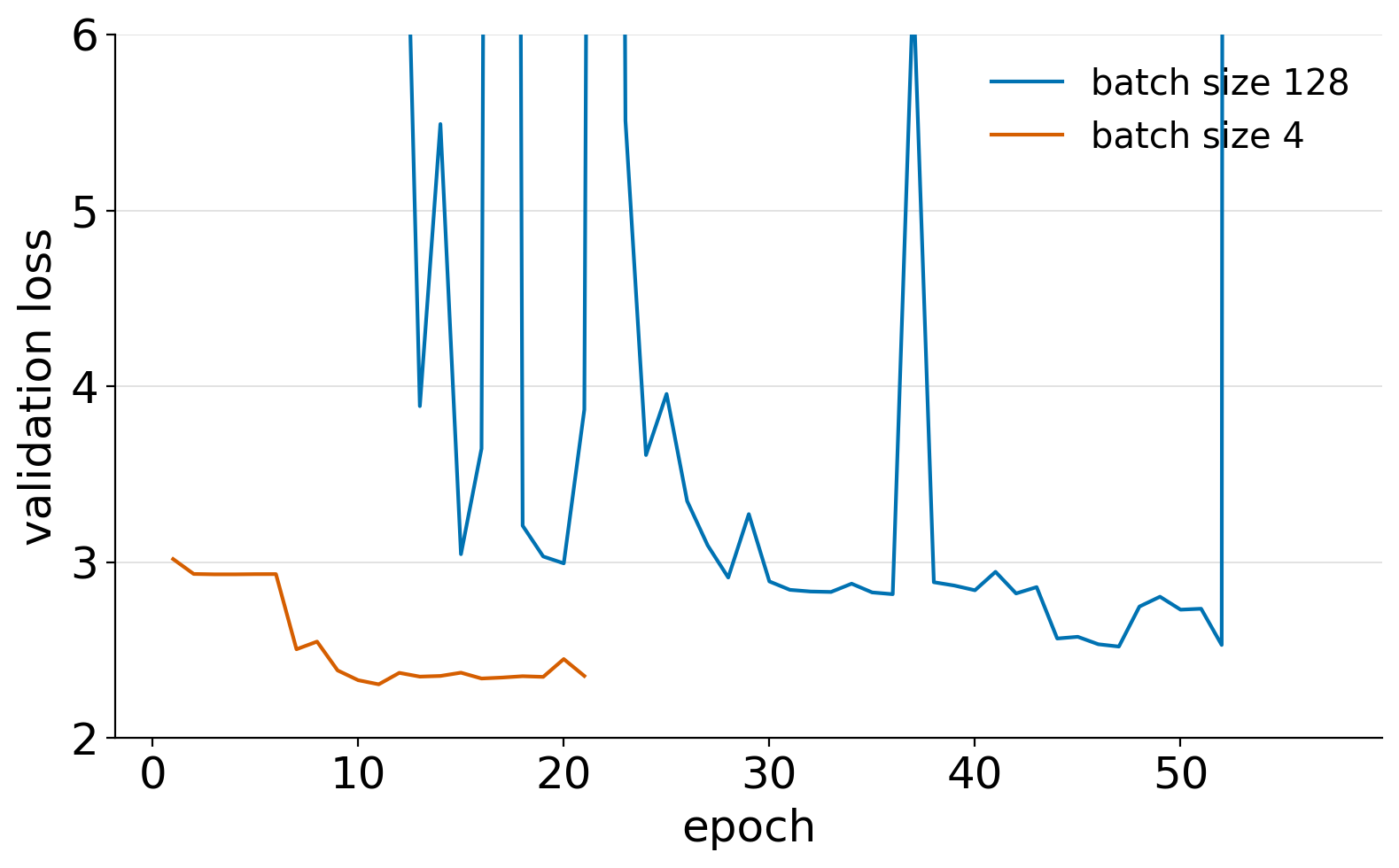} \\[2pt]

        \includegraphics[width=0.28\textwidth]{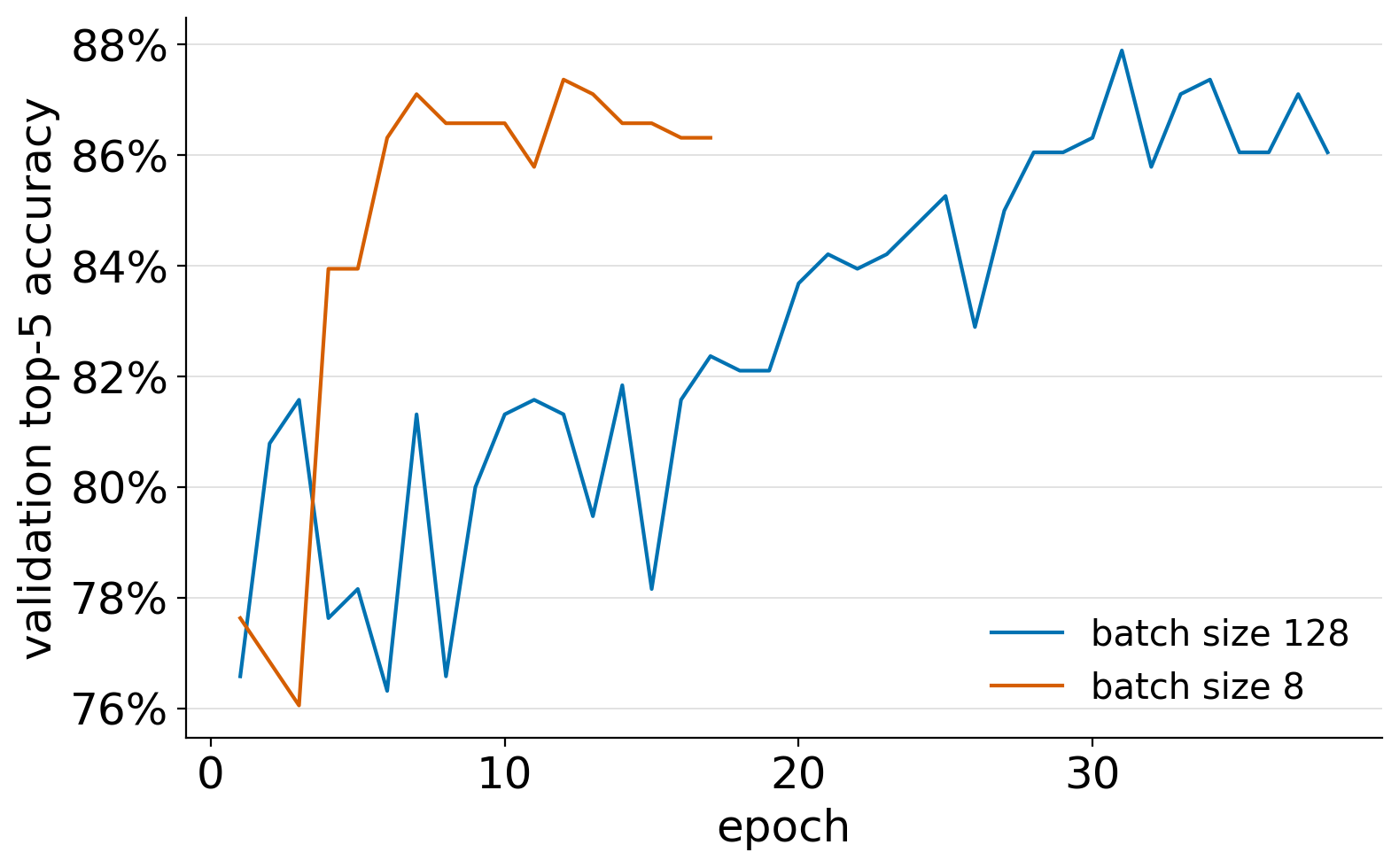} &
        \includegraphics[width=0.28\textwidth]{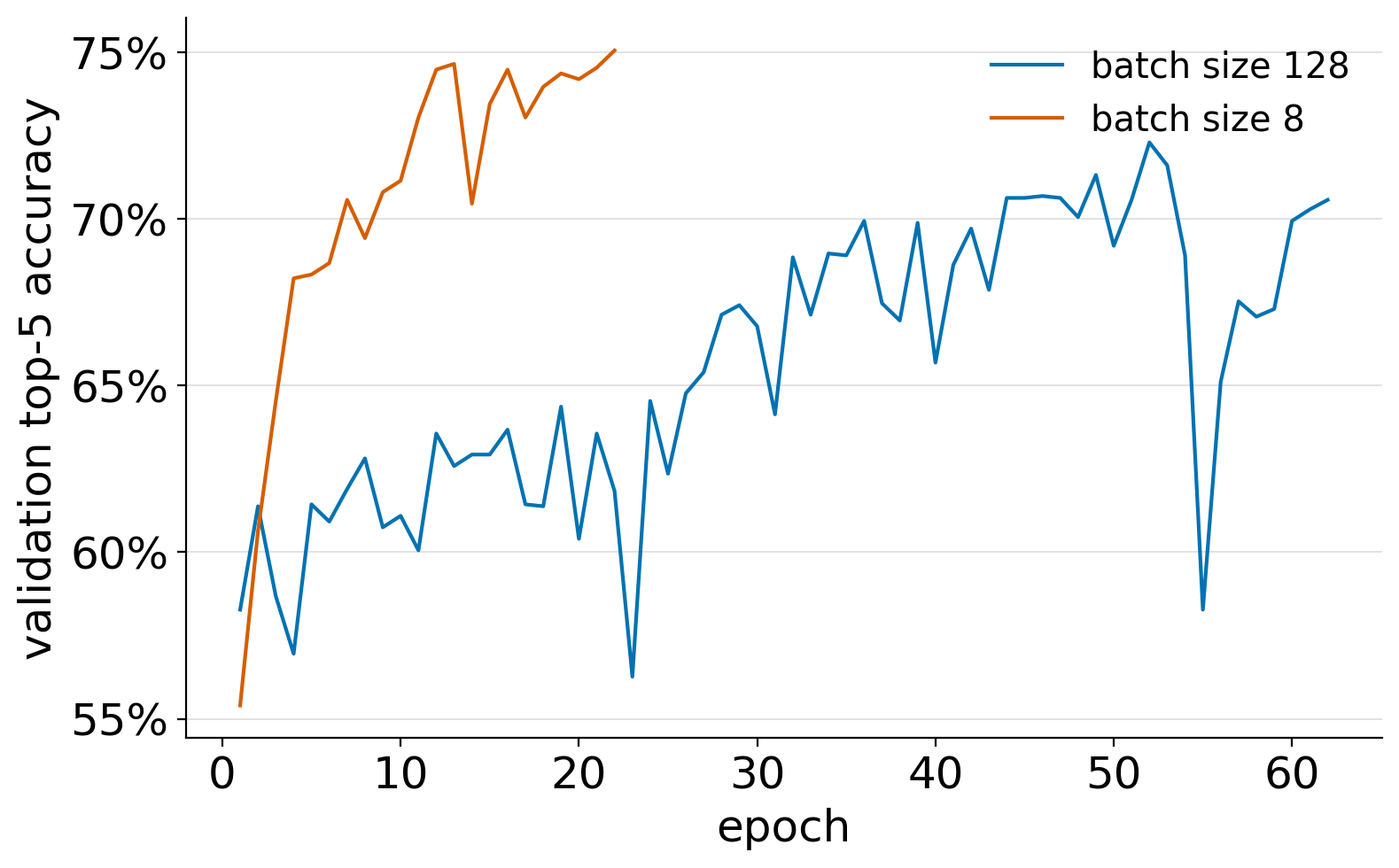} &
        \includegraphics[width=0.28\textwidth]{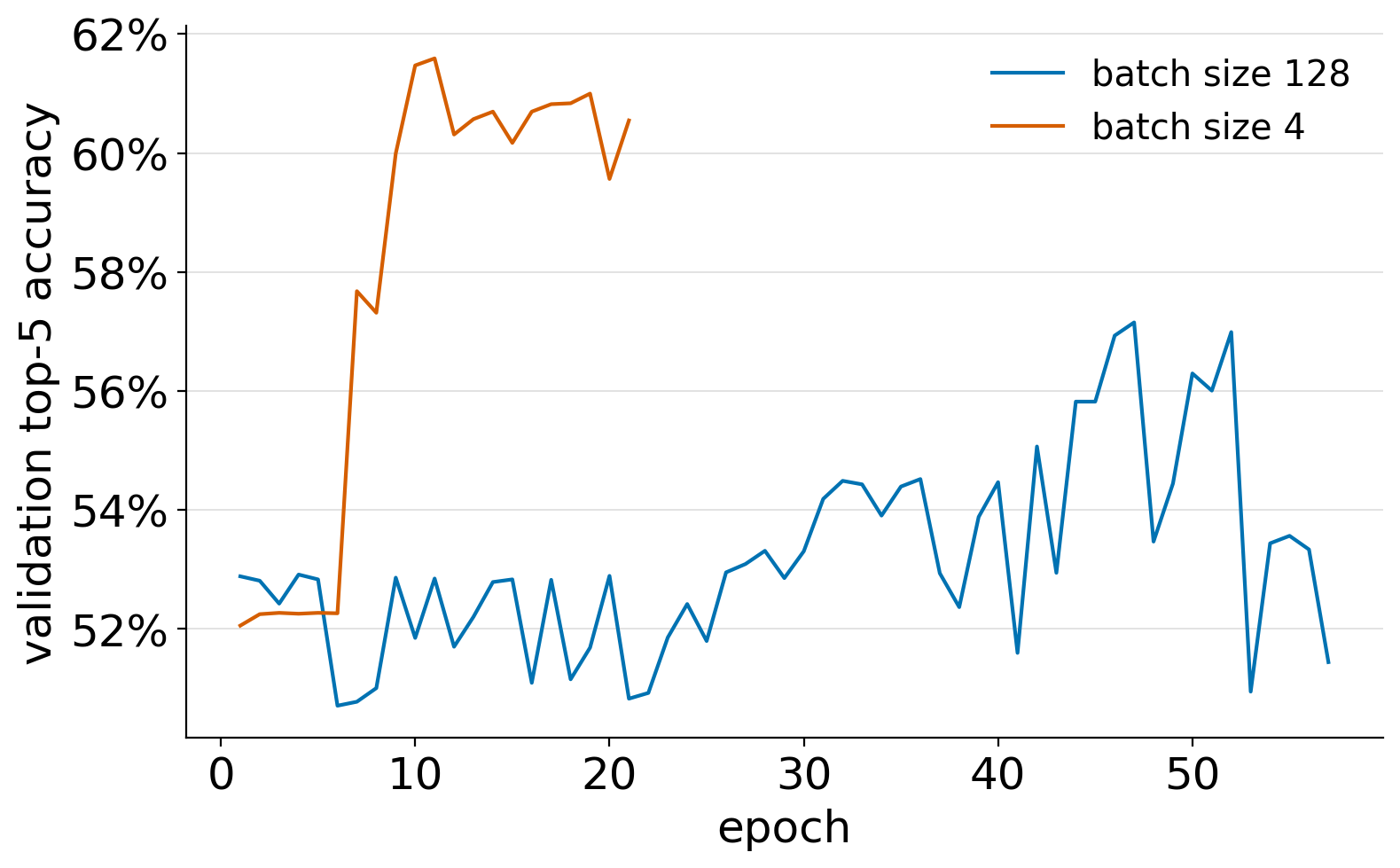} \\

    \end{tabular}
    \caption{BN validation loss (top row) and validation top-5 accuracy (bottom row) over training epochs, comparing the default batch size of 128 against a smaller batch size, on the three densest networks (Baboons, Boston, Copenhagen). On these networks, the training of BN is rather unstable with a batch size of 128; taking a smaller batch size while keeping the learning rate the same improves learning. A possible explanation is that the smaller batch size makes learning more exploratory through noisier gradients, which may help the optimizer find a better optimum on a rugged loss landscape. Validation top-5 accuracy comes from the 80/20 train/validation split used only for the early stopping criterion.}
    \label{fig:btnn_learning_curves_appendix}
\end{figure}

\begin{figure}[t]
    \centering
    \renewcommand{\arraystretch}{0.5}
    \setlength{\tabcolsep}{3pt}
    \begin{tabular}{r ccc c}
        &
        \footnotesize\textbf{$R_0 = 1.5$} &
        \footnotesize\textbf{$R_0 = 2.0$} &
        \footnotesize\textbf{$R_0 = 2.5$} &
        \\[4pt]

        \adjustbox{valign=c}{\rotatebox{90}{\footnotesize\textbf{Baboons}}} &
        \adjustbox{valign=c}{\includegraphics[width=0.275\textwidth]{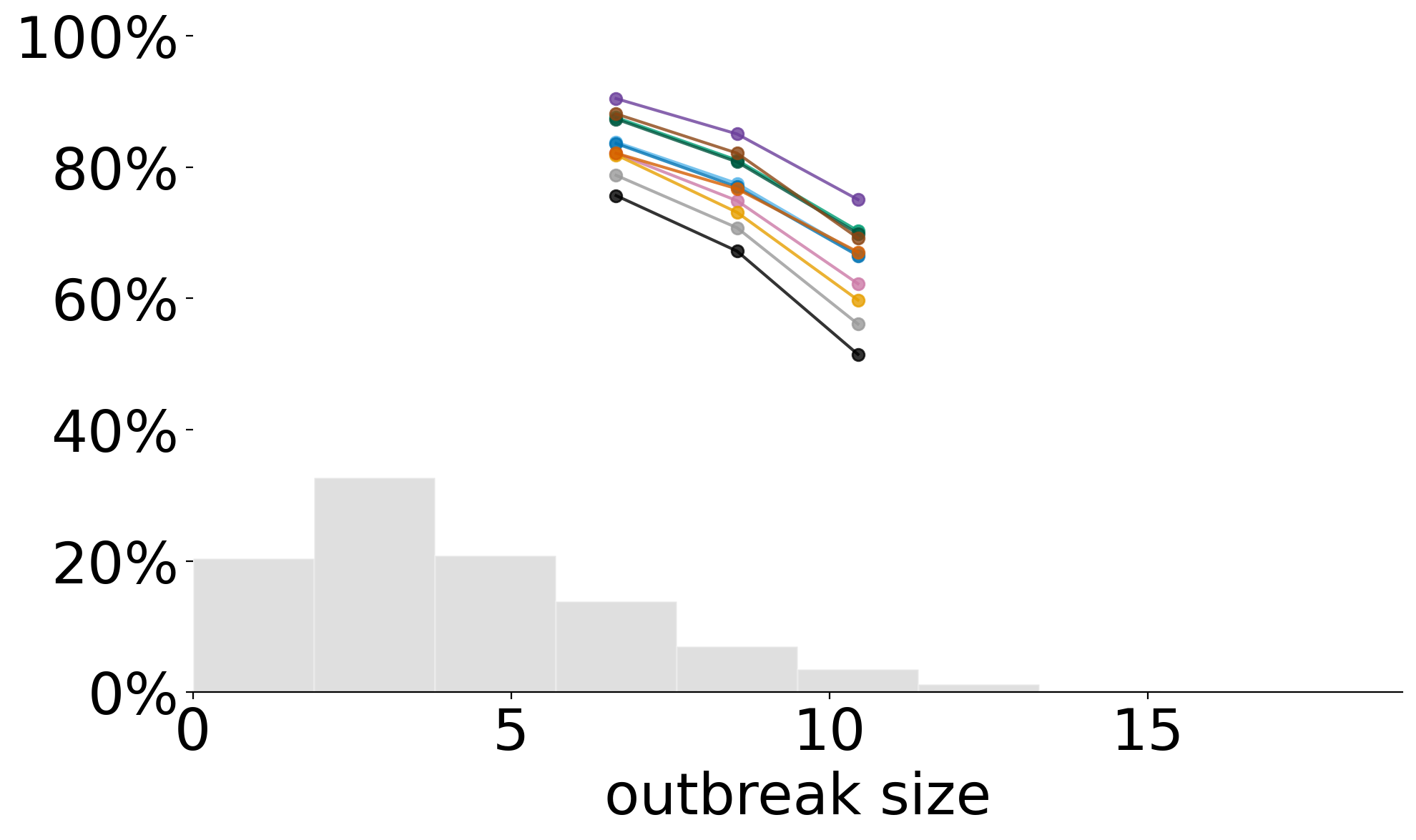}} &
        \adjustbox{valign=c}{\includegraphics[width=0.275\textwidth]{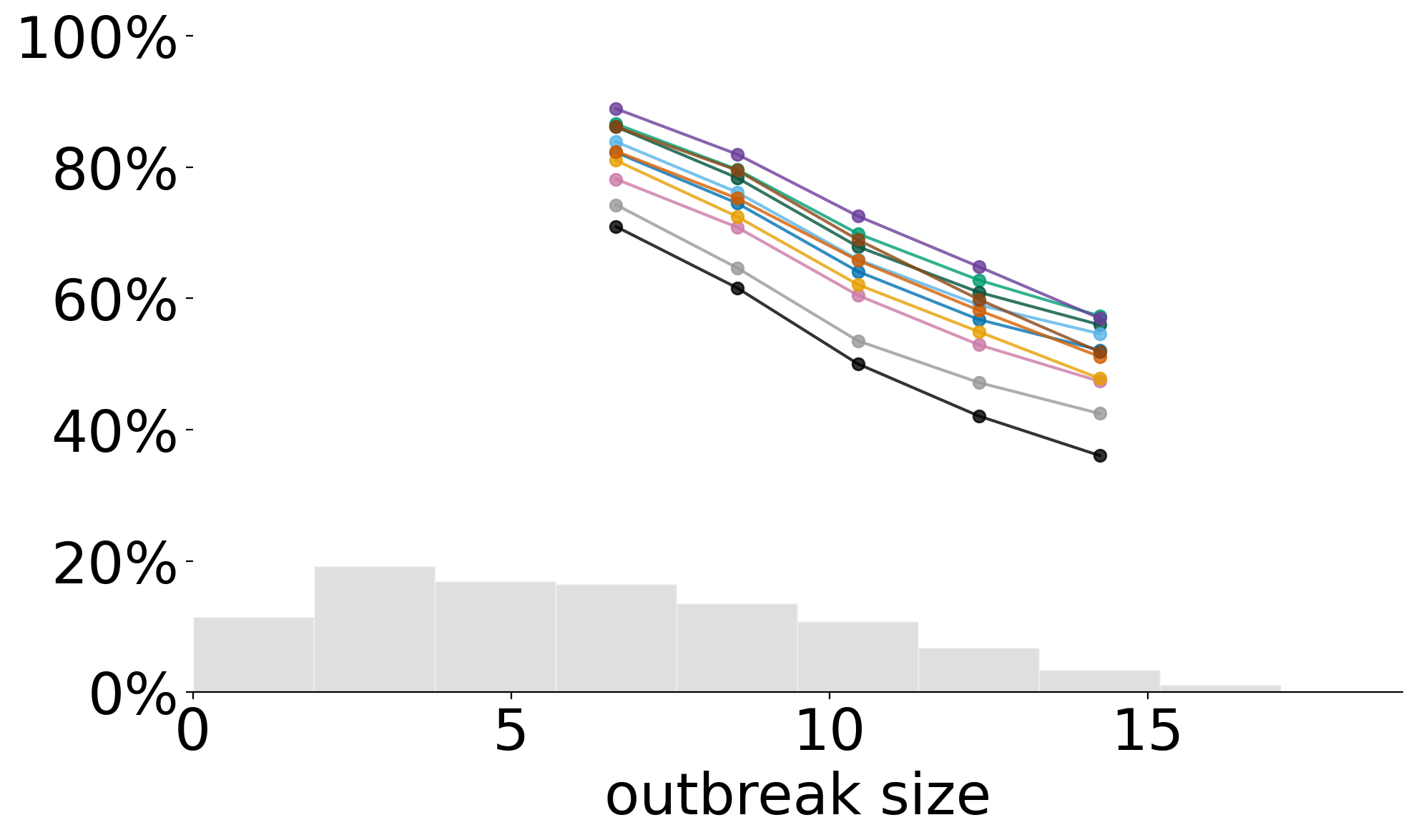}} &
        \adjustbox{valign=c}{\includegraphics[width=0.275\textwidth]{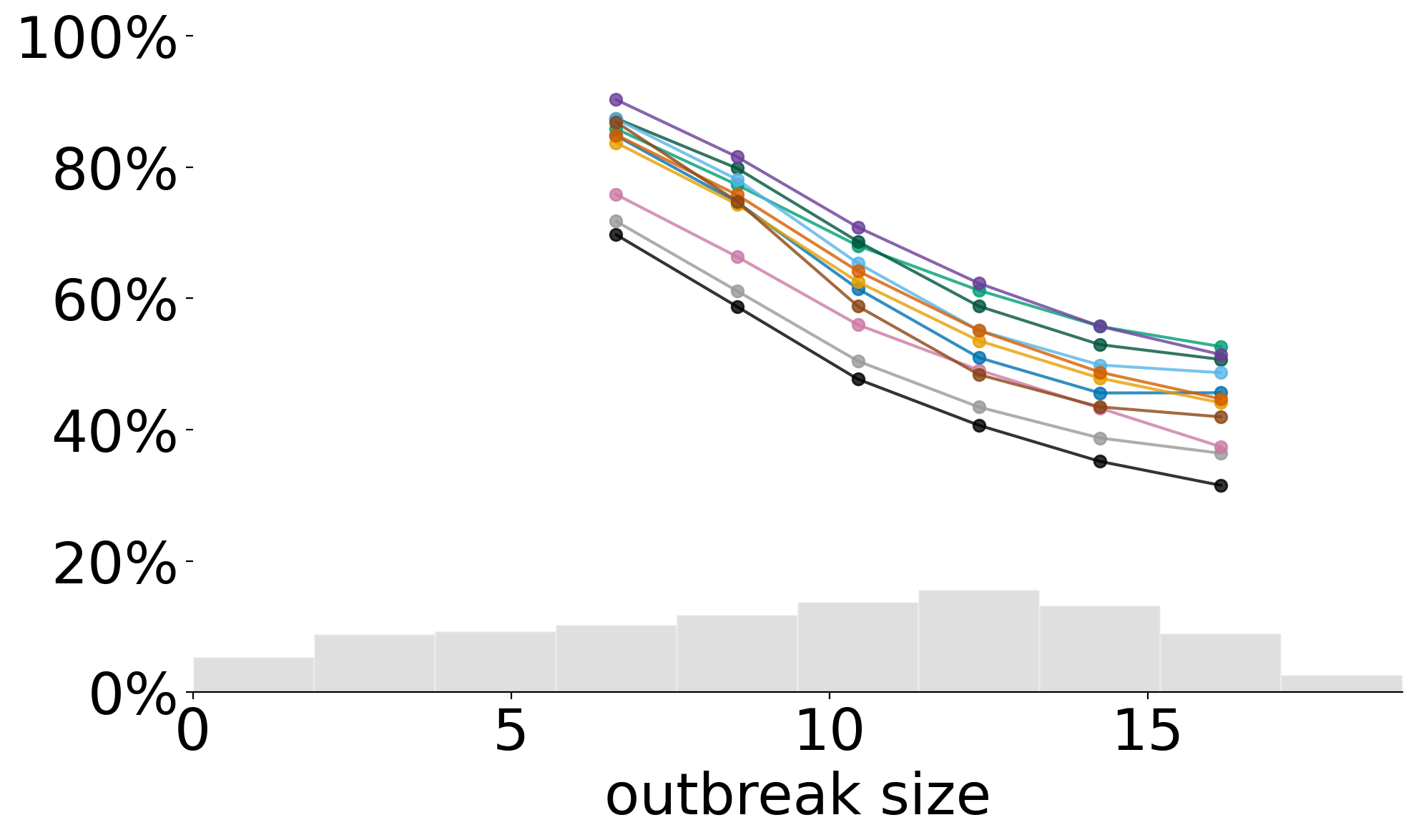}} &
        \makebox[0.0877\textwidth][l]{\raisebox{-0.5\height}[0pt][0pt]{\includegraphics[width=0.0981\textwidth]{plots/exp_2_vary_R0/legend_outbreak_size.png}}} \\[2pt]

        \adjustbox{valign=c}{\rotatebox{90}{\footnotesize\textbf{Malawi}}} &
        \adjustbox{valign=c}{\includegraphics[width=0.275\textwidth]{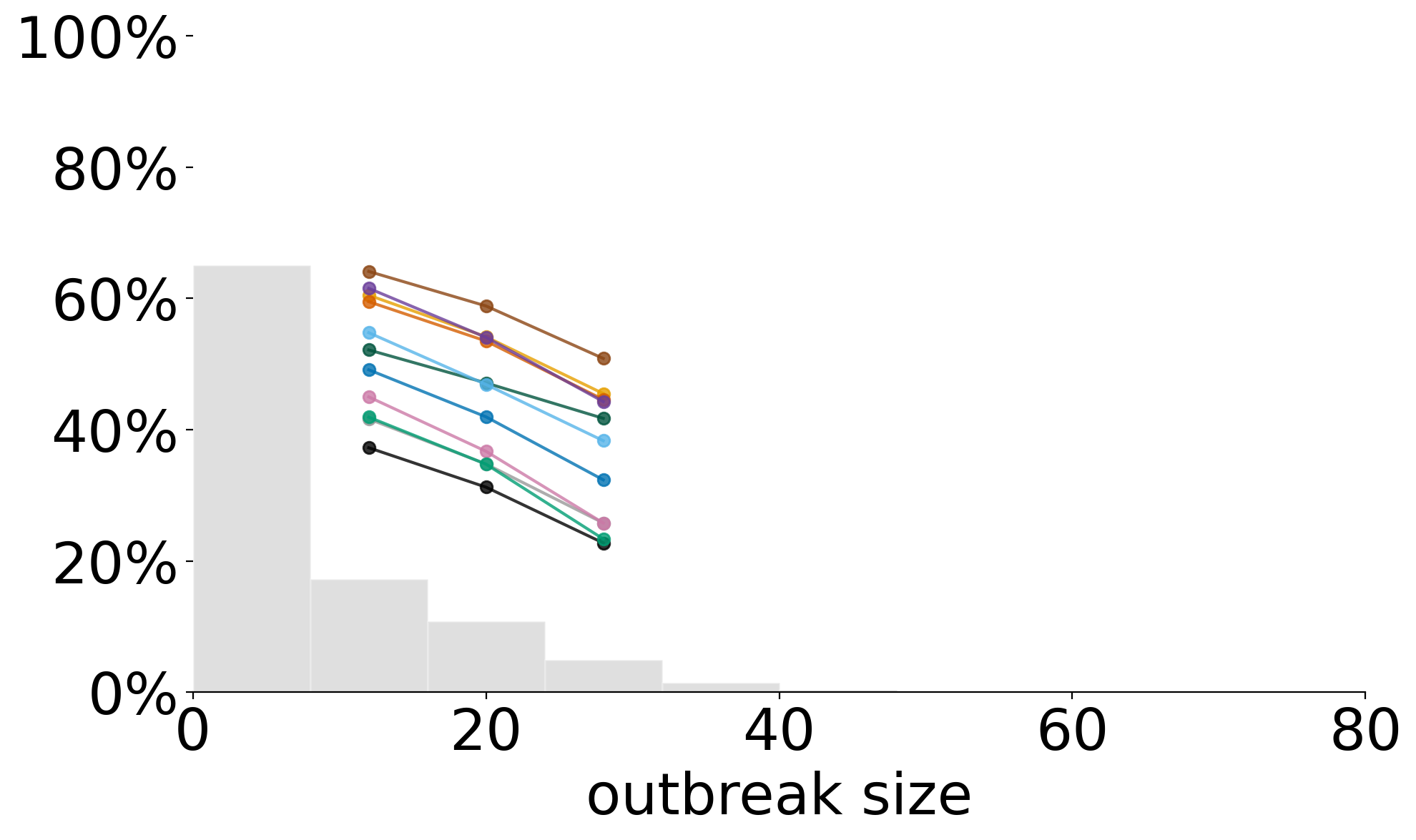}} &
        \adjustbox{valign=c}{\includegraphics[width=0.275\textwidth]{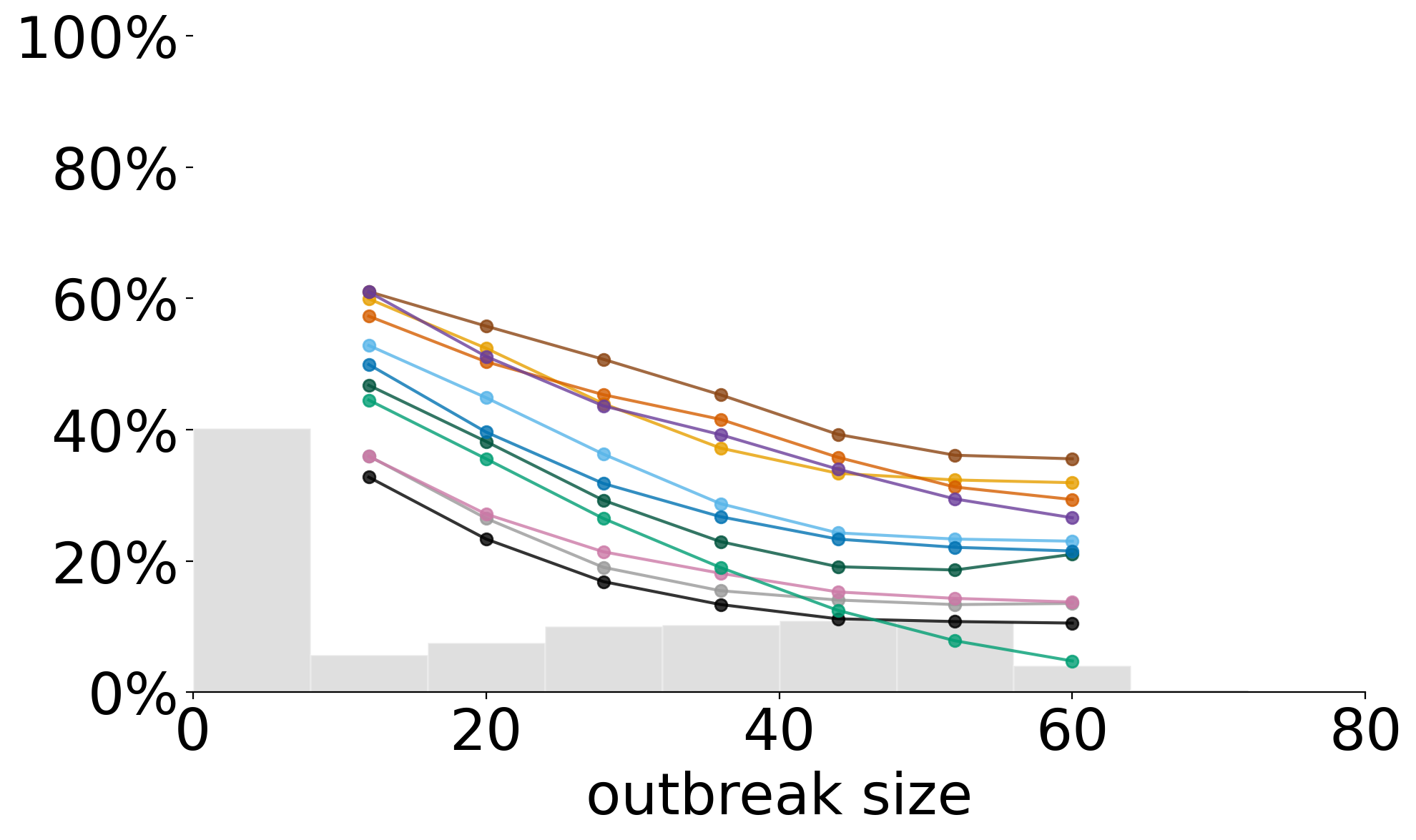}} &
        \adjustbox{valign=c}{\includegraphics[width=0.275\textwidth]{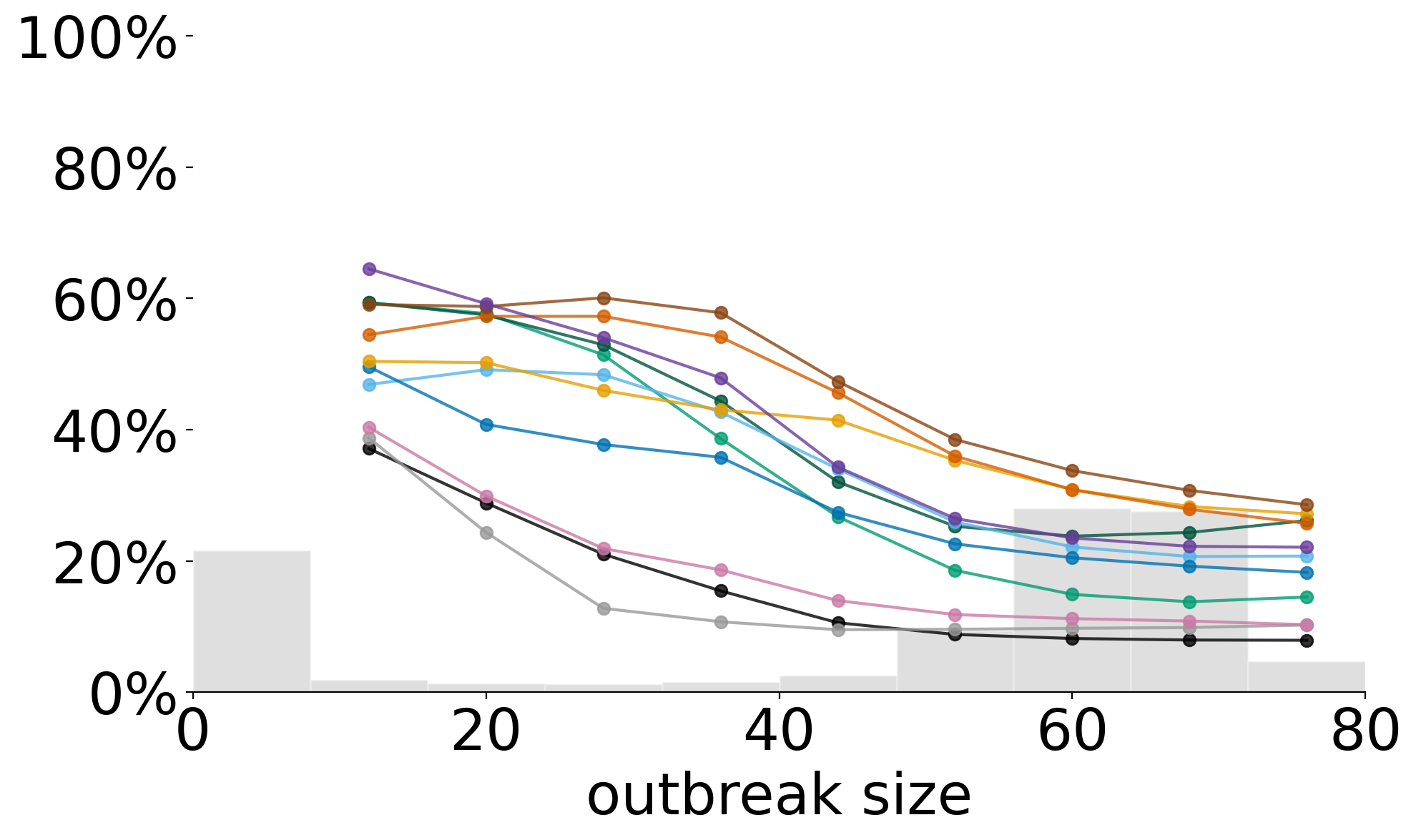}} &
         \\[2pt]

        \adjustbox{valign=c}{\rotatebox{90}{\footnotesize\textbf{Boston}}} &
        \adjustbox{valign=c}{\includegraphics[width=0.275\textwidth]{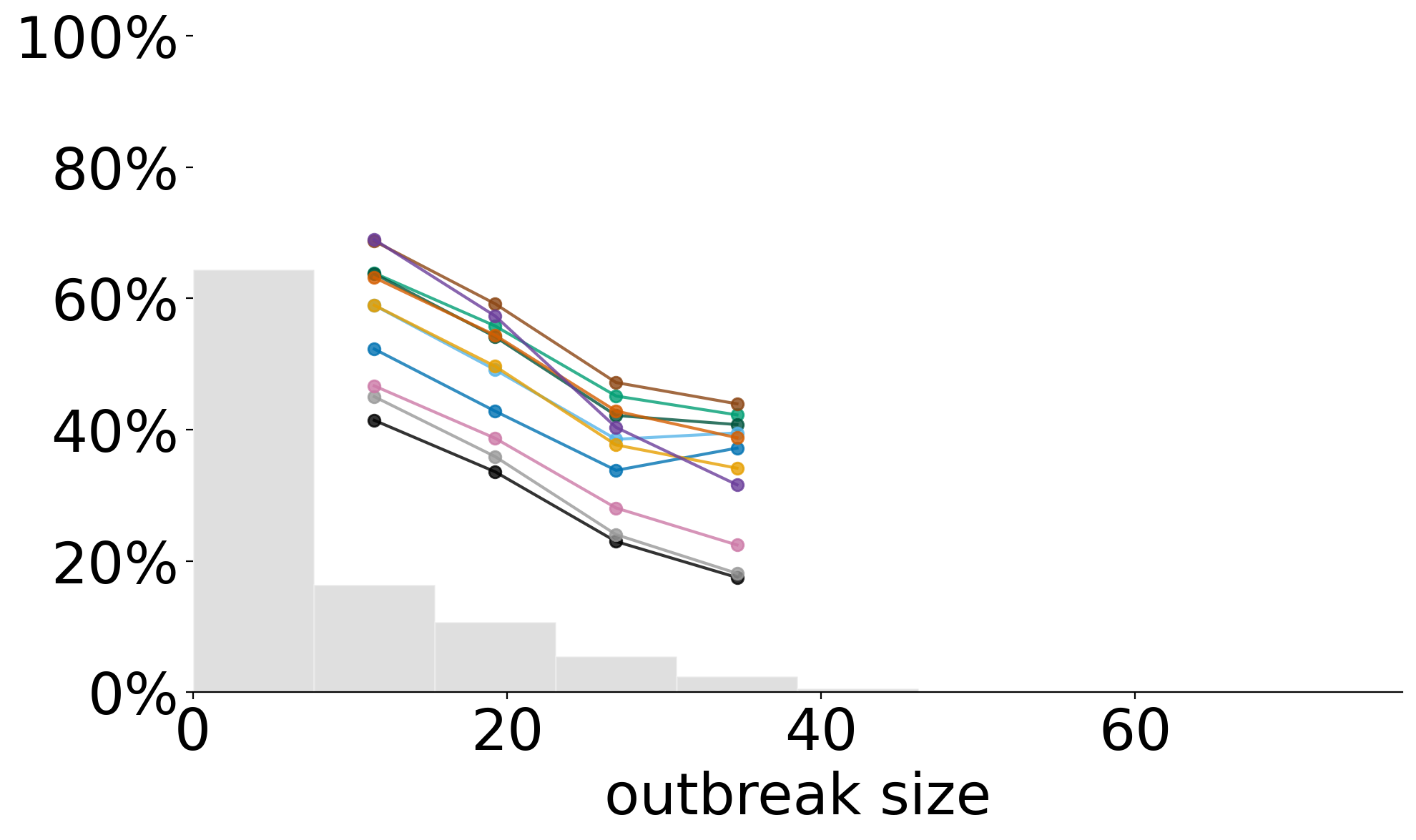}} &
        \adjustbox{valign=c}{\includegraphics[width=0.275\textwidth]{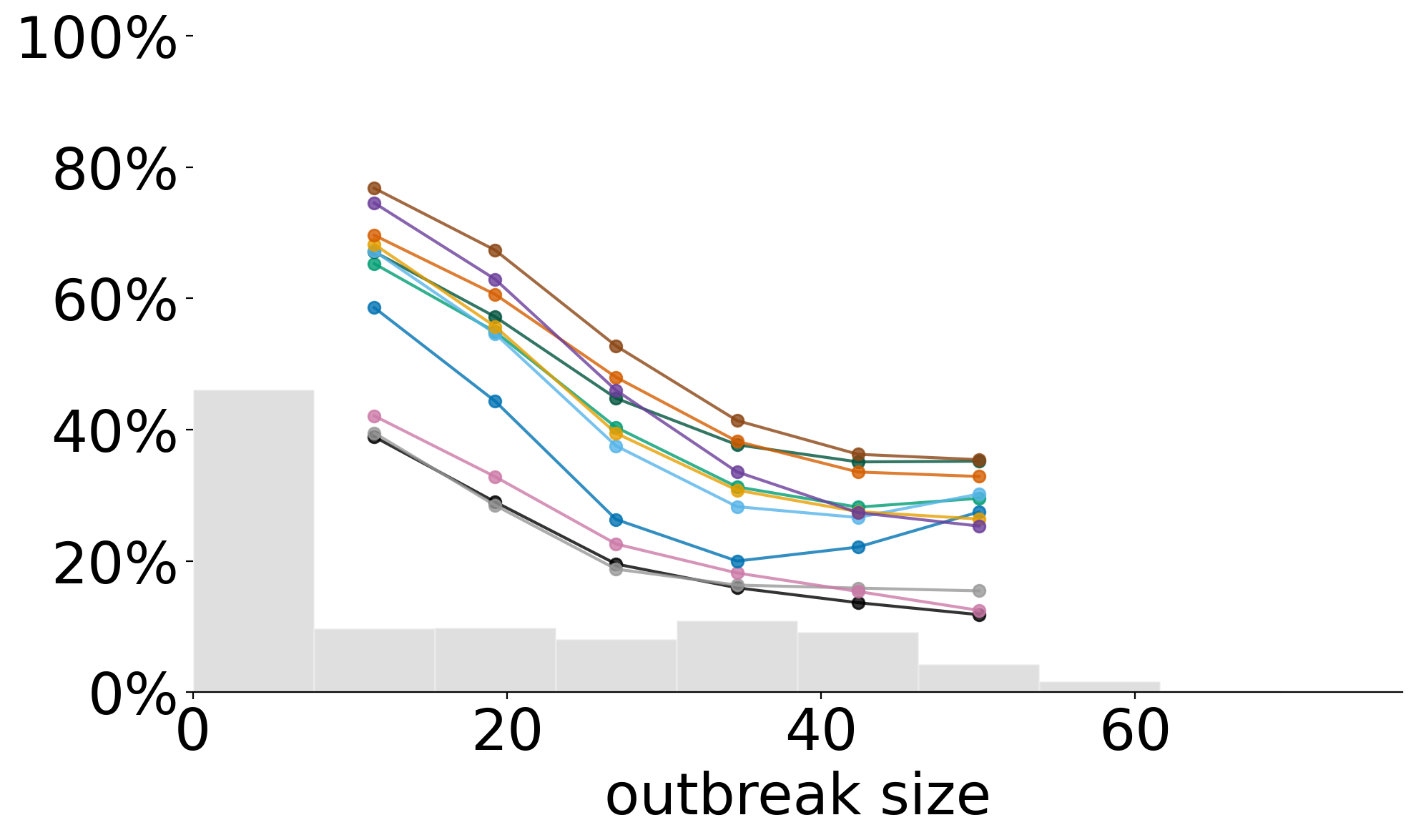}} &
        \adjustbox{valign=c}{\includegraphics[width=0.275\textwidth]{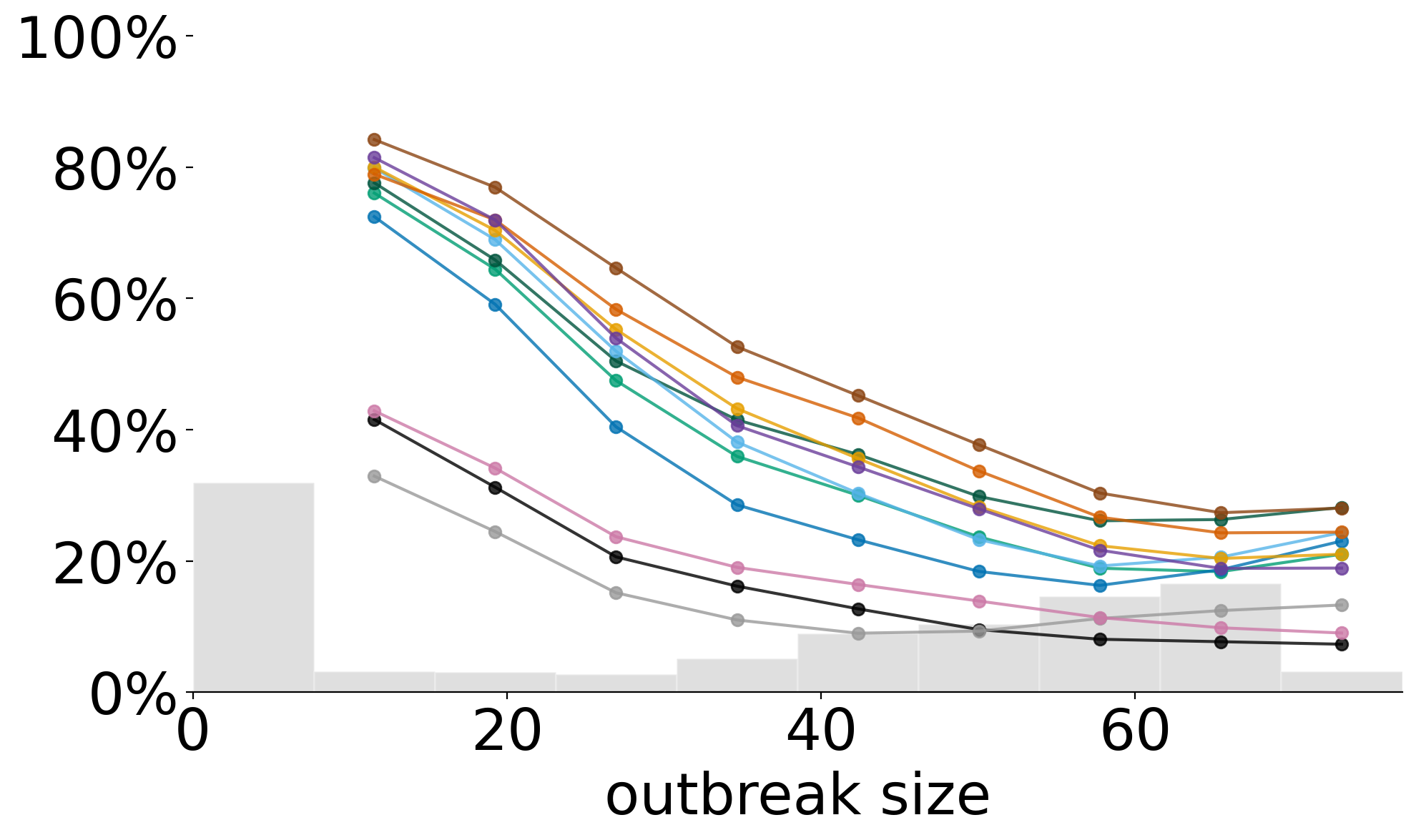}} &
         \\[2pt]

        \adjustbox{valign=c}{\rotatebox{90}{\footnotesize\textbf{New York}}} &
        \adjustbox{valign=c}{\includegraphics[width=0.275\textwidth]{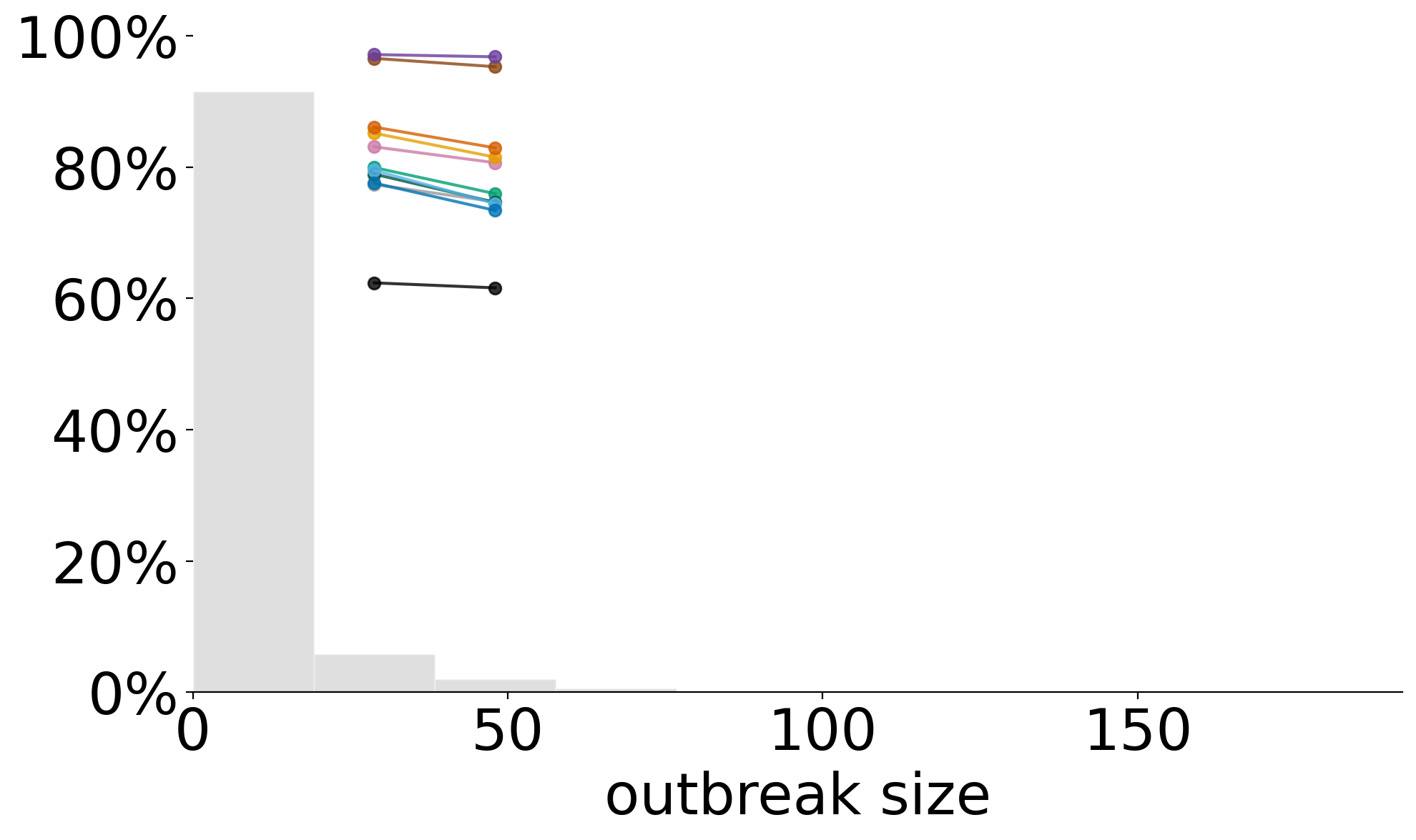}} &
        \adjustbox{valign=c}{\includegraphics[width=0.275\textwidth]{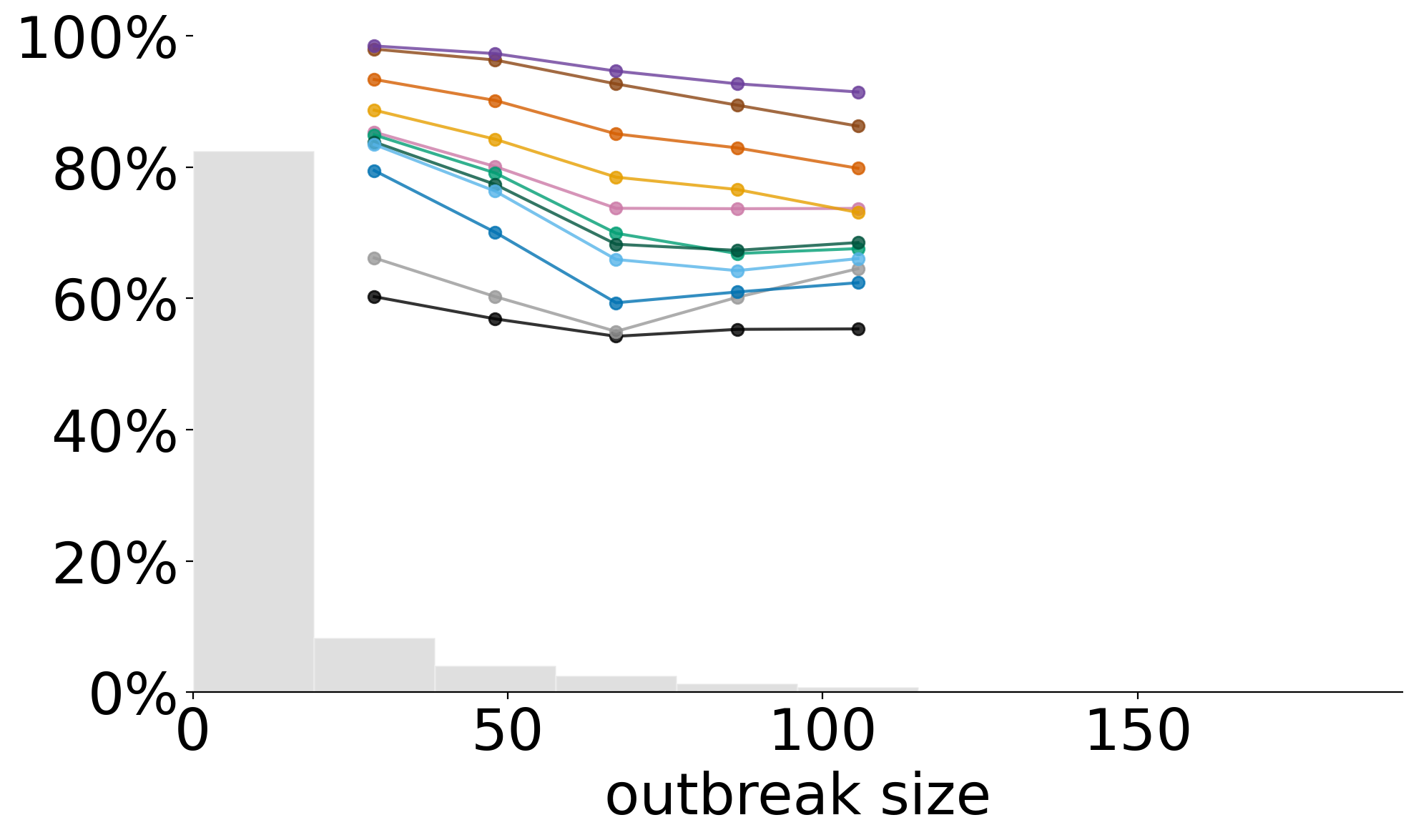}} &
        \adjustbox{valign=c}{\includegraphics[width=0.275\textwidth]{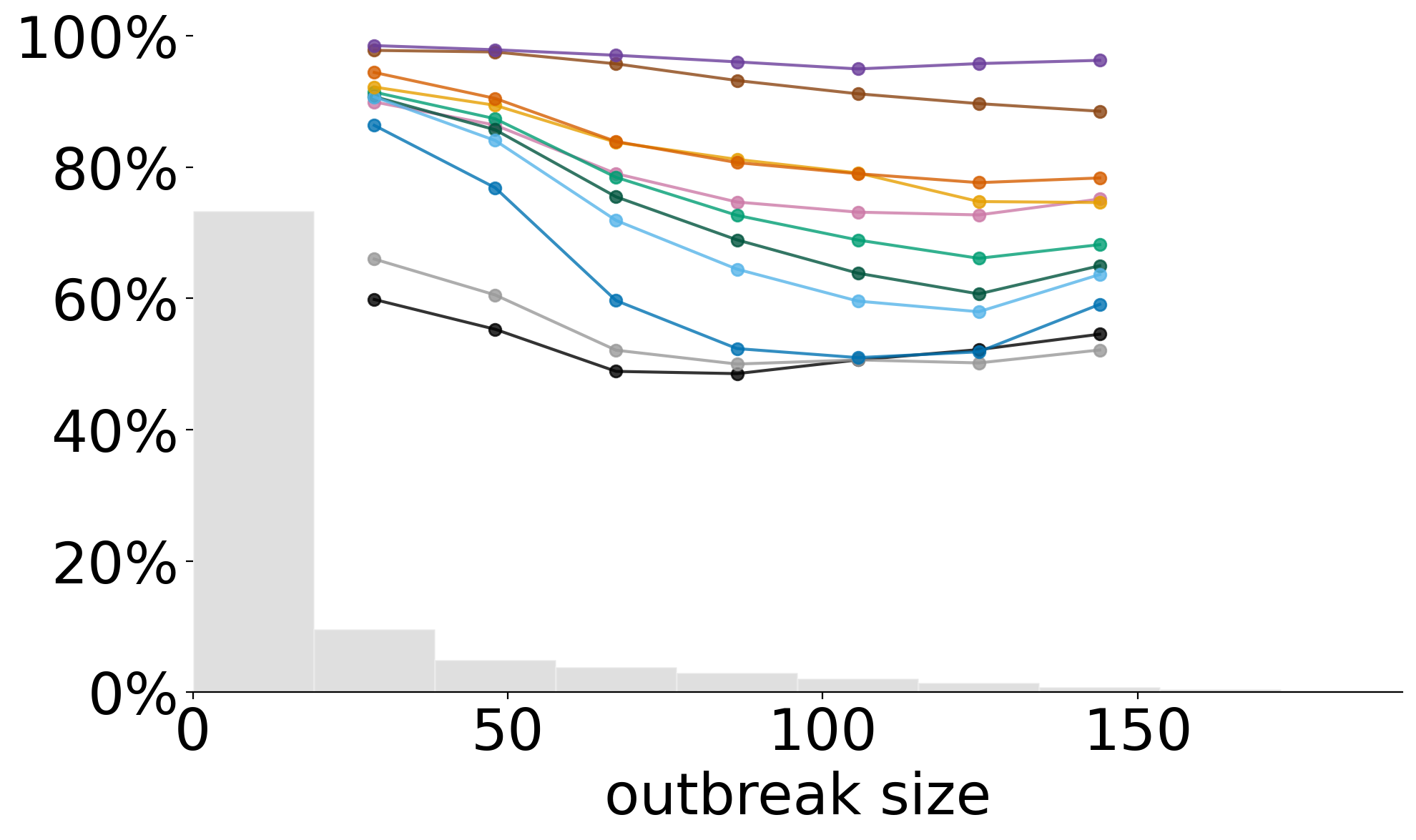}} &
         \\[2pt]

        \adjustbox{valign=c}{\rotatebox{90}{\footnotesize\textbf{Copenhagen}}} &
        \adjustbox{valign=c}{\includegraphics[width=0.275\textwidth]{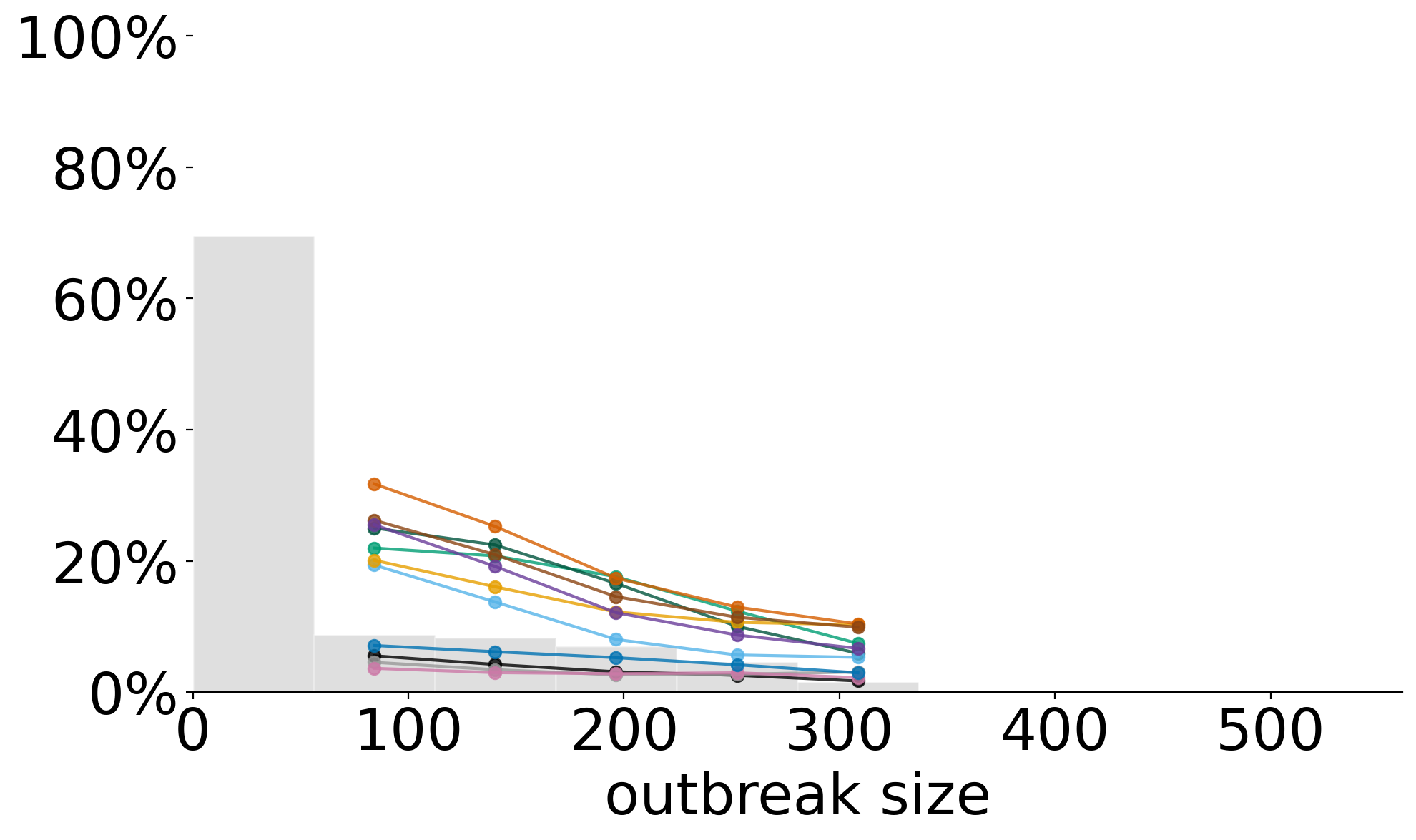}} &
        \adjustbox{valign=c}{\includegraphics[width=0.275\textwidth]{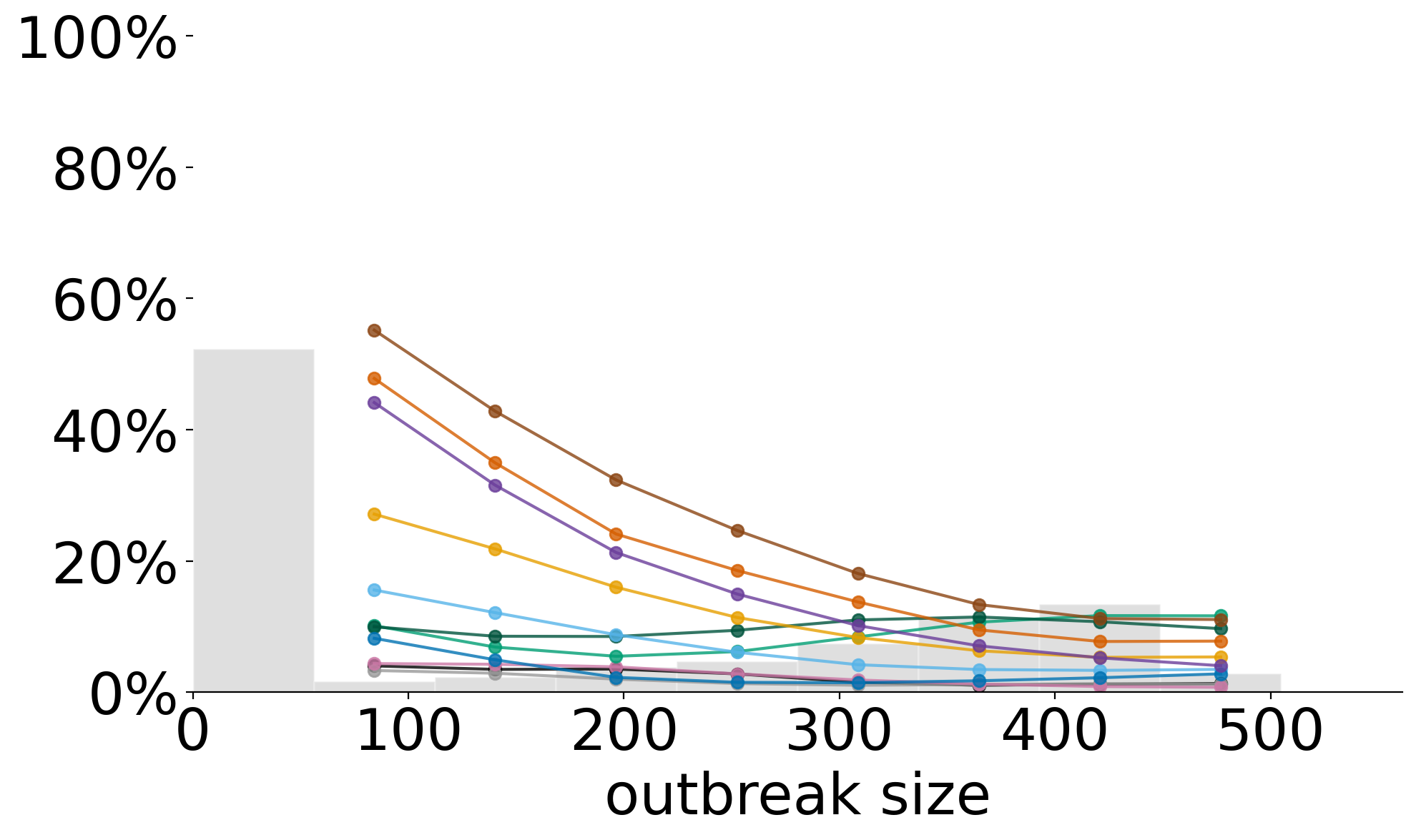}} &
        \adjustbox{valign=c}{\includegraphics[width=0.275\textwidth]{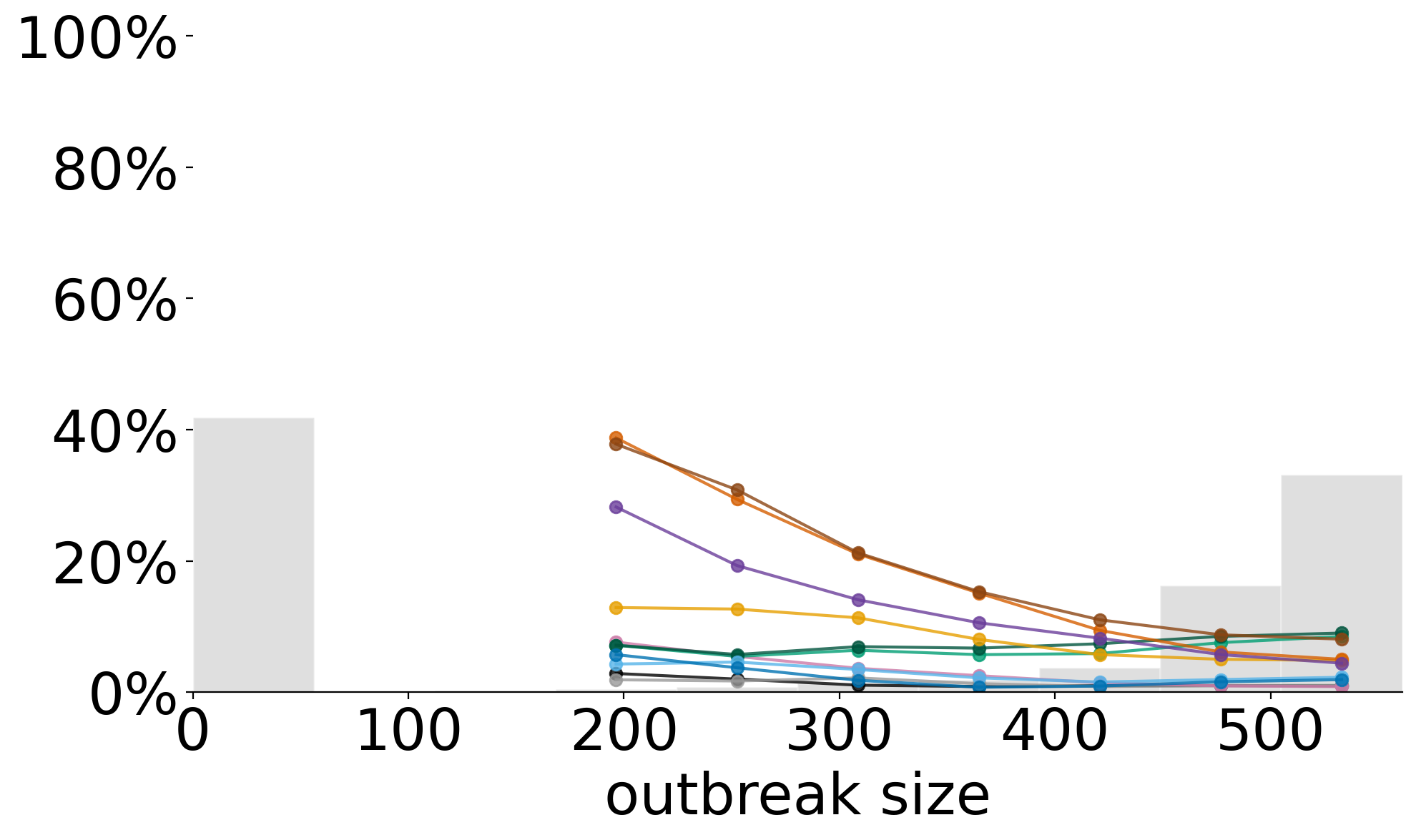}} &
         \\[2pt]

        \adjustbox{valign=c}{\rotatebox{90}{\footnotesize\textbf{Tokyo}}} &
        \adjustbox{valign=c}{\includegraphics[width=0.275\textwidth]{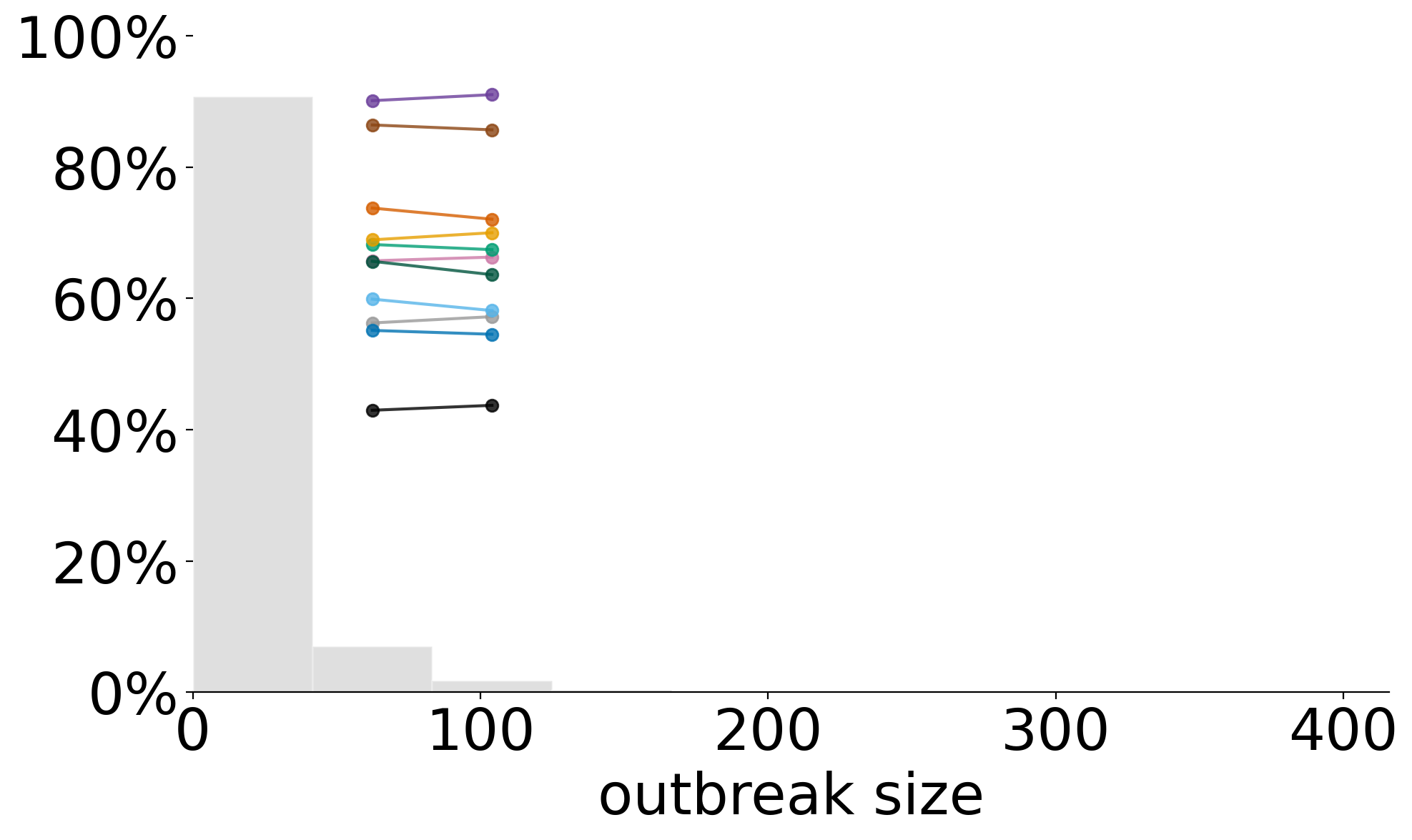}} &
        \adjustbox{valign=c}{\includegraphics[width=0.275\textwidth]{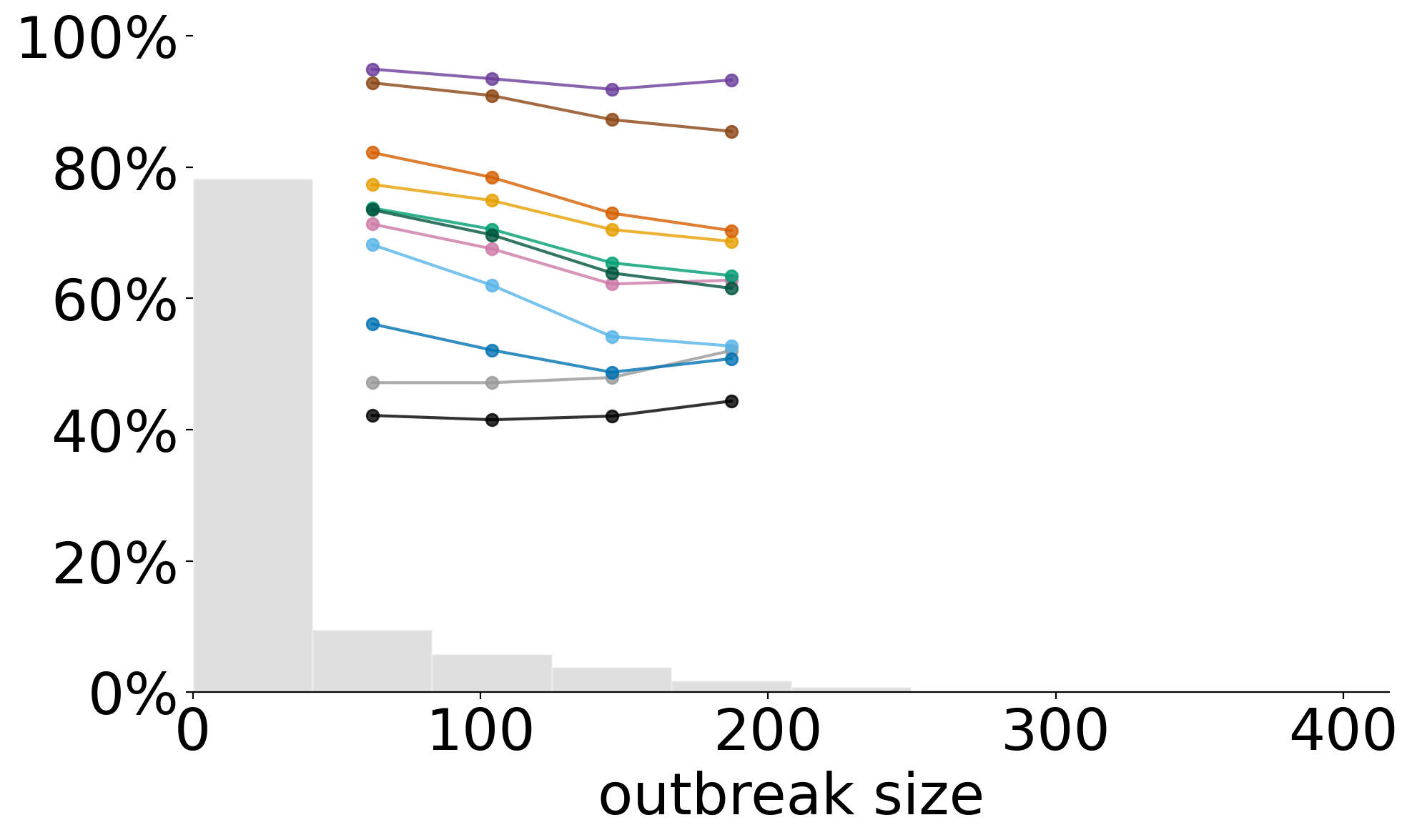}} &
        \adjustbox{valign=c}{\includegraphics[width=0.275\textwidth]{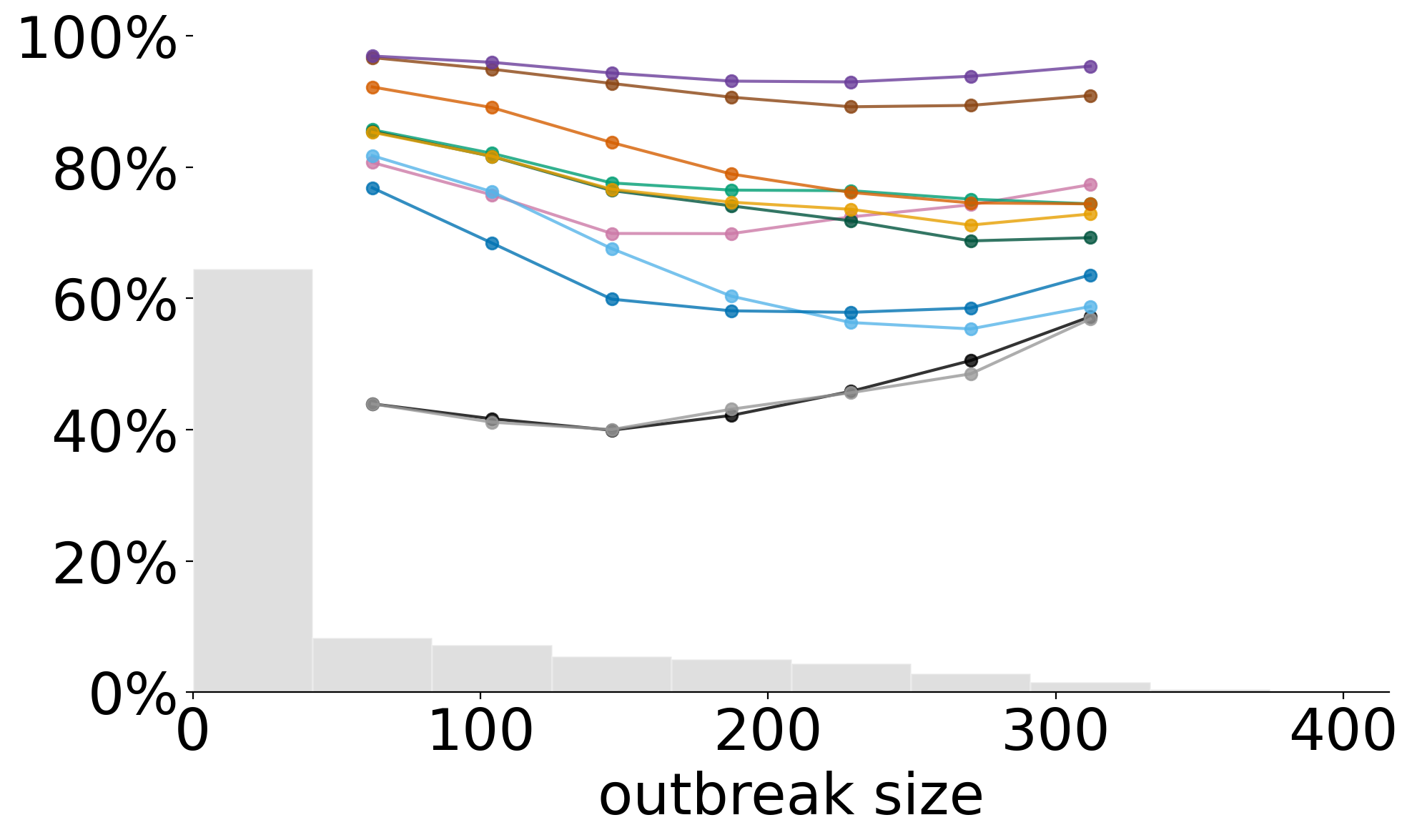}} &
         \\

    \end{tabular}
    \caption{Top-5 accuracy vs.\ outbreak size across contact networks and three values of $R_0$. Lines are smoothed with a three-bin weighted moving average with weights $(\tfrac{1}{4}, \tfrac{1}{2}, \tfrac{1}{4})$. The analogous figures for Escort and Office can be found in the main manuscript.}
    \label{fig:ranks_vs_outbreak_size_appendix}
\end{figure}

\begin{figure}[t]
    \centering
    \includegraphics[width=0.95\linewidth]{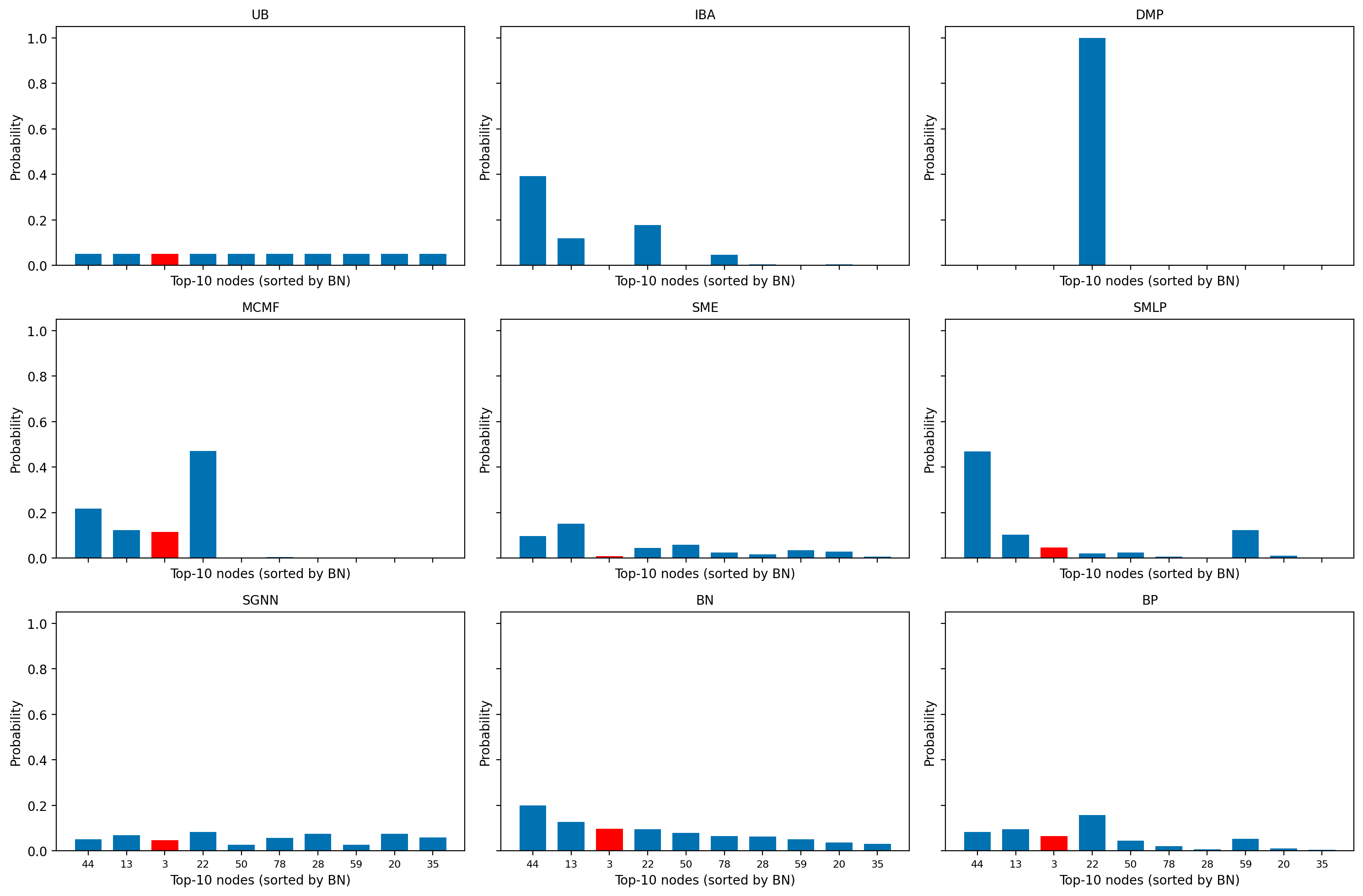}
    \caption{Predicted source probability distribution of each method for one example observed snapshot from the test data on the Malawi network. Nodes are sorted by the probabilities predicted by the Backtracking Network (BN); the true source is node 3, highlighted in red. TJC and ST are the only non-probabilistic methods and are therefore not shown in this figure. The figure illustrates the overconfidence of IBA, DMP, MCMF, and SMLP, which each place most of their probability mass on one or a few nodes.}
    \label{fig:source_probs_appendix}
\end{figure}

\begin{figure}[t]
    \centering
    \includegraphics[width=0.95\linewidth]{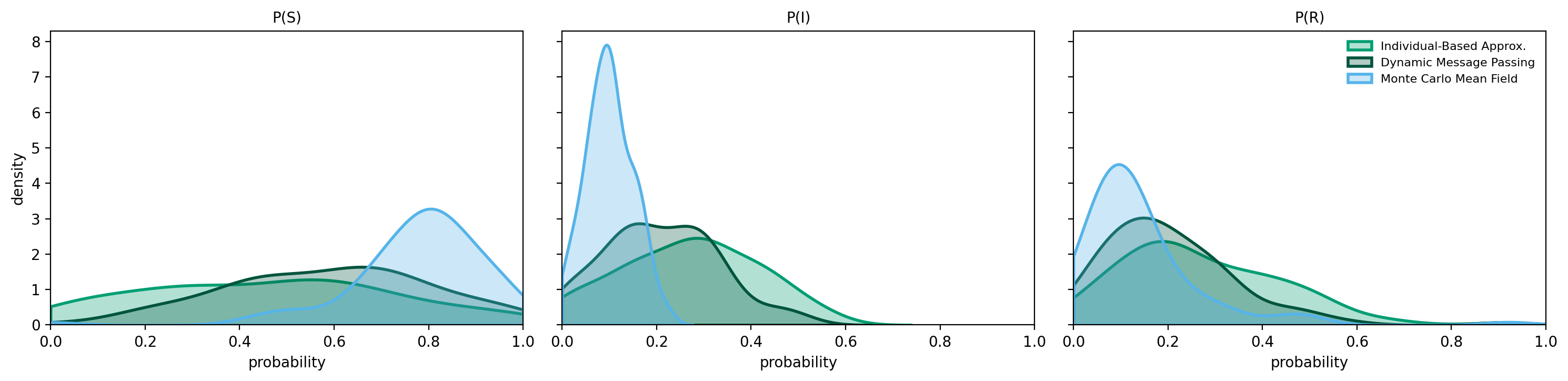}
    \caption{Predicted node-state probability distributions ($S$, $I$, $R$) from the factorized-likelihood methods (IBA, DMP, MCMF) on the Office network, with node 1 as the source. Distributions are smoothed using kernel density estimation. This illustrates the \emph{leakiness effect} discussed in the main experiment section: IBA and DMP tend to assign higher probabilities to the $I$ and $R$ states, whereas MCMF assigns most nodes a high probability of $S$ and lower probabilities of $I$ and $R$.}
    \label{fig:state_density_appendix}
\end{figure}

\end{document}